\documentclass[aps,prd,twocolumn,showpacs,superscriptaddress,groupedaddress,eqsecnum,nofootinbib,floatfix]{revtex4-2}

\usepackage{amsmath}
\usepackage{amsfonts}
\usepackage{amssymb}	
\usepackage{graphicx}
\usepackage{bm}
\usepackage{color}
\usepackage{commath}

\usepackage{graphicx}
\usepackage{dcolumn}

\usepackage{color}
\usepackage{xcolor}
\definecolor{linkcolor}{rgb}{0.0,0.3,0.5}
\usepackage[unicode,colorlinks=true,citecolor=linkcolor,linkcolor=linkcolor,urlcolor=linkcolor]{hyperref}

\usepackage{cleveref}

\graphicspath{{../figures/}}

\def\TEOBResumS{\texttt{TEOBResumS}}

\newcommand{\lso}{ISCO}

\begin{document}

\title{Effective one body Hamiltonian in scalar-tensor gravity at third post-Newtonian order}.

\author{Tamanna \surname{Jain}$^{1}$}
\email{tj317@cam.ac.uk}
\author{Piero \surname{Rettegno}$^{2}$}
\author{Michalis \surname{Agathos}$^{1,3}$}
\author{Alessandro \surname{Nagar}$^{2,4}$}
\author{Lorenzo \surname{Turco}$^{5}$}
\affiliation{${}^1$Department of Applied Mathematics and Theoretical Physics,University of Cambridge,Wilberforce Road CB3 0WA Cambridge, United Kingdom.}%
\affiliation{${}^2$INFN Sezione di Torino, Via P. Giuria 1, 10125 Torino, Italy}
\affiliation{${}^3$Kavli Institute for Cosmology Cambridge,University of Cambridge, Madingley Road CB3 0HA Cambridge, United Kingdom}
\affiliation{$^{4}$Institut des Hautes Etudes Scientifiques, 91440 Bures-sur-Yvette, France}
\affiliation{$^{5}$Dipartimento di Fisica, Universit\'a degli studi di Genova e INFN, Sezione di Genova, I-16146 Genova, Italy}

\date{\today}

\begin{abstract}
	
We determine the general local-in-time effective-one-body (EOB) Hamiltonian for massless Scalar-Tensor (ST) theories 
at third post-Newtonian (PN) order. Starting from the Lagrangian derived in [Phys. Rev. D 99, 044047 (2019)], we map 
it to the corresponding ordinary Hamiltonian describing the two-body interaction in ST theories at 3PN level. 
Using a canonical transformation, we then map this onto an EOB Hamiltonian so as to determine
the ST corrections to the 3PN-accurate EOB potentials $(A,B,Q_e)$ at 3PN\@. We then focus on circular orbits 
and compare the effect of the newly computed 3PN terms, also completed with finite-size and nonlocal-in-time 
contributions, on predictions for the frequency at the innermost stable circular orbit.
Our results will be useful to build high-accuracy waveform models in ST theory, which could be used to perform 
precise tests against General Relativity using gravitational wave data from coalescing compact binaries.
\end{abstract}

\maketitle


\section{\label{sec:intro}Introduction}
Among the gravitational theories alternative to Einstein's General Relativity (GR), scalar-tensor (ST) theories with massless scalar fields are 
those most thoroughly studied and tested~\cite{Damour:1992we,Damour:1993hw,Damour:1995kt,Freire:2012mg,Khalil:2022sii,Gautam:2022cpb}.
Besides their manifestly well-posed nature, the presence of an additional gravitational scalar field arises as a natural feature in theories 
of gravity designed to serve as UV completions of GR\@.
The nonminimally coupled scalar field also gives interesting phenomenology, both in gravitating astrophysical 
systems and in cosmology~\cite{Doneva:2022ewd}.

The recent breakthrough in experimental gravity, with the first direct observation of gravitational waves by the LIGO-Virgo Collaboration in 2015~\cite{Abbott:2016blz}, opened new unexplored pathways towards probing the dynamics of gravity at extreme conditions~\cite{Arun:2006yw,Mishra:2010tp,Li:2011cg,Agathos:2013upa,Cornish:2011ys,Berti:2015itd,Yunes:2016jcc} and led to the first bounds on high-order post-Newtonian coefficients~\cite{TheLIGOScientific:2016src}.
With the sensitivity of the LIGO~\cite{TheLIGOScientific:2014jea}, Virgo~\cite{TheVirgo:2014hva} and now KAGRA~\cite{KAGRA:2020agh} detector network continuously improving~\cite{Aasi:2013wya}, a wealth of detected GW signals emitted by coalescing black-hole and neutron-star binaries has been thoroughly analysed in order to test the strong-field dynamics of GR and probe the nature of the observed compact objects~\cite{LIGOScientific:2019fpa,LIGOScientific:2018dkp,LIGOScientific:2020tif,LIGOScientific:2021sio}.
So far, in their vast majority, these are either null-hypothesis tests or searches for generic GR-violating features (dispersion, non-tensorial polarizations, etc.), due to the lack of accurate and complete waveform models in any gravitational theory alternative to GR, that are reliable from the early binary inspiral all the way to merger.
Nevertheless, much progress has been achieved for a selective set of promising theories, including ST as we shall see in detail below.
With this work, we make the next step towards obtaining a model sufficiently accurate for performing apples-to-apples comparisons between ST and GR\@.
This is already important for reliably interpreting observational bounds using GW data from the current network of detectors, and will become even more crucial for probing the strong-field dynamics of gravity to much greater accuracy with the next generation of detectors, such as the Einstein Telescope~\cite{Maggiore:2019uih} and Cosmic Explorer~\cite{Evans:2021gyd}.

For ST theories in particular, although currently the most stringent constraints come from binary pulsar observations, studying the effects of scalarized neutron stars on the orbital evolution~\cite{Freire:2012mg},
there is hope that future gravitational wave (GW) detections of coalescing compact 
binaries will complement current knowledge by placing additional constraints on the 
ST parameters using genuine strong-field information from the binary inspiral.
GWs from compact binary inspirals may also reveal effects of ST gravity in scenarios where the 
scalarization process is suppressed for weakly gravitating systems (therefore circumventing binary 
pulsar bounds) but still arises during the late stages of the inspiral where gravity is strong, a phenomenon known as dynamical scalarization~\cite{Palenzuela:2013hsa,Sennett:2016rwa,Khalil:2022sii}.

The interpretation of detected GW signals relies on Bayesian data analysis techniques, where the theoretical prediction of a waveform is matched against the data.
To this aim, there is an increasing effort in improving the analytical knowledge of the two-body problem in ST theories, both for what concerns the dynamics~\cite{Lang:2013fna,Lang:2014osa,Bernard:2018hta,Bernard:2018ivi,Bernard:2019yfz,Schon:2021pcv} and the waveform generation~\cite{Sennett:2016klh,Bernard:2022noq}
through post-Newtonian (PN) theory.
Finite-size effects in ST theories further modify the binary dynamics by the contribution of dipole-sourced scalar radiation to the outgoing energy flux~\cite{Bernard:2019yfz}.
In addition, tidal Love numbers might be very different from their GR
counterparts~\cite{Brown:2022kbw}.
This can eventually impact the measurement of tidal polarizability
and related constraints put on the equation of state of cold matter at extreme densities~\cite{Hinderer:2009ca,Damour:2012yf,Agathos:2015uaa,Abbott:2018exr,Gamba:2020ljo}.

To robustly describe the binary dynamics and waveform in strong field up to merger, PN results should
be recast within the effective-one-body (EOB) description of the two-body problem~\cite{Buonanno:1998gg,Buonanno:2000ef,Damour:2000we,Damour:2008qf,Damour:2014jta,Damour:2015isa,Damour:2016abl}.
The generalization of the EOB method to ST theories at the 2PN level has been recently worked out~\cite{Julie:2017ucp,Julie:2017pkb}.
The aim of this paper is to extend the results of Refs.~\cite{Julie:2017ucp,Julie:2017pkb} to 3PN order
building upon the 3PN Lagrangian in ST of~\cite{Bernard:2018hta,Bernard:2018ivi}.
Additionally during this work, the authors became aware of an independent parallel effort by a different group on 
the calculation performed here, which is due to appear shortly~\cite{Julieprep}.

The paper is organized as follows. In Sec.~\ref{sec:STReminder} we briefly recall the definition of massless ST theories.
Then, in Sec.~\ref{sec:3PNLag} we derive the order-reduced Lagrangian for ST theory at 3PN, and from this we  obtain the center 
of mass ordinary Hamiltonian in Sec.~\ref{sec:COM3PNH}. Finally, in Sec.~\ref{sec:ST3PN-EOB}, we map the ordinary Hamiltonian
into an EOB Hamiltonian at 3PN order and in Sec.~\ref{sec:isco} we explore the relevance of the 3PN ST terms
(including nonlocal and finite-size contributions) by studying the innermost stable circular orbit ({\lso}).
We use geometric units throughout the paper, with $c=G=1$.

\section{Scalar-Tensor Theory Reminder}
\label{sec:STReminder}
We consider mono-scalar massless ST theories and mostly adopt the notations
and conventions of Damour and Esposito-Farese (DEF hereafter, see Table~\ref{LauravsDEF})~\cite{Damour:1992we,Damour:1995kt}.
The theory is defined by the following action in the Einstein frame,
\begin{align}
S&=\frac{c^4}{16 \pi G_{*}}\int d^4x \sqrt{-g}(R-2g^{\mu \nu} \partial_\mu \varphi \partial_\nu \varphi)\nonumber\\
  &\qquad\qquad\qquad\qquad+S_m[\Psi, {\mathcal{A}(\varphi)}^2g_{\mu\nu}]~,
\end{align}
where $g_{\mu\nu}$ is the Einstein metric, $R$ is the Ricci scalar, $\varphi$ is the scalar field, $\Psi$ collectively denotes the matter fields, $g \equiv \det(g_{\mu\nu})$ and $G_*$ is the bare Newton's constant.
In the Einstein frame, the scalar field is minimally coupled to the Einstein metric $g_{\mu\nu}$, 
and the dynamics of the latter is governed only by the usual Einstein-Hilbert action.
The dynamics of the scalar field arises from its coupling to the matter fields $\Psi$.
The scalar field couples non-minimally to the metric in the Jordan frame (physical frame),
\begin{equation}
\tilde{g}_{\mu\nu}={\mathcal{A}(\varphi)}^2 g_{\mu\nu} \ ,
\end{equation}
where $\tilde{g}_{{\mu\nu}}$ is the metric in Jordan frame. The function $\mathcal{A}(\varphi)$ uniquely fixes the 
ST theory, and General Relativity is recovered when $\mathcal{A}(\varphi)=cst$. 
The Einstein frame field equations can be found in Ref.~\cite{Damour:1992we}. The parameter 
\begin{equation}
\alpha(\varphi)=\frac{\partial \ln \mathcal{A}}{\partial \varphi} \ ,
 \end{equation}
arising in the equations of motion measures the coupling between the matter and the scalar field.
For compact, self-gravitating objects in ST theories, we follow the approach 
suggested by~\cite{1975ApJ...196L..59E} to ``skeletonize'' the extended bodies as point particles.
The skeletonized matter action is then given by
\begin{equation}
S_{m}=-\sum_{I=A,B}\int \sqrt{-\tilde{g}_{\mu \nu}\frac{dx^{\mu}}{d\lambda}\frac{d x^{\nu}}{d\lambda}} \tilde{m}_{I}(\varphi)~,
\end{equation}
where the Jordan frame mass $\tilde{m}_I(\varphi)$ of body $I$ is dependent on the local value of the scalar field, and $\lambda$
is the affine parameter.
Since $ \tilde{g}_{\mu\nu}={\mathcal{A}(\varphi)}^2 g_{\mu\nu}$, the Einstein-frame mass is defined as
\begin{equation}
m(\varphi)=\mathcal{A}(\varphi)\tilde{m}(\varphi) \ .
\end{equation}
The mass function $m(\varphi)$ is used to define dimensionless body-dependent parameters to
encompass the scalar field effect~\cite{Damour:1992we,Damour:1995kt,Julie:2017pkb} \ i.e.
\begin{align}
\alpha_I&=\frac{d\ln m(\varphi)_I}{d\varphi},\\
\beta_I&=\frac{d\alpha_I}{d\varphi},\\
\beta'_I&=\frac{d\beta_I}{d\varphi},\\
\beta''_I&=\frac{d\beta'_I}{d\varphi}~.
\end{align} 
Reference~\cite{Bernard:2018hta}, defines the two-body Lagrangian ST parameters at 3PN in the Jordan frame.  
Table~\ref{LauravsDEF} converts these Jordan-frame parameters into the Einstein-frame ones (DEF conventions),
and the corresponding ones used in this paper.

\begin{table}[t]
\begin{ruledtabular}
\caption{\label{LauravsDEF} Relation between the ST parameters used in the two-body Lagrangian of Ref.~\cite{Bernard:2018hta},
  the DEF ones and the slightly simplified notation that we are using here. 
  The index ``0'' signifies a quantity evaluated at $\varphi=\varphi_0$, where $\varphi_0$ is the asymptotic constant value of scalar field.  }
   \centering
   \renewcommand{\arraystretch}{1.7}
\begin{tabular}{c l c}
   LB\cite{Bernard:2018hta} & DEF~\cite{Damour:1992we,Damour:1995kt} & This paper \\
  \hline
  $m_{1}$ & $m^0_{A}/\mathcal A_0$ & $ m^0_{A}/\mathcal A_0 \equiv \tilde{m}^0_{A}$ \\
  $m_{2}$ & $m^0_{B}/\mathcal A_0$ & $ m^0_{B}/\mathcal A_0 \equiv \tilde{m}^0_{B}$ \\
  $\alpha$ & $\frac{1+\alpha^0_A\alpha^0_B}{1+\alpha_0^2}$ & $\alpha_{AB}$\\
  $\tilde{G}\alpha$ & $(1+\alpha^0_A\alpha^0_B)\mathcal A^2_0\equiv G_{AB}\mathcal A_0^2$ & $G_{AB}\mathcal A_0^2\equiv\tilde{G}_{AB}$ \\
  $\bar{\gamma}$ & $-2\frac{\alpha^0_A\alpha^0_B}{1+\alpha^0_A\alpha^0_B}\equiv\bar{\gamma}_{AB}$ & $\bar{\gamma}_{AB}$ \\
  $\bar{\beta}_1$ & $\frac{1}{2}\frac{(\beta_A\alpha_B^2)_0}{(1+\alpha_A^0\alpha_B^0)^2}\equiv\bar{\beta}^A_{BB}$ & $\bar{\beta}_A$\\
  $\bar{\beta}_2$ & $\frac{1}{2}\frac{(\beta_B\alpha_A^2)_0}{(1+\alpha_A^0\alpha_B^0)^2}\equiv\bar{\beta}^B_{AA}$ & $\bar{\beta}_B$\\
  $\bar{\delta}_1$ & $\frac{(\alpha_A^0)^2}{(1+\alpha_A^0\alpha_B^0)^2}$  & $\delta_A$ \\
  $\bar{\delta}_2$ & $\frac{(\alpha_B^0)^2}{(1+\alpha_A^0\alpha_B^0)^2}$  & $\delta_B$\\
  $\bar{\chi}_1$ & $-\frac{1}{4}\frac{(\beta'_A\alpha_B^3)_0}{(1+\alpha_A^0\alpha_B^0)^3}\equiv-\frac{1}{4}\epsilon^A_{BBB}$ & $-\frac{1}{4}\epsilon_A$\\
  $\bar{\chi}_2$ & $-\frac{1}{4}\frac{(\beta'_B\alpha_A^3)_0}{(1+\alpha_A^0\alpha_B^0)^3}\equiv-\frac{1}{4}\epsilon^B_{AAA}$ & $-\frac{1}{4}\epsilon_B$\\
  $\bar{\beta}_1\bar{\beta}_2/\bar{\gamma}$ & $-\frac{1}{8}\frac{\beta^0_A\alpha^0_A\beta^0_B\alpha^0_B}{(1+\alpha_A^0\alpha_B^0)^3}\equiv-\frac{1}{8}\zeta_{ABAB}$ & $-\frac{1}{8}\zeta$\\
  $\bar{\kappa}_1$& $\frac{(\alpha^4_B\beta''_A)_0}{8(1+\alpha_A^0\alpha_B^0)^4}$ &$\bar{\kappa}_{A}$\\
    $\bar{\kappa}_2$&  $\frac{(\alpha^4_A\beta''_B)_0}{8(1+\alpha_A^0\alpha_B^0)^4}$  &$\bar{\kappa}_{B}$\\
\end{tabular}
\end{ruledtabular}
\end{table}
\section[The Two-Body 3PN Hamiltonian]{Scalar-Tensor 3PN Ordinary Lagrangian}
\label{sec:3PNLag}
The ST two-body 3PN Lagrangian was obtained in Ref.~\cite{Bernard:2018ivi} in harmonic coordinates. As such, it depends
(linearly) on the acceleration of the two bodies. In this Section, we transform the Lagrangian of~\cite{Bernard:2018ivi} into
an ordinary Lagrangian, that only depends on positions and velocities. To do so, we can either boldly replace the acceleration 
by the equations of motion, or use the contact transformation (modulo a total time derivative which is irrelevant in a Lagrangian)~\cite{Schafer:1984mr}.
Here we choose the second approach, i.e.\ we use the contact transformation to make a 4-dimensional coordinate change
to eliminate the acceleration dependence. The contact transformation at 3PN order is based on the algorithm presented 
in Ref.~\cite{Damour:1990jh}, constructed so to eliminate higher order derivative terms from the Lagrangian. 
As the dependence on accelerations starts at 2PN order, the contact transformation also starts at 2PN order. 
However, the 2PN method presented in~\cite{Julie:2017pkb} for ST theory can not be extended at the 3PN because 
of the presence of the acceleration dependent terms in the functional derivative $\delta \mathcal{L}/\delta Z_{A}^i$ 
of the contact transformation. By contrast, following Ref.~\cite{deAndrade:2000gf}, we have to introduce some sort of  
`counter term' at 3PN for  ST theory in order to eliminate the acceleration dependence from the Lagrangian 
(an overview of contact transformation at 3PN order is given in Appendix~\ref{App:A}).
To construct a general contact transformation, we can freely add a term of the type $\partial_{V_A^i} \mathcal{F}/m^0_A$ to 
the contact transformation~(see, Ref.~\cite{Damour:1985mt}), where $\mathcal{F}$ is an arbitrary function that starts at 2PN and only
depends on positions and velocities.  The total time derivative of $\mathcal{F}$ eliminates the acceleration produced by the addition 
of this term, without affecting the dynamics of the system. Therefore, the ordinary (reduced) Lagrangian, $\mathcal{L}^{\rm{red}}_{f}$, 
is given by
\begin{equation}
\mathcal{L}^{\rm{red}}_{f}=\mathcal{L} + \frac{d\mathcal{F}}{dt}+\sum_{A}\delta Z_{A}^i\frac{\delta \mathcal{L}}{\delta Z_{A}^i} ~,
\end{equation}
where $\mathcal{L}$ is the Lagrangian in the harmonic coordinate system, and $\delta Z_{A}^i$ indicates the contact transformation.
Since $\mathcal{F}$ starts at 2PN order, it can be formally written as
\begin{equation}
\mathcal{F} =\mathcal{F}_{\text{2PN}}+\mathcal{F}_{\text{3PN}}~,
\end{equation}
where $\mathcal{F}_{\text{2PN}}$ is the 2PN contribution given in Ref.~\cite{Julie:2017pkb}, and $\mathcal{F}_{\text{3PN}}$ 
is the 3PN contribution, the most general form of which is given in Eq.~(\ref{eq:f3pn}).
This 3PN contribution depends on 56 parameters, and the factor
$G_{AB}$ appears in the definition of parameters $F_i$ of $\mathcal{F}_{\text{3PN}}$ for dimensional convenience.
We first derive the contact transformation at 3PN order for ST theory (see Appendix~\ref{App:B}), and then use it to derive the 
reduced Lagrangian,  $\mathcal{L}_{f}^{\rm{red}}$,  at 3PN order. This reduced Lagrangian 
is an ordinary Lagrangian, dependent on $F_{i}$ and $f_{i}$\footnote{By $f_{i}$, we denote the parameters of the generic function at 2PN order given in Ref.~\cite{Julie:2017pkb}.}.

Hence, we have a whole class of coordinate systems (dependent on the parameters $F_{i}$ and $f_{i}$) 
in which our derived Lagrangian, $\mathcal{L}_{f}^{\rm{red}}$, is ordinary, while harmonic coordinates 
do not belong to this class. For completeness, let us give here the explicit expression of $\mathcal{F}^{\rm 3PN}$, 
that reads
\begin{widetext}
\begin{align}
\mathcal{F}_{\text{3PN}}= &G_{AB}m^0_Am^0_B\left\{\left[F_{1}(\vec{V}_{A}\cdot\vec{V}_{A})^2+F_2 (\vec{V}_{A}\cdot\vec{V}_{A})\vec{V}_{A}\cdot\vec{V}_{B}+F_3(\vec{V}_{A}\cdot\vec{V}_{B})^2+F_{4}(\vec{V}_{A}\cdot\vec{V}_{A})(\vec{V}_{B}\cdot\vec{V}_{B})+F_5(\vec{V}_{B}\cdot\vec{V}_{B})\vec{V}_{A}\cdot\vec{V}_{B}\right.\right.\nonumber \\
&\left.\left.+F_6(\vec{V}_{B}\cdot\vec{V}_{B})^2\right] \vec{n}_{AB}\cdot\vec{V}_{A}-\left[F_{7}(\vec{V}_{A}\cdot\vec{V}_{A})^2+F_8 (\vec{V}_{A}\cdot\vec{V}_{A})\vec{V}_{A}\cdot\vec{V}_{B}+F_9(\vec{V}_{A}\cdot\vec{V}_{A})(\vec{V}_{B}\cdot\vec{V}_{B})+F_{10}(\vec{V}_{A}\cdot\vec{V}_{B})^2\right.\right.\nonumber \\
&\left.\left.+F_{11}(\vec{V}_{B}\cdot\vec{V}_{B})\vec{V}_{A}\cdot\vec{V}_{B}+F_{12}(\vec{V}_{B}\cdot\vec{V}_{B})^2\right]\vec{n}_{AB}\cdot\vec{V}_{B}+[F_{13}\vec{V}_{A}\cdot\vec{V}_{A}+F_{14}\vec{V}_{A}\cdot\vec{V}_{B}+F_{15}\vec{V}_{B}\cdot\vec{V}_{B}](\vec{n}_{AB}\cdot\vec{V}_{A})^3\right.\nonumber \\
&\left.+[F_{16}\vec{V}_{A}\cdot\vec{V}_{A}+F_{17}\vec{V}_{A}\cdot\vec{V}_{B}+F_{18}\vec{V}_{B}\cdot\vec{V}_{B}](\vec{n}_{AB}\cdot\vec{V}_{A})^2(\vec{n}_{AB}\cdot\vec{V}_{B})\right.\nonumber\\
&\left.-[F_{19}\vec{V}_{A}\cdot\vec{V}_{A}+F_{20}\vec{V}_{A}\cdot\vec{V}_{B}+F_{21}\vec{V}_{B}\cdot\vec{V}_{B}](\vec{n}_{AB}\cdot\vec{V}_{A})(\vec{n}_{AB}\cdot\vec{V}_{B})^2\right.\nonumber \\
&\left.-[F_{22}\vec{V}_{A}\cdot\vec{V}_{A}+F_{23}\vec{V}_{A}\cdot\vec{V}_{B}+F_{24}\vec{V}_{B}\cdot\vec{V}_{B}](\vec{n}_{AB}\cdot\vec{V}_{B})^3+F_{25}(\vec{n}_{AB}\cdot\vec{V}_{A})^5+F_{26}(\vec{n}_{AB}\cdot\vec{V}_{A})^4\vec{n}_{AB}\cdot\vec{V}_{B}\right.\nonumber\\
&\left.+F_{27} (\vec{n}_{AB}\cdot\vec{V}_{A})^3(\vec{n}_{AB}\cdot\vec{V}_{B})^2-F_{28} (\vec{n}_{AB}\cdot\vec{V}_{A})^2(\vec{n}_{AB}\cdot\vec{V}_{B})^3-F_{29} (\vec{n}_{AB}\cdot\vec{V}_{A})(\vec{n}_{AB}\cdot\vec{V}_{B})^4-F_{30} (\vec{n}_{AB}\cdot\vec{V}_{B})^5\right\}\nonumber \\
&+\frac{G_{AB}^2 (m^0_A)^2m^0_B}{R}\left\{\left[F_{31}\vec{V}_{A}\cdot\vec{V}_{A}+F_{32}\vec{V}_{A}\cdot\vec{V}_{B}+F_{33}\vec{V}_{B}\cdot\vec{V}_{B}\right]\vec{n}_{AB}\cdot\vec{V}_{A}+F_{37}(\vec{n}_{AB}\cdot\vec{V}_{A})^3+F_{38} \vec{n}_{AB}\cdot\vec{V}_{B} (\vec{n}_{AB}\cdot\vec{V}_{A})^2\right.\nonumber \\
&\left.-\left[F_{34}\vec{V}_{A}\cdot\vec{V}_{A}+F_{35}\vec{V}_{A}\cdot\vec{V}_{B}+F_{36}\vec{V}_{B}\cdot\vec{V}_{B}\right]\vec{n}_{AB}\cdot\vec{V}_{B}-F_{39} (\vec{n}_{AB}\cdot\vec{V}_{B})^2 (\vec{n}_{AB}\cdot\vec{V}_{A})-F_{40} (\vec{n}_{AB}\cdot\vec{V}_{B})^3\right\}\nonumber \\
&+\frac{G_{AB}^2 m^0_A (m^0_B)^2}{R} \left\{(F_{41} \vec{V}_{A}\cdot\vec{V}_{A}+F_{42}\vec{V}_{A}\cdot\vec{V}_{B}+F_{43} \vec{V}_{B}\cdot\vec{V}_{B}) \vec{n}_{AB}\cdot\vec{V}_{A}-(F_{44}\vec{V}_{A}\cdot\vec{V}_{A} +F_{45}\vec{V}_{A}\cdot\vec{V}_{B} +F_{46} \vec{V}_{B}\cdot\vec{V}_{B})\vec{n}_{AB}\cdot\vec{V}_{B}\right. \nonumber \\
&\left.+F_{47} (\vec{n}_{AB}\cdot\vec{V}_{A})^3+F_{48} \vec{n}_{AB}\cdot\vec{V}_{B} (\vec{n}_{AB}\cdot\vec{V}_{A})^2-F_{49} \vec{n}_{AB}\cdot\vec{V}_{A} (\vec{n}_{AB}\cdot\vec{V}_{B})^2-F_{50} (\vec{n}_{AB}\cdot\vec{V}_{B})^3\right\}\nonumber \\
&+\frac{ G_{AB}^3 (m^0_A)^2 (m^0_B)^2}{R^2} \left\{F_{51} \vec{n}_{AB}\cdot\vec{V}_{A}-F_{52} \vec{n}_{AB}\cdot\vec{V}_{B}\right\}+\frac{G_{AB}^3 (m^0_A)^3 m^0_B}{R^2} \left\{F_{53} \vec{n}_{AB}\cdot\vec{V}_{A}-F_{54} \vec{n}_{AB}\cdot\vec{V}_{B}\right\}\nonumber\\
&+\frac{G_{AB}^3 m^0_A (m^0_B)^3}{R^2} \left\{F_{55} \vec{n}_{AB}\cdot\vec{V}_{A}-F_{56} \vec{n}_{AB}\cdot\vec{V}_{B}\right\} \ ~,
\label{eq:f3pn}
\end{align}
\end{widetext}
where, $G_{AB}$ is the effective Newton's gravitational constant in DEF coordinates (see, Table~\ref{LauravsDEF}), $\vec{V}_{A,B}$ are
the velocities of the two bodies,  $\vec{n}_{AB}\equiv(\vec{Z}_A-\vec{Z}_B)/R$ the unit vector of the relative separation, 
where $\vec{Z}_{A,B}$ indicate the positions of the two bodies and $R\equiv |\vec{Z}_A-\vec{Z}_B|$.

\section{Center of Mass frame two-body Ordinary Hamiltonian at 3PN}
\label{sec:com_Ham}
Let us now derive the ordinary  Hamiltonian, in the center of mass (COM) frame, corresponding to the class of ordinary Lagrangians of Sec.~\ref{sec:3PNLag}.
We do so by ordinary Legendre transformation. In the COM frame, the total momentum vanishes, $\vec{P}_{A}+\vec{P}_{B}=0$, where $\vec{P}_{A,B}$ are
the momenta of the two bodies, so the conjugate variables are $\vec{Z} = \vec{Z}_{A} - \vec{Z}_{B}$ and $\vec{P} = \vec{P}_{A} = -\vec{P}_{B}$.
Since we are considering nonspinning bodies, the motion is planar and we use polar coordinates $(R, \phi)$ with conjugate 
momenta $(P_R,P_{\phi})$,  setting $\theta =\pi/2$.
The general structure of the isotropic, time-translation invariant Hamiltonian at 2PN in the COM frame is presented in Eqs.~(III.15) - (III.16) of Ref.~\cite{Julie:2017pkb}.
By defining $M\equiv m^0_A+m^0_B$ the total mass of the system and $\mu\equiv m^0_A m^0_B/M$ its reduced mass, it is convenient to use mass reduced variables (always indicated with hat superscript, here and below)
$(\hat{P},\hat{P}_R,\hat{R})$, where $\hat{P}^2\equiv P^2/\mu^2=\hat{P}_R^2+\hat{P}_{\phi}^2/\hat{R}^2$,
$ \hat{P}_R \equiv P_R/\mu$, $\hat{P}_{\phi}\equiv P_{\phi}/(\mu M)$
and $\hat{R}\equiv R/M$, so that the 3PN contribution formally reads
\begin{align}
\hat{H}^{\text{3PN}}=& \left(h_1^{\rm3PN}\hat{P}^8+h_2^{\rm3PN}\hat{P}^6\hat{P}_R^2+h_3^{\rm3PN}\hat{P}^4\hat{P}_R^4\right.\nonumber\\
&\left.+h_4^{\rm3PN}\hat{P}^2\hat{P}_R^6+h_5^{\rm3PN}\hat{P}_R^8\right)+\frac{1}{\hat{R}}\left(h_6^{\rm 3PN}\hat{P}^6\right.\nonumber \\
&\left. + h_7^{\rm 3PN}\hat{P}^4\hat{P}_R^2+h_8^{\rm 3PN}\hat{P}^2\hat{P}_R^4+h_9^{\rm 3PN}\hat{P}_R^6\right)\nonumber\\
&+\frac{1}{\hat{R}^2}\left(h_{10}^{\rm 3PN}\hat{P}^4+h_{11}^{\rm 3PN}\hat{P}^2\hat{P}_R^2+h_{12}^{\rm 3PN}\hat{P}_R^4\right)\nonumber\\
&+\frac{1}{\hat{R}^3}\left(h_{13}^{\rm 3PN}\hat{P}^2+h_{14}^{\rm 3PN}\hat{P}_R^2\right)+\frac{1}{\hat{R}^4}h_{15}^{\rm3PN}~,
\label{eq:genH}
\end{align}
where the $h^{\rm 3PN}_i$'s formally indicate the numerical coefficients we are going to calculate.
Before doing so explicitly and introducing our results, let us recall an important technical fact. 
The ordinary Hamiltonian directly obtained from the class of Lagrangians of Sec.~\ref{sec:3PNLag} 
via Legendre transformation contains  two undetermined constants $r^{\prime}_1$ and $r^{\prime}_2$.
These constants parametrize logarithmic terms and are directly inherited from the harmonic-coordinates 
Lagrangian of Ref.~\citep{Bernard:2018ivi}, where they arise due to the regularization procedure 
(via Hadamard partie finie technique). However, Ref.~\cite{Damour:1999cr} found that these constants
are absent in the 3PN ordinary Hamiltonian in GR\@.
As shown in Ref.~\cite{Blanchet:2000ub}, the reason why it is so is
that the two constants, and the related logarithmic terms,  can be gauged away from the harmonic coordinate 
Lagrangian, in accordance with the fact that these are pure gauge quantities.  More precisely,
Ref.~\cite{deAndrade:2000gf} showed that, in the GR case, the regularization constants can also be eliminated by 
including a logarithmic dependence in the function $\mathcal{F}$ of the contact transformation. 
Therefore, to gauge away the dependence on $r^{\prime}_1$ and $r^{\prime}_2$ from the ordinary Hamiltonian of ST theory, 
we take analogy with the approach of Ref.~\cite{deAndrade:2000gf} in GR and add the most generic logarithmic dependent terms,
\begin{align}
\frac{G^3_{AB} (m^0_A)^3 m^0_B}{R^2}(F_{57} \log \frac{r^\prime_1}{R})\nonumber\\+\frac{G^3_{AB} m^0_A (m^0_B)^3}{R^2}( F_{58}\log \frac{r^\prime_2}{R})~,
\end{align}
to our arbitrary function $\mathcal{F}_{3PN}$.  In ST theory, the coefficients
 $F_{57}$ and $F_{58}$ must be
\begin{align}
F_{57}&=\left[-\frac{11}{4}(2+\bar{\gamma}_{AB})^2+\delta_A\right], \\
F_{58}&=\left[-\frac{11}{4}(2+\bar{\gamma}_{AB})^2+\delta_B \right],
\end{align}
in order to remove the gauge dependence on $r^{\prime}_1$ and $r^{\prime}_2$ from the ordinary class of 
Hamiltonians. The so obtained complete expression of the ordinary Hamiltonian coefficients 
of Eq.~(\ref{eq:genH}) at 3PN is given in Appendix~\ref{app:C}.
\label{sec:COM3PNH}

\section{Scalar-Tensor Deformation of 3PN Effective One-Body Hamiltonian}
\label{sec:ST3PN-EOB}
Let us now turn to discussing the main result of this work, i.e.\ the instantaneous 3PN contribution to the EOB potentials in ST theory.
The mapping between the two-body ordinary (ADM-like) Hamiltonian and the EOB Hamiltonian can be done using different procedures
(e.g., Delaunay Hamiltonian, canonical transformation, comparison of the periastron advance, see 
e.g.~\cite{Buonanno:1998gg,Damour:2000we,Hinderer:2013uwa,Blumlein:2021txe}). Here we will use the canonical transformation
approach, adapting the procedure of Ref.~\cite{Damour:2000we}.

\subsection{Canonical Transformation at 3PN}
\label{sec:CanTransf}
To fix notation, we indicate with $(Q,P)$ the conjugate variables of the real, ordinary, two-body Hamiltonian, while with $(q,p)$ those of the 
EOB Hamiltonian. We start from the 2PN-accurate canonical transformation ins ST theory of Ref.~\cite{Julie:2017pkb} 
(see Eqs.~(III.23) and (III.25) therein) and augment it with 3PN terms. We do so by modifying the generating function of the 
canonical transformation, whose 3PN contribution reads
\begin{align}
\frac{G(Q,p)^{\text{3PN}}}{\mu M}=\hat{R}\, \hat{p}_r \Bigg[ &\left(\alpha_3\mathcal{P}^6+\beta_3\mathcal{P}^4\hat{p}_r^2+\gamma_3\mathcal{P}^2\hat{p}_r^4+\delta_3\hat{p}_r^6\right) \nonumber\\
&\left.+\frac{1}{\hat{R}}(\epsilon_3 \mathcal{P}^4+\eta_3\mathcal{P}^2\hat{p}_r^2+\theta_3\hat{p}_r^4)\right.\nonumber\\
&+\frac{1}{\hat{R}^2}(\lambda_3\mathcal{P}^2+\rho_3\hat{p}_r^2)+\frac{\sigma_3}{\hat{R}^3}\Bigg]~,
\end{align} 
where $\mathcal{P}$, $\hat{p}_r$, and $\hat{R}$ are the dimensionless variables with $\mathcal{P}^2\equiv\hat{p}_r^2+\frac{\hat{p}_{\phi}^2}{\hat{R}^2}$,
and $(\alpha_3$,$\beta_3$,..,$\sigma_3)$ formally indicate the ten 3PN coefficients.
We will then follow the same procedure as Ref.~\cite{Julie:2017pkb}, i.e.\ we will express the
real and effective Hamiltonian in an intermediate coordinate system $(Q,p)$ in order to match the two.

\subsection{Scalar-Tensor Effective One-Body Hamiltonian at 3PN\@: instantaneous part}
\label{sec:STH3PN}
Within the EOB approach, the real EOB Hamiltonian is related to the effective Hamiltonian as
\begin{equation}
\label{eq:Hreal}
\hat{H}_{\rm real}\equiv \dfrac{H_{\rm real}}{\mu}=\dfrac{1}{\nu}\sqrt{1+2\nu\left(\hat{H}_{\rm eff}-1\right)},
\end{equation}
where $\nu=\mu/M$ is the symmetric mass ratio and $\hat{H}_{\rm eff}\equiv H_{\rm eff}/\mu$ is
the reduced-mass effective Hamiltonian. This relation was originally proved to be correct up to
3PN in GR~\cite{Damour:2000we}. Recently, within the Post-Minkowskian scheme,  Ref.~\cite{Damour:2016gwp}
proved it to hold at \textit{all} PN orders, both in GR and in ST theories. Here we choose to incorporate the 3PN
terms within the EOB Hamiltonian following the scheme of Ref.~\cite{Damour:2000we}, i.e.\ by writing the
effective Hamiltonian $\hat{H}_{\rm eff}$ as
\begin{equation}
\hat{H}_{\rm eff}=\sqrt{A(\hat{r})\left[1+\frac{\hat{p}_{\phi}^2}{\hat{r}^2}+\frac{\hat{p}_r^2}{B(\hat{r})}+\hat{Q}_e(\hat{r},\hat{p})\right]} \ .
\label{eq:genhe}
\end{equation}
Here, we indicated with $(\phi,\hat{p}_{\phi},\hat{r}, \hat{p}_r)$ the canonical variables, with $\hat{p}^2=\hat{p}_r^2+\hat{p}_\phi^2/\hat{r}^2$,
while $(A,B,Q_e)$ are the EOB potentials. The structure of the nongeodesic term $\hat{Q}_e \equiv Q_e/\mu^2$ at 3PN reads
\begin{align}
\hat{Q}_e(\hat{r},\hat{p}) &= \frac{1}{\hat{r}^2}\left(q_1\hat{p}^4+q_2\hat{p}^2\hat{p}_r^2+q_3\hat{p}_r^4\right)\ .
\end{align}
Following Ref.~\cite{Damour:2000we}, we use the gauge freedom at our disposal to set $q_1=q_2=0$, 
so that the $Q_e$ function only depends on the radial momentum. This choice is known as Damour-Jaranowski-Sch\"afer (DJS) gauge,
first introduced in Ref.~\cite{Damour:2000we}. We recall, in passing, that this is just one among the many
(actually infinite) possibilities of devising an effective dynamics based on a generalized mass shell condition
$g^{ \mu \nu}_{\rm eff}p_\mu p_\nu + Q_e(p)=0$ and the relation between the effective and real Hamiltonian 
given by Eq.~\eqref{eq:Hreal}~\cite{Damour:2016gwp,Damour:2017zjx,Damour:2019lcq,Antonelli:2019ytb,Antonelli:2019fmq,Khalil:2022ylj,Damour:2022ybd}.
In the DJS gauge, the three EOB potentials at 3PN formally read
\begin{align}
A(\hat{r}) &= 1 - \frac{2}{\hat{r}}+\frac{a_2}{\hat{r}^2}+\frac{a_3}{\hat{r}^3}+\frac{a_4}{\hat{r}^4} \ ,\label{eq:Aexpansion}\\
 B(\hat{r}) &= 1 + \frac{b_1}{\hat{r}}+\frac{b_2}{\hat{r}^2}+\frac{b_3}{\hat{r}^3}\ , \\
 \hat{Q}_e(\hat{r},\hat{p}) &= \frac{q_3}{\hat{r}^2}{\hat{p}_r^4} \ ,
 \label{eq:genAB}
\end{align}
where the $a_i$ and $b_i$ terms are the $\nu$-dependent deformations of the Schwarzschild 
metric potentials, that take into account both GR and ST 
corrections\footnote{Let us remember in this respect that the 1PN term in the $a_{i}$ function, $a_2$, 
is identically zero in GR~\cite{Buonanno:1998gg}, while it is nonzero in ST theory~\cite{Julie:2017pkb}.}.
The GR and ST contributions are separated as
\begin{align}
	A &= A^{\rm{GR}}+\delta A^{\rm{ST}}, \\
	B &= B^{\rm{GR}}+\delta B^{\rm{ST}}, \\
	\hat{Q}_e&=\hat{Q}_e^{\rm GR}+\delta \hat{Q}_e^{\rm ST},
\end{align}
which reflects on the $\nu$-dependent contributions as
\begin{align}
	a_i  &= a_{i}(\nu) = a_i^{\rm{GR}}(\nu)+\delta a_i^{\rm{ST}}(\nu), \label{eq:a-exp} \\
	b_i  &= b_i(\nu) = b_i^{\rm{GR}}(\nu)+\delta b_i^{\rm{ST}}(\nu),  \\
	q_3 &= q_3(\nu) = q_3^{\rm GR}(\nu)+\delta q_3^{\rm ST}(\nu)~.
\end{align}

The GR terms are known analytically up to 6PN~\cite{Bini:2020nsb,Bini:2020hmy},
except for some yet unknown coefficients proportional to $\nu^2$.
As mentioned above,
the 2PN ST corrections to the EOB potentials have been computed in Refs.~\cite{Julie:2017pkb,Julie:2017ucp}.

Starting from the 3PN, nonlocal-in-time contributions have to be added to the local terms.
In this section, we will focus on these latter and we postpone the computation of the complete nonlocal terms to future work.
See, however, next section for the nonlocal terms restricted to the circular case.

Let us now compute the local ST contributions to the EOB potentials at 3PN,
i.e. $(\delta a_{\rm 4, loc}^{\rm ST},\delta b_{\rm 3, loc}^{\rm ST},\delta q_{\rm 3, loc}^{\rm ST})$.
We do so by first applying the the canonical transformation of Sec.~\ref{sec:CanTransf} to both the real two body ordinary Hamiltonian ($\hat{H}_{\rm real})$ of Sec.~\ref{sec:COM3PNH} and the 3PN EOB Hamiltonian ($\hat{H}_{\rm eff}$) to express them in the intermediate coordinate system $(Q,p)$.
Then, the canonically transformed $\hat{H}_{\rm real}$ is matched with the PN expanded $\hat{H}_{\rm real}$ obtained using Eq.~(\ref{eq:Hreal}) to find the unique solution of $a_i,~b_i$, and $q_i$ up to 3PN\@.
For the new 3PN-order ST contributions, we get:
\begin{widetext}
\begin{align}
\delta a_{\rm 4, loc}^{\text{ST}}=&-2 \bar{\gamma}_{AB}-\frac{13}{2}\bar{\gamma}_{AB}^2-5\bar{\gamma}_{AB}^3+\left[-\frac{7}{6}\left(2+\bar{\gamma}_{AB}\right)^2+9\bar{\gamma}_{AB}^2\right]\langle \bar{\beta}\rangle-\bar{\gamma}_{AB}\langle \epsilon \rangle +\frac{2}{3}\langle \bar{\kappa} \rangle +\frac{2}{3}\left(1+2\bar{\gamma}_{AB}\right)\langle {\delta}\rangle +6(\beta_+^2+\beta_-^2)\nonumber \\
&-12 X_{AB}\beta_-\beta_+ +\frac{2}{3}\left(\beta_-(\delta_-+X_{AB}\delta_+)-\langle {\delta}\rangle  \beta_+\right) + \nu \Bigg\{\frac{11}{4\,\alpha_{AB}}\left(2+\bar{\gamma}_{AB}\right)\bar{\gamma}_{AB}-\frac{\langle {\delta}\rangle}{\alpha_{AB}(2+\bar{\gamma}_{AB})} +\frac{32}{\bar{\gamma}_{AB}^2}\langle \bar{\beta}\rangle\left(\beta_+^2-\beta_-^2\right) \nonumber \\
&+\frac{4}{\bar{\gamma}_{AB}}\left[\beta_-\left(\epsilon_--X_{AB} \epsilon_+ + \frac{4}{3}\left(2 \delta_- - X_{AB}\delta_+\right)\right)+\beta_+\left(\frac{4}{3}\left(2\delta_+-X_{AB}\delta_-\right)- \langle {\epsilon}\rangle\right)\right]\nonumber\\
&+\bar{\gamma}_{AB}\left[\frac{581}{18}-\frac{75}{64}\pi^2 - 8 \zeta-\frac{32}{3}\langle \bar{\beta}\rangle -20 \beta_++\frac{1}{2}\langle{\delta}\rangle + \left(\frac{4}{3}+\frac{7}{32}\pi^2\right)\delta_++2\epsilon_+\right]+\bar{\gamma}_{AB}^2 \left(\frac{239}{18}-\frac{5}{32}\pi^2-\frac{2}{3}\langle \bar{\beta}\rangle \right) \nonumber\\
&+\bar{\gamma}_{AB}^3\left(-\frac{3}{8}+\frac{7}{128}\pi^2\right)+3 \zeta-6 \left(\beta_+^2+\beta_-^2\right)-\frac{5}{3}\langle\bar{\beta}\rangle+\frac{3}{2}X_{AB}\beta_- -12 X_{AB}\beta_-\beta_+ - \frac{8}{3}\left(\delta_+ \langle \bar{\beta}\rangle +\beta_- \delta_-\right) \nonumber \\
&-\delta_+\left(\frac{92}{9}-\frac{7}{16}\pi^2\right)+\frac{1}{3} X_{AB}\delta_- -\frac{2}{3}\langle\bar{\kappa}\rangle -\frac{4}{3}\kappa_+ +\langle{\epsilon}\rangle \Bigg\}+\nu^2 \left(-4 \beta_-^2\right)~,
\label{eq:Ast}
\end{align}
\begin{align}
\delta b_{\rm 3, loc}^{\text{ST}}=&~ 29\bar{\gamma}_{AB}+\frac{131}{4}\bar{\gamma}_{AB}^2+\frac{47}{4}\bar{\gamma}_{AB}^3+22\left(1+\bar{\gamma}_{AB}\right)\langle\bar{\beta}\rangle-5 \left(1+\bar{\gamma}_{AB}\right)\langle{\delta}\rangle -\langle {\epsilon}\rangle \nonumber\\
&+\nu\Bigg[-93 \bar{\gamma}_{AB}-42\bar{\gamma}_{AB}^2-3\bar{\gamma}_{AB}^3-7\zeta +2\left(3+2 \bar{\gamma}_{AB}\right) \delta_+ +2 \langle {\delta}\rangle+\epsilon_+ +X_{AB}\epsilon_- \nonumber\\
&-2\left(\frac{57}{4}+ \bar{\gamma}_{AB}\right) X_{AB}\beta_--2\left(4-9 \bar{\gamma}_{AB}\right) \langle \bar{\beta}\rangle \Bigg]+\nu^2 \left(10\bar{\gamma}_{AB}+\bar{\gamma}_{AB}^2-6 \zeta - 18\beta_++4 \delta_+ + 2 \epsilon_+ \right)~,
\label{eq:Bst}
\end{align}
\begin{align}
\delta q_{\rm 3, loc}^{\text{ST}}=&~\nu\left(-\frac{26}{3} \bar{\gamma}_{AB}-\frac{5}{2} \bar{\gamma}_{AB}^2-\frac{2}{3} \langle \bar{\beta} \rangle+\frac{2}{3} \langle {\delta} \rangle\right) + \nu^2 \left(4\bar{\gamma}_{AB}-2\langle \bar{\beta} \rangle\right)~,
\label{eq:qst}
\end{align}
\end{widetext}
where we combine the notations of Refs.~\cite{Julie:2017pkb,Sennett:2016klh}. More precisely, introducing $X_{A,B}\equiv m^0_{A,B}/M$, we have
\begin{align}
&X_{AB}\equiv X_A-X_B \ , \\
&\langle\bar{\beta}\rangle\equiv -X_{AB}\beta_-+\beta_+,\\
&\langle\bar{\kappa}\rangle\equiv -X_{AB}\kappa_-+\kappa_+ \ , \\
&\langle{\delta}\rangle\equiv X_{AB}\delta_-+\delta_+~,\\
&\langle\bar{\chi}\rangle\equiv -X_{AB}\chi_-+\chi_+\equiv-\frac{\langle{\epsilon}\rangle}{4}~,\\
&\langle{\epsilon}\rangle\equiv -X_{AB}\epsilon_-+\epsilon_+ \ ,
\end{align}
with the $``{\pm}''$ subscript denoting the symmetric and anti-symmetric parts of the ST parameters, e.g. \hbox{$x_{\pm}\equiv(x_A\pm x_B)/2$}.

As expected, the functions $(A,B,\hat{Q}_e)$ do not depend on the function $\mathcal{F}$ of Sec.~\ref{sec:3PNLag} since,
similarly to the 2PN case of Ref.~\cite{Julie:2017pkb}, it is absorbed by the canonical transformation.
As a consistency check of our results, we verify that the binding energy of the system along circular orbits, obtained from the condition 
$\partial_{\hat r}\hat{H}_{\rm eff}=0$, exactly matches the corresponding function given in Eq.~(5.4) of~\cite{Bernard:2018ivi}.
Similarly, we correctly obtain the GR terms up to 3PN as well as the ST ones at 2PN calculated
in Refs.~\cite{Julie:2017pkb,Julie:2017ucp}.

\subsection{Completing the circular conservative 3PN dynamics with contributions from tails and tides}
\label{sec:circular}

Throughout this paper, we focused on the instantaneous contributions to the EOB Hamiltonian at 3PN\@.
However, at 3PN, there are two more contributions entering the PN computations: (i) the (nonlocal-in-time) tail terms~\cite{Bernard:2018hta,Bernard:2018ivi}; and (ii) the finite-size (tidal) effects~\cite{Bernard:2019yfz}. 
The coordinate invariant circular case real two body energy for tail and finite-size contributions at 3PN in ST theory are given in Eq.~(5.5) of Ref.~\cite{Bernard:2018ivi}, and Eq.~(8) of Ref.~\cite{Bernard:2019yfz}, respectively. 
In this section, we will restrict ourselves to circular orbits and compute the corrections to the EOB $A$ potential due to both tail and finite-size effects.
For the circular case, the \textit{complete} 3PN term $\delta_{4}^{\rm ST}$ is decomposed as
\begin{equation}
\delta a_{4}^{\rm ST} = \delta a_{\rm 4, loc}^{\rm ST} + \delta a_{\rm 4, nonloc}^{\rm ST} + \delta a_{\rm 4, tidal}^{\rm ST}\ .
\end{equation}
When only considering circular systems, the EOB metric potential $A$ is simply computed by comparing the circular 
case real two body ordinary Hamiltonian with the circular case EOB Hamiltonian using Eq.~(\ref{eq:Hreal}).

Let us now compute the \textit{complete} 3PN metric potential $A$. We do so, first by computing the gauge invariant 
circular case EOB energy. Then, this gauge invariant EOB energy is matched with the gauge invariant real 
two body energy using Eq.~(\ref{eq:Hreal}) to  find ST corrections to $A$.
 
We leave the extension to noncircular orbits with the \textit{complete} computation of the $B$ and $Q_e$ corrections to future works.

\subsubsection{Tail effects}
\label{tailst}
Following the procedure discussed above, the tail contribution to\ $A$ reads:
 \begin{equation}
\delta a_{\rm 4, nonloc}^{\rm ST}=\delta a_{\rm 4, nonloc,0}^{\rm ST} +\delta a_{\rm 4, nonloc,log}^{\rm ST} ~ \log\left(\frac{1}{\hat{r}}\right)~,
\label{atail}
\end{equation}
where 
\begin{align}
\label{atail0}
\delta a_{\rm 4, nonloc,0}^{\rm ST}&= \nu\left(2\delta_{+}+\frac{\bar{\gamma}_{AB}(2+\bar{\gamma}_{AB})}{2}\right)\left[\frac{8}{3}(\log2 +\gamma_E)\right] \ ,
\end{align}
and
\begin{align}
\delta a_{\rm 4, nonloc,log}^{\rm ST}=\frac{4}{3} \nu\left(2\delta_{+}+\frac{\bar{\gamma}_{AB}(2+\bar{\gamma}_{AB})}{2}\right)~,
\label{ataillog}
\end{align}
with $\delta_{+}$ defined as above. For the equal-mass case, the tail contribution, Eq.~(\ref{atail}), vanishes as
 the common factor of Eqs.~\eqref{atail0}-\eqref{ataillog} for the equal mass case is zero. 
 It can be seen from Table~\ref{LauravsDEF}, and that $\alpha_A^0=\alpha_B^0$ as scalar charge for 
 both the bodies are same for equal-mass case.

\subsubsection{Finite-size effects}
\label{tidalst}
When considering extended bodies, tidal effects have an impact on the binary dynamics. 
Therefore, by considering the finite-size addition to the 3PN circular case energy, Eq.~(8) of Ref.~\cite{Bernard:2019yfz}, the procedure discussed above yields:
\begin{align}
\label{eq:tidal}
\delta a_{\rm 4, tidal}^{\rm ST} &=-\frac{16}{5}\nu\alpha_0^2\left[\frac{m_A^0}{M}\delta_A k_A^{(s)}+\frac{m_B^0}{M}\delta_B k_B^{(s)}\right]~,
\end{align}
where $k_{A,B}^{(s)}$ are the dimensionless scalar $\ell=2$ tidal Love numbers of the two bodies.

To estimate the magnitude of the the scalar-mediated finite size effect, we observe that the pre-factor in Eq.~\eqref{eq:tidal} 
for the equal-mass case is $\mathcal{O}(10^{-8})$ in the dynamical scalarization regime, i.e.\ for $\alpha \simeq 10^{-1}$.
The correction to $A$ is then linear in the scalar Love numbers, which for neutron stars typically range
up to $\lesssim 10^{-2}$~\cite{Brown:2022kbw}, depending on the equation of state. This results to a correction of at 
most $\mathcal{O}(10^{-10})$, thus rendering the effect of scalar tides on the {\lso} frequency practically unmeasurable.

\section{Modifications of the conservative binary dynamics at the Innermost Stable Circular Orbit}
\label{sec:isco}

Let us now study the impact of the 3PN ST corrections to the circular dynamics. 
We do so by evaluating the frequency at {\lso}.

For the GR part, we rely on the NR-informed $A$ potential used
within the \TEOBResumS{} waveform model~\cite{Nagar:2020pcj,Schmidt:2020yuu,Riemenschneider:2021ppj}.
More precisely, the $A^{\rm GR}$ is based on (formal) 5PN information where the non-logarithmic
5PN coefficient, $a_6^c(\nu)$, is informed by NR simulation \textit{after} that the full potential
has been resummed using a $(1,5)$ Pad\'e approximant, and is given by Eq.~(33) of Ref.~\cite{Nagar:2020pcj}.
In practice, we have
\begin{equation}
	\label{A:pade}
	A_{\rm GR}(\hat{r};\nu)\equiv P^1_5\left[A_{\rm GR}^{\rm 5PN}(\hat{r};\nu) \right]
\end{equation}
where $A_{\rm GR}^{\rm 5PN}$ indicates the GR $A$ function expanded at 5PN\@.
Focusing on the sequence of circular orbits ($\hat{p}_r=0$), the {\lso} orbital radius
and angular momentum $(\hat{r}^{\rm \lso},p_\phi^{\rm \lso})$ are defined by the
conditions  $\partial_{\hat r}\hat{H}_{\rm eff}=\partial^2_{\hat r}\hat{H}_{\rm eff}=0$,
and $\Omega_{\rm \lso}$ is obtained from the corresponding Hamilton's equation.
Reference~\cite{Julie:2017pkb} already considered two ways of flexing the GR potential 
so to include the scalar-tensor contribution. First, $\delta A^{\rm ST}$ was just considered
as a contribution simply added to the Pad\'e resummed GR potential of Eq.~\eqref{A:pade},
but this choice was not found to be robust versus the ST coupling constant. 
As an alternative, Ref.~\cite{Julie:2017pkb} proposed to resum, with a Pad\'e approximant,
the full PN-expanded function with GR and ST contributions. We follow here this 
approach and define
\begin{align}
	\label{eq:AGR-ST}
	A(\hat{r})= P^1_5\left[A^{\rm GR}_{\rm 5PN}+\delta A^{\rm{ST}}\right] ,
\end{align}
where now $\delta A^{\rm{ST}}$ includes up to the 3PN instantaneous 
correction of Eq.~\eqref{eq:Ast} and tail correction of Eq.~\eqref{atail}. Focusing on the equal-mass case, 
Ref.~\cite{Julie:2017pkb} noted that the 1PN and 2PN ST corrections 
are numerically of the same order, so that, for simplicity, in the numerical 
analysis of the {\lso} frequency behavior they were considered to be
exactly the same.
For completeness we stick here to using the correct analytical expression
without any approximation.

\begin{figure}[t]
  \begin{center}
    \includegraphics[width=0.48\textwidth]{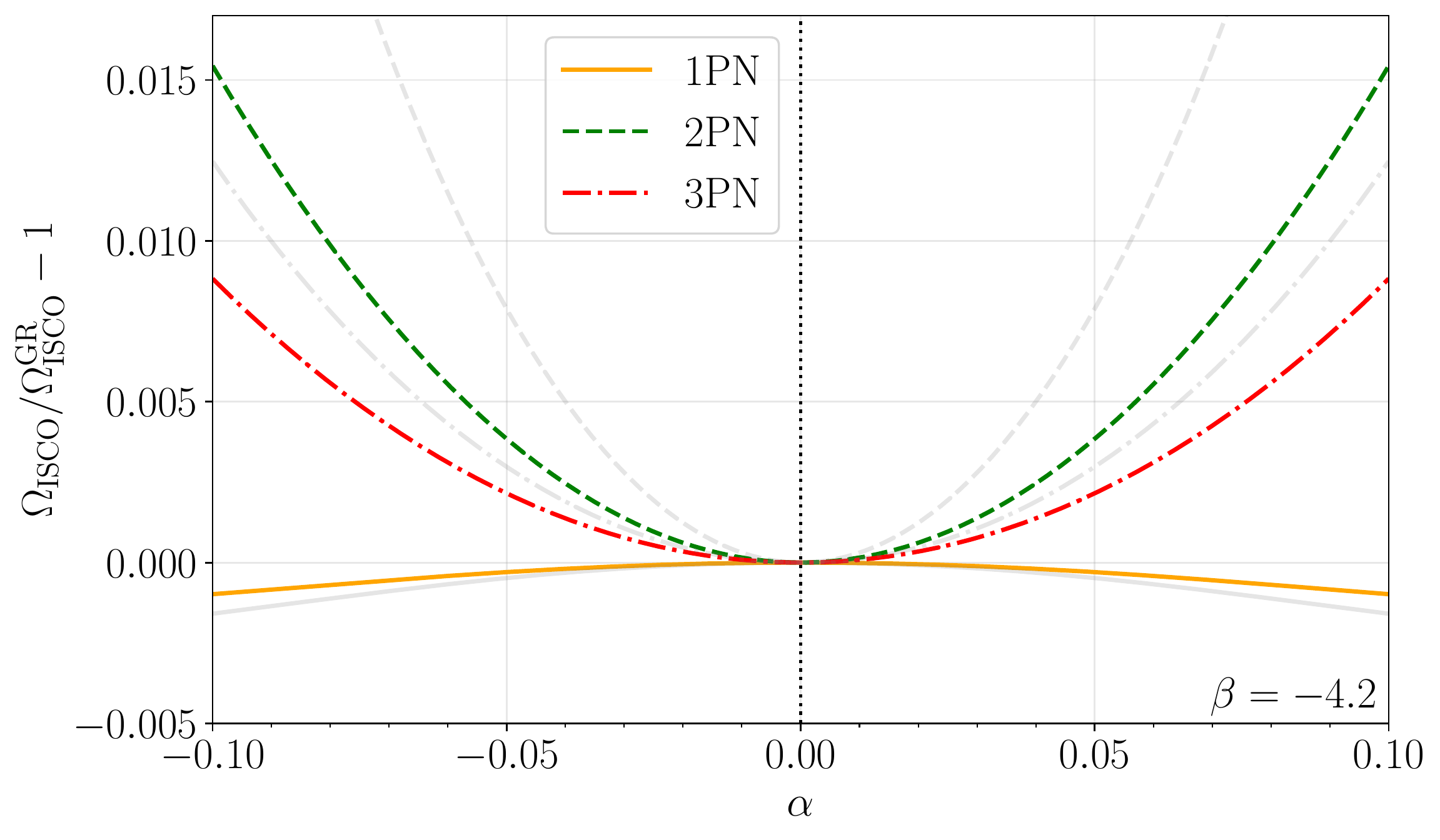}
  \end{center}
  \caption{
    Fractional change in {\lso} frequency w.r.t. GR for equal-mass systems, $\nu = 1/4$, as a function of the ST parameter $\alpha$, with $\beta$ fixed to $\beta = -4.2\,$.
    We show different PN orders for the ST corrections [included in the overall Pad\'e resummation, Eq.~\eqref{eq:AGR-ST}].
    They all reduce to the GR limit when $\alpha \rightarrow 0$.
    The results obtained by simply adding the nonresummed ST terms are reported using dashed lines.}
\label{fig:Freq-lso-frac}
\end{figure}

Figure~\ref{fig:Freq-lso-frac} focuses on an equal-mass binary\footnote{The ST tail correction vanishes for equal-mass case as mentioned in Sec.~\ref{tidalst}.} and depicts $\Omega_{\rm \lso}$ as a function of the ST coupling constant $\alpha$ considered in a reasonable range of values compatible with the experimental constraints, where we fix $\beta=-4.2$.
Here, for simplicity we neglect the corrections from the ST parameters $\beta'$ and $\beta''$, by fixing them to zero.
Their impact on the 2PN and 3PN ST corrections to $A$, $\delta a^{\mathrm{ST}}_3$ and $\delta a^{\mathrm{ST}}_4$,  when varying $\beta'$ and $\beta''$ within a reasonable range of values $[-10, 10]$, is at the level of $5\%$ and $1\%$ respectively.
Therefore, their overall contribution to the potential $A$ (see Eqs.~\eqref{eq:Aexpansion} and~\eqref{eq:a-exp}), that controls 
the circular conservative dynamics and thus the correction to the {\lso} frequency, will indeed be negligible.

\begin{figure}
  \begin{center}
    \includegraphics[width=0.48\textwidth]{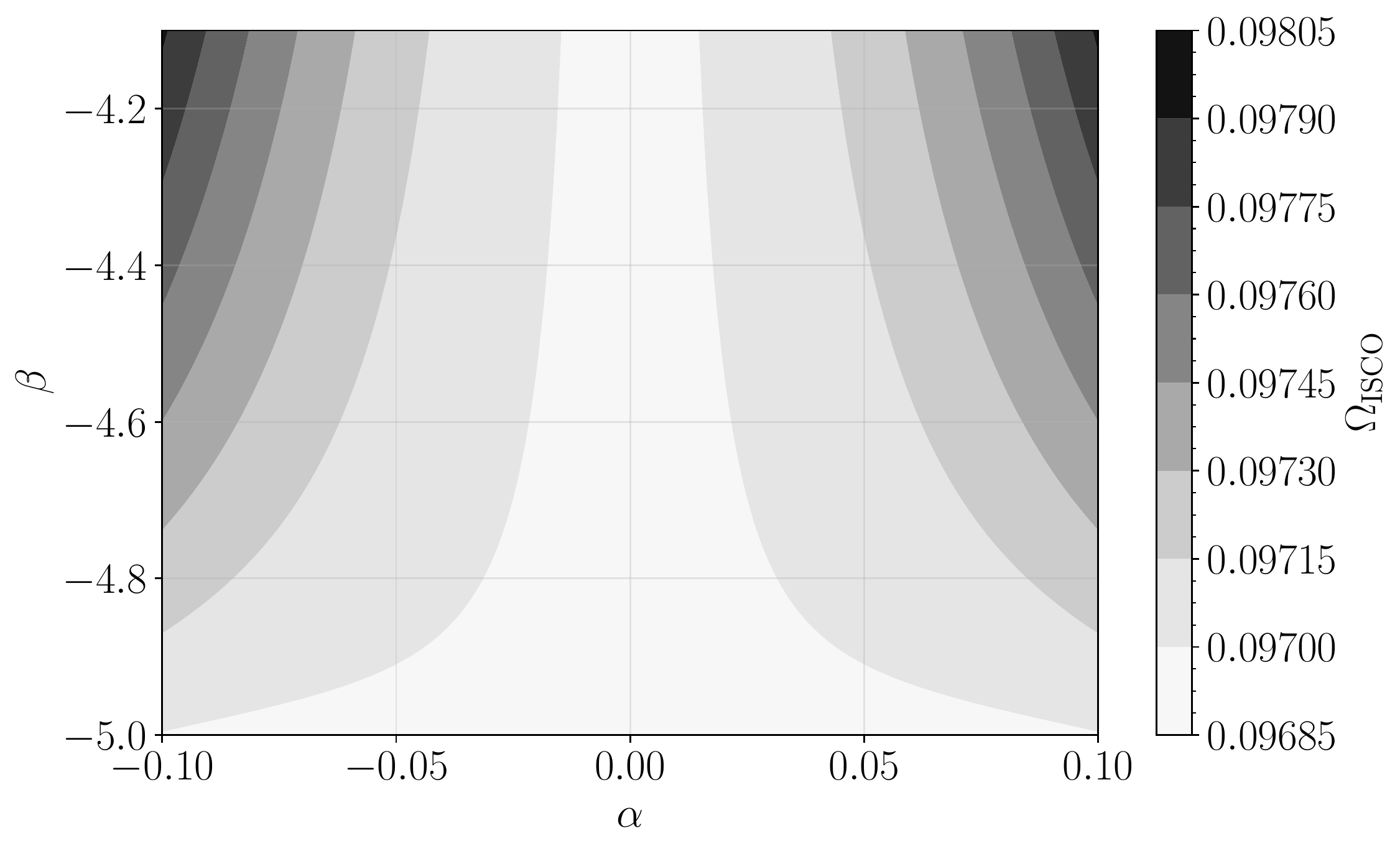}
  \end{center}
  \caption{
    Frequency at the innermost stable circular orbit at 3PN order for equal-mass systems, as a function of the ST parameters $\alpha$ and $\beta$.
    }
\label{fig:freq-lso-2d}
\end{figure}

For comparison, the figure also shows, as dashed lines, the curves obtained when not applying a resummation on the ST part of the potential.
In Fig.~\ref{fig:freq-lso-2d} we also show the modification of $\Omega_{\mathrm{\lso}}$ as a function of both $\alpha$ and $\beta$ within a viable range, again for the equal-mass case.


\section{Conclusions}
\label{sec:conclusions}
Building upon recent results~\citep{Bernard:2018ivi} for massless scalar-tensor theory, we have generalized the 2PN EOB Hamiltonian 
of Ref.~\cite{Julie:2017pkb} to 3PN order, though restricting ourselves to the local-in-time contribution. First, we derived a class 
of two-body ordinary Hamiltonians (i.e.\ dependent only on position and momenta) for ST theory by: (i) reducing the two-body
harmonic coordinate local-in-time Lagrangian of~\citep{Bernard:2018ivi} to an ordinary class of Lagrangians by constructing 
a general contact transformation for ST theories; (ii) augmenting the general contact transformation with the logarithm dependent 
terms to gauge away regularization constants from the ordinary Lagrangian by taking analogy with General Relativity results
of~\cite{deAndrade:2000gf}; and (iii) performing Legendre transformation of this ordinary Lagrangian to derive the ordinary 
Hamiltonian. We then recasted this local-in-time ordinary Hamiltonian into equivalent, 3PN-accurate, EOB potentials $(A,B,Q_e)$,
see Eqs.~\eqref{eq:Ast}-\eqref{eq:qst}. As a test of our results, we checked that that the energy along circular orbits computed 
from the EOB Hamiltonian precisely coincides with the the one given in Eq.~(5.4) of Ref.~\cite{Bernard:2018ivi}. We also computed the corrections due to the tail and finite-size effects at 3PN for ST theories only for the circular conservative dynamics using the gauge-invariant circular case energy of Refs.~\cite{Bernard:2018ivi,Bernard:2019yfz}.
We additionally studied the shift in the orbital frequency of the innermost stable circular orbit induced by scalar tensor corrections.

This paper must be seen as a first step within the effort of incorporating massless scalar-tensor corrections within the EOB
waveform model {\tt TEOBResumS-GIOTTO}~\cite{Gamba:2021ydi,Gamba:2022mgx} for coalescing, precessing, 
black-hole-neutron star and neutron star binaries. In future work we will address the problem of computing factorized and
resummed scalar tensor corrections to the waveform and radiation reaction, building upon the results of~\cite{Sennett:2016klh,Bernard:2022noq}.

\begin{acknowledgments}
T.~J., P.~R. thank the hospitality and the stimulating environment of the Institut des Hautes Etudes Scientifiques.
The authors are grateful to Thibault Damour for fruitful discussions and useful suggestions in the preparation of this paper.
A.~N. and P.~R. acknowledge useful discussions with Carlos Palenzuela.
T.~J. is jointly funded by the University of Cambridge Trust, Department of Applied Mathematics and 
Theoretical Physics (DAMTP), University of Cambridge, and Centre for Doctoral Training, University of Cambridge.
M.~A. is supported by the Kavli Foundation.
P.~R. is supported by the Italian Minister of University and Research (MUR) via the 
PRIN 2020KB33TP, {\it Multimessenger astronomy in the Einstein Telescope Era (METE)}.
The present research was also partly supported by the ``\textit{2021 Balzan Prize for 
Gravitation: Physical and Astrophysical Aspects}'', awarded to Thibault Damour. 
\end{acknowledgments}

\appendix
\section{Reminder of 3PN-accurate contact tranformation.}
\label{App:A}
The 4-dimensional coordinate change through the contact transformation at 3PN order is based on the application 
of the method introduced in Ref.~\cite{Damour:1999cr} to eliminate higher order derivative terms. 
The new Lagrangian based on the coordinate change $\delta Z_{A}$ is 
\begin{equation}
\mathcal{L}\rightarrow\mathcal{L}+ \  \sum_{I=A,B} \delta Z^{i}_{I} \frac{\delta \mathcal{L}}{\delta Z^{i}_{I}} \ + \frac{d \mathcal{F}}{dt} = \mathcal{L}_{f}^{red}~,
\label{eq:CT1}
\end{equation}
where $\mathcal{F}$ is the generic function introduced in Sec.~\ref{sec:3PNLag}, and $\frac{\delta \mathcal{L}}{\delta Z_{I}^{i}}$ 
is the functional derivative term. Since the contact transformation coordinate change $\delta Z_{I}^{i}$ starts at 2PN order, 
according to Eq.~\eqref{eq:CT1}, the fractional derivative term at 3PN should be considered up to 1PN order, i.e.,
\begin{equation}
\frac{\delta \mathcal{L}}{\delta Z_{I}^{i}}= m^0_I \left(-a^{i}_{I}-\sum_{I \neq J} \frac{G_{AB}m_{J}}{R^2} n^i_{AB} + \frac{1}{c^2}C_{I}^{i} \right)~,
\label{eq:CT2}
\end{equation}
where $C_{I}^{i}$ can be derived using the harmonic-coordinate Lagrangian given in Ref.~\cite{Bernard:2018ivi}. 

The 1PN term $C_{I}^{i}$ depends on accelerations, therefore it will give additional acceleration dependent 
terms (on multiplying with 2PN contribution of contact transformation $\delta Z_{I}^{i}$) in the Lagrangian at 3PN order. 
Therefore, as shown in Ref.~\cite{deAndrade:2000gf} in the GR case, we will have to introduce some sort 
of `counter term', $X_{I}^i$, for Scalar Tensor theory as well to eliminate the acceleration dependence from 
the Lagrangian. Hence, the contact transformation at 3PN order becomes,~\cite{deAndrade:2000gf},
\begin{align}
\delta Z_{I}^i=\frac{1}{m^0_I}\left(q_{A}^i+\frac{\partial \mathcal{F}}{\partial V_{I}^{i}}+X_{I}^i \right)~,
\label{eq:CT3}
\end{align}
with the counter term, $X_{I}^i $ as defined in Eq.~(3.17) of Ref.~\cite{deAndrade:2000gf},
and $q_{I}^i = \frac{\partial \mathcal{L}}{\partial a_I^i}$ is the conjugate momenta of acceleration.

\section{Contact transformation for scalar-tensor theory at 3PN}
\label{App:B}
The final result of our contact transformation dependent on the parameters of the function $\mathcal{F}$ are as follows. 
The first and second term of the Eq.~\eqref{eq:CT3} are readily obtained by differentiating Lagrangian and function 
$\mathcal{F}$ w.r.t acceleration, and velocity, respectively. The third term, i.e.\ the counter term $X_{A,B}^i$,
purely 3PN order term, is: 
\small{\begin{widetext}
\begin{align}
X_{A}^i=&\vec{V_B^i}\left(G_{AB} m^0_A m^0_B (\vec{V}_{A}\cdot\vec{V}_{A})\left[-\frac{1}{2} f_2  (\vec{n}_{AB}\cdot\vec{V}_{A})+\left(\frac{1}{2} f_5 +\frac{1}{2} \bar{\gamma}_{AB}+\frac{7}{8} \right) \vec{n}_{AB}\cdot\vec{V}_{B} \right]\right.\nonumber\\
&\left.+\frac{G_{AB}^2 (m^0_A)^2 m^0_B}{R}\left[f_3\left(7+4 \bar{\gamma}_{AB}\right)\vec{n}_{AB}\cdot\vec{V}_{A}-f_6\left(7+4 \bar{\gamma}_{AB}\right)\vec{n}_{AB}\cdot\vec{V}_{B}\right]\right.\nonumber\\
&\left.+\frac{G_{AB}^2 m^0_A (m^0_B)^2}{R}\left[f_2(-3-2\bar{\gamma}_{AB}) \vec{n}_{AB}\cdot\vec{V}_{A}+\left(2 \bar{\gamma}_{AB} ^2+\frac{13 \bar{\gamma}_{AB} }{2}+2 \bar{\gamma}_{AB}  f_5+3 f_5+\frac{21}{4}\right)\vec{n}_{AB}\cdot\vec{V}_{B}\right]\right)\nonumber \\
&+ \vec{V}_{A}^i\left(G_{AB} m^0_A m^0_B \left[\left(-f_3-\frac{1}{2}\bar{\gamma}_{AB}-\frac{7}{8}\right) \vec{n}_{AB}\cdot\vec{V}_{A}(\vec{V}_{B}\cdot\vec{V}_{B})+ \left(-2 f_2 \vec{n}_{AB}\cdot\vec{V}_{A}+\left(f_5+\bar{\gamma}_{AB}+\frac{7}{4}\right)\vec{n}_{AB}\cdot\vec{V}_{B}\right)\vec{V}_{A}\cdot\vec{V}_{B}\right.\right.\nonumber\\
&\left.\left.+\left(3 f_4 \vec{n}_{AB}\cdot\vec{V}_{B}-4 f_1 \vec{n}_{AB}\cdot\vec{V}_{A}\right)\vec{V}_{A}\cdot\vec{V}_{A}-3 f_7(\vec{n}_{AB}\cdot\vec{V}_{A})^3-2 f_8 (\vec{n}_{AB}\cdot\vec{V}_{A})^2  \vec{n}_{AB}\cdot\vec{V}_{B}+(f_9+\frac{1}{8}) \vec{n}_{AB}\cdot\vec{V}_{A} (\vec{n}_{AB}\cdot\vec{V}_{B})^2\right]\right.\nonumber\\
&\left.+\frac{G_{AB}^2 (m^0_A)^2 m^0_B}{R}\left[\left(2 \bar{\gamma}_{AB} f_2+\frac{7}{2}f_2-f_{11} +  2 \bar{\gamma}_{AB}^2+7 \bar{\gamma}_{AB}+\frac{49}{8}\right)\vec{n}_{AB}\cdot\vec{V}_{A}+f_5(-2 \bar{\gamma}_{AB} - \frac{7}{2}) \vec{n}_{AB}\cdot\vec{V}_{B}\right]\right.\nonumber\\
&\left.+\frac{G_{AB}^2 m^0_A (m^0_B)^2}{R}\left[(-f_{12}-6 f_1- 4 \bar{\gamma}_{AB} f_1)  \vec{n}_{AB}\cdot\vec{V}_{A}+ f_4(6+4\bar{\gamma}_{AB}) \vec{n}_{AB}\cdot\vec{V}_{B}\right]\right)\nonumber \\
&+\vec{n}_{AB}^i \left( G_{AB} m^0_A m^0_B\left[\left(-\frac{3}{2} f_7 (\vec{n}_{AB}\cdot\vec{V}_{A})^2-f_8(\vec{n}_{AB}\cdot\vec{V}_{A})(\vec{n}_{AB}\cdot\vec{V}_{B})+\left(\frac{f_9}{2}+\frac{1}{16}\right) (\vec{n}_{AB}\cdot\vec{V}_{B})^2 + \left(-\frac{\bar{\gamma}_{AB} }{4}-\frac{f_3}{2}-\frac{7}{16}\right)\vec{V}_{B}\cdot\vec{V}_{B}\right.\right.\right.\nonumber\\
&\left.\left.\left.-\frac{1}{2} f_1(\vec{V}_{A}\cdot\vec{V}_{A})-\frac{1}{2} f_2(\vec{V}_{A}\cdot\vec{V}_{B}) \right)\vec{V}_{A}\cdot\vec{V}_{A}\right]+\frac{G_{AB}^2 (m^0_A)^2m^0_B}{R}\left[\left(-\frac{7}{2}-\bar{\gamma}_{AB} ^2-\frac{15 \bar{\gamma}_{AB} }{4}-2 \bar{\gamma}_{AB}  f_4-4 f_4-\frac{f_{11}}{2}\right)\vec{V}_{A}\cdot\vec{V}_{A}\right.\right.\nonumber\\
&\left.\left.+f_5(-4-2\bar{\gamma}_{AB})\vec{V}_{A}\cdot\vec{V}_{B}+f_6(-4-2\bar{\gamma}_{AB})\vec{V}_{B}\cdot\vec{V}_{B}+\left(\frac{11}{8}+\frac{3 \bar{\gamma}_{AB}}{2}+\frac{1}{2}f_2+f_8(4+2\bar{\gamma}_{AB})\right)(\vec{n}_{AB}\cdot\vec{V}_{A})^2\right.\right.\nonumber \\
&\left.\left.+\left(f_3-\frac{1}{2}f_5-4 f_9(2+\bar{\gamma}_{AB})\right)(\vec{n}_{AB}\cdot\vec{V}_{A})(\vec{n}_{AB}\cdot\vec{V}_{B})+\left(-f_6-6f_{10}(2+\bar{\gamma}_{AB})\right)(\vec{n}_{AB}\cdot\vec{V}_{B})^2\right]\right.\nonumber\\
&\left.+\frac{G_{AB}^2 m^0_A (m^0_B)^2}{R}\left[\left(f_1(-3-2\bar{\gamma}_{AB})-\frac{1}{2}f_{12}\right)\vec{V}_{A}\cdot\vec{V}_{A}+f_2(-3-2\bar{\gamma}_{AB})\vec{V}_{A}\cdot\vec{V}_{B}+\left(-\frac{21}{8}-\frac{13 \bar{\gamma}_{AB}}{4}-\bar{\gamma}_{AB}^2+f_3(-3-2\bar{\gamma}_{AB})\right)\vec{V}_{B}\cdot\vec{V}_{B}\right.\right.\nonumber\\
&\left.\left.+3f_7(-3-2\bar{\gamma}_{AB})(\vec{n}_{AB}\cdot\vec{V}_{A})^2+2f_8(-3-2\bar{\gamma}_{AB})(\vec{n}_{AB}\cdot\vec{V}_{A})(\vec{n}_{AB}\cdot\vec{V}_{B})+\left(\frac{3}{8}+\frac{\bar{\gamma}_{AB}}{4}+f_9(3+2\bar{\gamma}_{AB})\right)(\vec{n}_{AB}\cdot\vec{V}_{B})^2 \right] \right.\nonumber\\
&\left.+\frac{G_{AB}^3 (m^0_A)^3 m^0_B}{R^2}\left[f_{13}(-4-2\bar{\gamma}_{AB})\right]+\frac{G_{AB}^3(m^0_A)^2(m^0_B)^2}{R^2}\left[f_{11}(-3-2\bar{\gamma}_{AB})+f_{14}(-4-2\bar{\gamma}_{AB})\right]\right.\nonumber\\
&\left.+\frac{G_{AB}^3m^0_A (m^0_B)^3}{R^2}\left[f_{12}(-3-2\bar{\gamma}_{AB})\right]\right)
\label{eq:FCT1}
\end{align}
\end{widetext}}
\small{\begin{widetext}
\begin{align}
X_{B}^i&=\vec{V}_{B}^i \left(G_{AB} m^0_A m^0_B \left[ \left(f_4  +\frac{1}{2} \bar{\gamma}_{AB}+\frac{7}{8} \right)\vec{n}_{AB}\cdot\vec{V}_{B}(\vec{V}_{A}\cdot\vec{V}_{A})+\left(2 f_5 \vec{n}_{AB}\cdot\vec{V}_{B}-\left(f_2+\bar{\gamma}_{AB}+\frac{7}{4}\right) \vec{n}_{AB}\cdot\vec{V}_{A}\right)\vec{V}_{A}\cdot\vec{V}_{B} \right.\right.\nonumber\\
&\left.\left.+\left(4 f_6 \vec{n}_{AB}\cdot \vec{V}_{B}-3 f_3 \vec{n}_{AB}\cdot\vec{V}_{A}\right)\vec{V}_{B}\cdot\vec{V}_{B} +3 f_{10} \left(\vec{n}_{AB}\cdot\vec{V}_{B}\right)^3+2 f_9 (\vec{n}_{AB}\cdot\vec{V}_{A}) \left(\vec{n}_{AB}\cdot \vec{V}_{B}\right)^2-\left(f_8+\frac{1}{8}\right) \left(\vec{n}_{AB}\cdot\vec{V}_{A}\right)^2 \vec{n}_{AB}\cdot\vec{V}_{B}\right]\right.\nonumber\\
&\left.+\frac{G^2_{AB} (m^0_A)^2 m^0_B}{R}\left[\left(-4 \bar{\gamma}_{AB}  f_3 -6 f_3\right)\vec{n}_{AB}\cdot\vec{V}_{A}+\left(4 \bar{\gamma}_{AB}  f_6 + 6 f_6 + f_{13}\right) \vec{n}_{AB}\cdot\vec{V}_{B}\right]\right.\nonumber\\
&\left.+\frac{G^2_{AB} m^0_A (m^0_B)^2}{R}\left[\left(2 \bar{\gamma}_{AB}  f_2 +\frac{7}{2} f_2\right) \vec{n}_{AB}\cdot \vec{V}_{A}+\left(-2 \bar{\gamma}_{AB}  f_5 -\frac{7}{2} f_5 +f_{14} -2 \bar{\gamma}_{AB} ^2 -7 \bar{\gamma}_{AB} -\frac{49}{8}\right) \vec{n}_{AB}\cdot\vec{V}_{B}\right]\right)\nonumber\\
&+\vec{V}_{A}^i\left(G_{AB} m^0_A m^0_B~\vec{V}_{B}\cdot\vec{V}_{B} \left[\left(-\frac{1}{2} f_2 -\frac{1}{2} \bar{\gamma}_{AB}-\frac{7}{8}\right)\vec{n}_{AB}\cdot\vec{V}_{A}+\frac{1}{2} f_5~\vec{n}_{AB}\cdot\vec{V}_{B}\right]\right.\nonumber\\
&\left.+\frac{G_{AB}^2 (m^0_A)^2 m^0_B}{R} \left[\left(-2 \bar{\gamma}_{AB}  f_2 -3 f_2 -2 \bar{\gamma}_{AB} ^2 -\frac{13}{2} -\frac{21}{4}\right) \vec{n}_{AB}\cdot\vec{V}_{A}+\left(2 \bar{\gamma}_{AB}  f_5 +3 f_5\right) \vec{n}_{AB}\cdot\vec{V}_{B}\right]\right.\nonumber\\
&\left.+\frac{G_{AB}^2 m^0_A (m^0_B)^2}{R} \left[\left(4 \bar{\gamma}_{AB}  f_1 +7 f_1\right)\vec{n}_{AB}\cdot\vec{V}_{A}-\left(4 \bar{\gamma}_{AB}  f_4 +7 f_4\right) \vec{n}_{AB}\cdot \vec{V}_{B}\right]\right)\nonumber\\
&+\vec{n}_{AB}^i\left(G_{AB} m^0_A m^0_B \left[\left(-\frac{1}{2} f_8  \left(\vec{n}_{AB}\cdot\vec{V}_{A}\right)^2+f_9(\vec{n}_{AB}\cdot\vec{V}_{B})(\vec{n}_{AB}\cdot\vec{V}_{A})+\frac{3}{2} f_{10} \left(\vec{n}_{AB}\cdot\vec{V}_{B}\right)^2+\frac{1}{2} f_4 \vec{V}_{A}\cdot\vec{V}_{A}+\frac{1}{2} f_5 \vec{V}_{A}\cdot \vec{V}_{B}\right.\right.\right.\nonumber\\
&\left.\left.\left.+\frac{1}{2} f_6 \left(\vec{V}_{B}\cdot\vec{V}_{B}\right)-\frac{1}{16} \left(\vec{n}_{AB}\cdot\vec{V}_{A}\right)^2+\frac{1}{4} \bar{\gamma}_{AB}  \vec{V}_{A}\cdot \vec{V}_{A}+\frac{7}{16} \vec{V}_{A}\cdot\vec{V}_{A}\right) \vec{V}_{B}\cdot\vec{V}_{B}\right]\right.\nonumber\\
&\left.+\frac{G_{AB} (m^0_A)^2 m^0_B}{R} \left[\left(\bar{\gamma}_{AB} ^2+\frac{13 \bar{\gamma}_{AB} }{4}+(2 \bar{\gamma}_{AB}+3) f_4+\frac{21}{8}\right) \vec{V}_{A}\cdot\vec{V}_{A}+\left((2 \bar{\gamma}_{AB}+3) f_6+\frac{f_{13}}{2}\right) \vec{V}_{B}\cdot\vec{V}_{B}+f_5\left(2 \bar{\gamma}_{AB} +3 \right) \vec{V}_{A}\cdot\vec{V}_{B}\right.\right.\nonumber\\
&\left.\left.+\left((-2 \bar{\gamma}_{AB} -3)  f_8 -\frac{3}{8}-\frac{1}{4} \bar{\gamma}_{AB}\right)\left(\vec{n}_{AB}\cdot \vec{V}_{A}\right){}^2+f_9(4 \bar{\gamma}_{AB} +6) (\vec{n}_{AB}\cdot\vec{V}_{B}) (\vec{n}_{AB}\cdot\vec{V}_{A})+f_{10}(6 \bar{\gamma}_{AB}  +9) \left(\vec{n}_{AB}\cdot \vec{V}_{B}\right)^2\right]\right.\nonumber\\
&\left.+\frac{G_{AB}m^0_A (m^0_B)^2}{R} \left[f_1\left(2 \bar{\gamma}_{AB} +4 \right) \vec{V}_{A}\cdot\vec{V}_{A}+f_2\left(2 \bar{\gamma}_{AB} +4 \right) \vec{V}_{A}\cdot\vec{V}_{B}+\left(\bar{\gamma}_{AB} ^2+\frac{15 \bar{\gamma}_{AB} }{4}+(2 \bar{\gamma}_{AB}+4) f_3+\frac{f_{14}}{2}+\frac{7}{2}\right) \vec{V}_{B}\cdot\vec{V}_{B}\right.\right.\nonumber\\
&\left.\left.+\left((6 \bar{\gamma}_{AB} +12) f_7+f_1\right)\left(\vec{n}_{AB}\cdot\vec{V}_{A}\right)^2+\left((4 \bar{\gamma}_{AB} +8) f_8+\frac{f_2}{2}-f_4\right) (\vec{n}_{AB}\cdot\vec{V}_{A} \vec)({n_{12}}\cdot\vec{V}_{B})\right.\right.\nonumber\\
&\left.\left.+\left(-\frac{3 \bar{\gamma}_{AB} }{4}-(2 \bar{\gamma}_{AB}  +4) f_9-\frac{f_5}{2}-\frac{11}{8}\right) \left(\vec{n}_{AB}\cdot\vec{V}_{B}\right)^2\right]+\frac{G^3_{AB} (m^0_A)^3 m^0_B}{R^2}\left[f_{13}(2 \bar{\gamma}_{AB} +3)\right]\right.\nonumber\\
&\left.+\frac{G^3_{AB} (m^0_A)^2 (m^0_B)^2}{R^2}\left[f_{11}(2 \bar{\gamma}_{AB} +4)+f_{14}(2 \bar{\gamma}_{AB}+3)\right]+\frac{G^3_{AB}m^0_A (m^0_B)^3}{R^2} \left[f_{12}(2 \bar{\gamma}_{AB} +4 )\right]\right)
\label{eq:FCT2}
\end{align}
\end{widetext}}
Here, $f_{i}$ are the parameters of the generic function $\mathcal{F}$ at 2PN order given in Ref.~\cite{Julie:2017pkb}.

With these results, the full expression of the coordinate change (based on contact transformation) at 3PN, and that dependens 
on the parameters of the function $\mathcal{F}$, is derived. This is then used to obtain the class of ordinary Lagrangian in ST theory at 3PN\@.

\section{3PN two-body Hamiltonian}
\label{app:C}
For the whole class of ordinary Hamiltonian of Sec.~\ref{sec:COM3PNH}, the 15 coefficients $h_i^{\text{3PN}}$ 
of the Hamiltonian at 3PN for ST theory in COM frame are:

\begin{widetext}
\begin{align*}
h_1^{\text{3PN}} =& -\frac{5}{128}+\frac{35}{128}\nu-\frac{70}{128}\nu^2+\frac{35}{128}\nu^3, ~~~~~~~~~~~~h_2^{\text{3PN}}=h_3^{\text{3PN}}=h_4^{\text{3PN}}=h_5^{\text{3PN}}=0,\nonumber \\
h_6^{\text{3PN}}=&G_{AB}\left(-\frac{7}{16} -\frac{5}{16}\nu^3-\frac{3}{8}\bar{\gamma}_{AB}+\frac{m^0_A}{M}(2 f_6-F_{12})+\frac{m^0_B}{M}(2f_1-F_1)+\nu\left(4+\frac{23}{8}\bar{\gamma}_{AB}+ F_1+F_2-F_6-F_7+F_{11}+F_{12}-2[f_1+f_6]\right.\right.\nonumber\\
&\left.\left.-\frac{3}{2}[f_2+f_5]+\frac{3}{2}[f_3+f_4]+\frac{m^0_A}{M}\left[-F_2+F_7+3F_{12}+\frac{3}{2}f_2-\frac{3}{2}f_4-\frac{13}{2}f_6\right]+\frac{m^0_B}{M}\left[3F_1+F_6-F_{11}-\frac{13}{2}f_1-\frac{3}{2}f_3+\frac{3}{2}f_5\right]\right)\right.\nonumber\\
&\left.+\nu^2\left(-\frac{331}{16}-\frac{93}{8}\bar{\gamma}_{AB}-2[F_1+F_2-F_6-F_7+F_{11}+F_{12}]+\frac{9}{2}[f_1+f_6]+\frac{7}{2}[f_2-f_3-f_4+f_5]\right.\right.\nonumber\\
&\left.\left.+\frac{m^0_A}{M}\left[F_2+F_5-F_7-F_9-F_{10}-F_{12}-3f_2+3f_4+3f_6\right]+\frac{m^0_B}{M}\left[-F_1-F_3-F_4-F_6+F_8+F_{11}+3f_1+3f_3-3f_5\right]\right)\right),\nonumber\\ 
h_7^{\text{3PN}}=&G_{AB}\left(-\frac{3}{16}\nu^3+\frac{m^0_A}{M}[F_{12}-3F_{24}-2f_6+6f_{10}]+\frac{m^0_B}{M}[F_1-3F_{13}-2f_1+6f_7]+\nu\left(-\frac{27}{16}-\bar{\gamma}_{AB}-F_1-F_2+F_6+F_7-F_{11}\right.\right.\nonumber\\
&\left.\left.-F_{12}+3[F_{13}+F_{14}+F_{16}+F_{21}+F_{23}+F_{24}]+2[f_1+f_6]+\frac{3}{2}[f_2+f_5-f_3-f_4]-6[f_7+f_{10}]-\frac{9}{2}[f_8+f_9]\right.\right.\nonumber\\
&\left.\left.+\frac{m^0_A}{M}\left[F_2-F_7+\frac{13}{2}(f_6-3f_{10})-3\left(F_{12}+F_{14}+F_{16}-3F_{24}+\frac{1}{2}(f_2-f_4-3f_8)\right)\right]\right.\right.\nonumber\\
&\left.\left.+\frac{m^0_B}{M}\left[F_{11}-F_6-3\left(F_{1}+F_{21}+F_{23}-3F_{13}+\frac{1}{2}(f_5-f_3-3f_9)\right)+\frac{13}{2}(f_1-3f_{7})\right]\right)+\nu^2\left(\frac{169}{4}+\frac{89}{4}\bar{\gamma}_{AB}+2\left[F_1+F_2-F_6\right.\right.\right.\nonumber\\
&\left.\left.\left.-F_7+F_{11}+F_{12}-3F_{13}-3F_{14}-3F_{16}-3F_{21}-3F_{23}-3F_{24}\right]-\frac{9}{2}[f_1+f_6-3f_7-3f_{10}]+\frac{7}{2}[f_3-f_2-f_5+f_4+3f_8+3f_9]\right.\right.\nonumber\\
&\left.\left.+\frac{m^0_A}{M}\left[-F_2-F_5+F_7+F_9+F_{10}+F_{12}+3(F_{14}+F_{16}+F_{18}-F_{20}-F_{22}-F_{24}+f_2-f_4-f_6-3f_8+3f_{10})\right]\right.\right.\nonumber\\
&\left.\left.+\frac{m^0_B}{M}\left[F_1+F_3+F_4+F_6-F_{8}-F_{11}+3(-F_{13}-F_{15}-F_{17}+F_{19}+F_{21}+F_{23}-f_1-f_3+f_5+3f_7-3f_{9})\right]\right)\right)~,\nonumber\\ 
h_8^{\text{3PN}}=&G_{AB}\left(-\frac{3}{16}\nu^3+\frac{m^0_A}{M}[3F_{24}-5F_{30}-6f_{10}]+\frac{m^0_B}{M}[3F_{13}-5F_{25}-6f_7]+\nu\left(\frac{5}{16}+\frac{3}{8}\bar{\gamma}_{AB}-3[F_{13}+F_{14}+F_{16}+F_{21}+F_{23}+F_{24}]\right.\right.\nonumber\\
&\left.\left.+5[F_{25}+F_{26}+F_{29}+F_{30}]+6[f_7+f_{10}]+\frac{9}{2}[f_8+f_9]+3\frac{m^0_A}{M}\left[F_{14}+F_{16}-3F_{24}-\frac{5}{3}F_{26}+5F_{30}-\frac{3}{2}f_8+\frac{13}{2}f_{10}\right]\right.\right.\nonumber\\
&\left.\left.+3\frac{m^0_B}{M}\left[F_{21}+F_{23}-3F_{13}-\frac{5}{3}F_{29}+5F_{25}-\frac{3}{2}f_9+\frac{13}{2}f_{7}\right]\right)\right.\nonumber\\
&\left.+\nu^2\left(-\frac{117}{4}-\frac{129}{8}\bar{\gamma}_{AB}+6[F_{13}+F_{14}+F_{16}+F_{21}+F_{23}+F_{24}]-10[F_{25}+F_{26}+F_{29}+F_{30}]-\frac{21}{2}[f_8+f_9]-\frac{27}{2}[f_7+f_{10}]\right.\right.\nonumber\\
&\left.\left.+3\frac{m^0_A}{M}\left[-F_{14}-F_{16}-F_{18}+F_{20}+F_{22}+F_{24}+\frac{5}{3}(F_{26}-F_{28}-F_{30})+3(f_8-f_{10})\right]\right.\right.\nonumber\\
&\left.\left.+3\frac{m^0_B}{M}\left[F_{13}+F_{15}+F_{17}-F_{19}-F_{21}-F_{23}+\frac{5}{3}(F_{29}-F_{27}-F_{25})+3(f_9-f_7)\right]\right)\right)~,\nonumber\\ 
h_9^{\text{3PN}}=&G_{AB}\left(-\frac{5}{16}\nu^3+5\frac{m^0_A}{M}F_{30}+5\frac{m^0_B}{M}F_{25}+5 \nu\left(-F_{25}-F_{26}-F_{29}-F_{30}+\frac{m^0_A}{M}[F_{26}-3F_{30}]+\frac{m^0_B}{M}[F_{29}-3F_{25}]\right)\right.\nonumber\\
&\left.+5\nu^2\left(\frac{15}{16}+\frac{1}{2}\bar{\gamma}_{AB}+2[F_{25}+F_{26}+F_{29}+F_{30}]+\frac{m^0_A}{M}[F_{28}+F_{30}-F_{26}]+\frac{m^0_B}{M}[F_{25}+F_{27}-F_{29}]\right)\right)~,
\allowdisplaybreaks
\nonumber \\
h_{10}^{\text{3PN}}=&G_{AB}^2\left(-\frac{17 \bar{\gamma}_{AB} ^2}{16}-\frac{11 \bar{\gamma}_{AB} }{4}-\frac{29}{16}+\frac{m^0_A}{M} [F_{12}-F_{36}+(9 \bar{\gamma}_{AB}+12)  f_6+f_{13}]+\frac{m^0_B}{M} [F_1-F_{41}+(9 \bar{\gamma}_{AB}+12) f_1+f_{12}]\right.\nonumber\\
&\left.+\nu\left(\frac{41}{12}+\frac{199}{48} \bar{\gamma}_{AB} ^2+\frac{67 }{8} \bar{\gamma}_{AB}-F_1-F_{12}+F_{36}+F_{41}-(9 \bar{\gamma}_{AB} +12)[f_1+f_6]+\frac{1}{2}[f_{11}+f_{14}]-f_{12}-f_{13}\right.\right.\nonumber\\
&\left.\left.+\frac{m^0_A}{M} \left[F_6-F_{11}-3 F_{12}-F_{33}+F_{35}+2 F_{36}-F_{46}+9 \bar{\gamma}_{AB}  (f_3-  f_5- f_6)+\frac{1}{2}(27 f_3-26 f_5+f_{14}-11 f_6-5 f_{13})\right]
\right.\right.\nonumber\\
&\left.\left.+\frac{m^0_B}{M}\left[F_7-3 F_1-F_2-F_{31}+F_{42}+2F_{41}-F_{44}+9\bar{\gamma}_{AB}( f_4-f_1- f_2)+\frac{1}{2}(27 f_4-26 f_2+f_{11}-11 f_1-5 f_{12})\right]\right)\right.\nonumber\\
&\left.+\nu^2\left(\frac{395}{48}-\frac{77 \bar{\gamma}_{AB} ^2}{12}-\frac{29 \bar{\gamma}_{AB} }{4}+2 F_1+F_2-F_6-F_7+F_{11}+2 F_{12}+F_{31}+F_{32}+F_{33}-F_{34}-F_{35}-F_{36}-F_{41}-F_{42}-F_{43}\right.\right.\nonumber\\
&\left.\left.+F_{44}+F_{45}+F_{46}+\frac{1}{2}[f_3+f_4]-\frac{3}{2}[f_2+f_5]-\frac{13}{2}[f_1+f_6]+2[f_{12}+f_{13}-f_{11}-f_{14}]\right.\right.\nonumber\\
&\left.\left.+\frac{m^0_A}{M}\left[F_{12}-F_5-F_6+F_9+F_{10}+F_{11}+3(f_3-f_5-f_6)\right]+\frac{m^0_B}{M}\left[F_1+F_2+F_3+F_4- F_7-F_8+3(f_4-f_1-f_2)\right]\right)\right)~,
\nonumber \\
h_{11}^{\rm 3PN}=&G_{AB}^2\left(\frac{1}{16}(\bar{\gamma}_{AB}+2)^2+\frac{m_A^0}{M}\left[4F_{12}+3F_{24}+2F_{36}-3F_{40}-12f_6(2+\bar{\gamma}_{AB})+9f_{10}(4+3\bar{\gamma}_{AB})-2f_{13}\right]\right.\nonumber\\
&\left.+\frac{m^0_B}{M}\left[4F_1+3F_{13}+2F_{41}-3F_{47}-12f_1(2+\bar{\gamma}_{AB})+9f_7(4+3\bar{\gamma}_{AB})-2f_{12}\right]+\nu\left(\frac{211}{48}+\frac{71}{8}\bar{\gamma}_{AB}+\frac{137}{48}\bar{\gamma}_{AB}^2-4[F_1+F_{12}]\right.\right.\nonumber\\
&\left.\left.-3[F_{13}+F_{24}-F_{40}-F_{47}]-2[F_{36}+F_{41}]+12[2+\bar{\gamma}_{AB}](f_1+f_6)-9[4+3\bar{\gamma}_{AB}](f_7+f_{10})-[f_{11}+f_{14}]+2[f_{12}+f_{13}]\right.\right.\nonumber\\
&\left.\left.+\frac{m^0_A}{M}\left[4(F_6-F_{11}-3F_{12}-F_{36}]-3[F_{21}+F_{23}+3F_{24}-F_{39}-2F_{40}+F_{50}]+2[F_{33}-F_{35}+F_{46}]-12(2+\bar{\gamma}_{AB})[f_3-f_5-f_6]\right.\right.\right.\nonumber\\
&\left.\left.\left.-\frac{5}{2}f_5+\frac{19}{2}f_6-\frac{27}{2}f_9(3+2\bar{\gamma}_{AB})-\frac{27}{2}f_{10}(1+2\bar{\gamma}_{AB})+5f_{13}-f_{14}\right]\right.\right.\nonumber\\
&\left.\left.+\frac{m^0_B}{M}\left[4(F_7-F_2-3F_1-F_{41})-3(3F_{13}+F_{14}+F_{16}+F_{37}+2F_{47}+3F_{48})+2(F_{31}-F_{42}+F_{44})+12(2+\bar{\gamma}_{AB})[f_1+f_2-f_4]\right.\right.\right.\nonumber\\
&\left.\left.\left.+\frac{19}{2}f_1-\frac{5}{2}f_2-\frac{27}{2}f_7(1+2\bar{\gamma}_{AB})-\frac{27}{2}f_8(3+2\bar{\gamma}_{AB})-f_{11}+5f_{12}\right]\right)+\nu^2\left(\frac{889}{12}+63\bar{\gamma}_{AB}+\frac{149}{12}\bar{\gamma}_{AB}^2\right.\right.\nonumber\\
&\left.\left.+4[2F_1+F_2-F_6-F_7+F_{11}+2F_{12}]+3[2F_{13}+F_{14}+F_{16}+F_{21}+F_{23}+2F_{24}+F_{37}+F_{38}-F_{39}-F_{40}-F_{47}-F_{48}+F_{49}+F_{50}]\right.\right.\nonumber\\
&\left.\left.-2[F_{31}+F_{32}+F_{33}-F_{34}-F_{35}-F_{36}-F_{41}-F_{42}-F_{43}+F_{44}+F_{45}+F_{46}]-\frac{19}{2}[f_1+f_6]+5[f_3+f_4]-\frac{45}{2}[f_7+f_{10}]\right.\right.\nonumber\\
&\left.\left.-\frac{9}{2}[f_8+f_9]+4[f_{11}-f_{12}-f_{13}+f_{14}]+\frac{m_A^0}{M}\left[4(-F_5-F_6+F_9+F_{10}+F_{11}+F_{12})+3(-F_{18}+F_{20}+F_{21}+F_{22}+F_{23}+F_{24})\right.\right.\right.\nonumber\\
&\left.\left.\left.+\frac{9}{2}(f_3-f_5-f_6-2f_9-2f_{10})\right]+\frac{m_B^0}{M}\left[4(F_1+F_2+F_3+F_4-F_7-F_8)+3(F_{13}+F_{14}+F_{15}+F_{16}+F_{17}-F_{19})\right.\right.\right.\nonumber\\
&\left.\left.\left.+\frac{9}{2}(f_4-f_1-f_2-2F_7-2f_8)\right]\right)\right)~,\nonumber\\
h_{12}^{\text{3PN}}=&G_{AB}^2\left(\frac{m^0_A}{M} [2 F_{24}+5 F_{30}+4 F_{40}-30 \bar{\gamma}_{AB}  f_{10}-48 f_{10}]+\frac{m^0_B}{M} [2 F_{13}+5 F_{25}+4 F_{47}-30 \bar{\gamma}_{AB}  f_7-48 f_7]\right.\nonumber\\
&\left.+\nu\left(\frac{5 \bar{\gamma}_{AB} ^2}{36}+\frac{19 \bar{\gamma}_{AB} }{12}-2 F_{13}-2 F_{24}-5 F_{25}-5 F_{30}-4 F_{40}-4 F_{47}+30 \bar{\gamma}_{AB}  f_7+30 \bar{\gamma}_{AB}  f_{10}+48 f_7+48 f_{10}+\frac{227}{72}\right.\right.\nonumber\\
&\left.\left.+\frac{m^0_B}{M} \left[-6 F_{13}-2 F_{14}-2 F_{16}-15 F_{25}-5 F_{26}+4 F_{37}-8 F_{47}-4 F_{48}+30 \bar{\gamma}_{AB}  f_7+30 \bar{\gamma}_{AB}  f_8-f_1+2 f_2+\frac{81 f_7}{2}+51 f_8\right]\right.\right.\nonumber\\
&\left.\left.+\frac{m^0_A}{M} \left[-2 F_{21}-2 F_{23}-6 F_{24}-5 F_{29}-15 F_{30}-4 F_{39}-8 F_{40}+4 F_{50}+30 \bar{\gamma}_{AB}  f_9+30 \bar{\gamma}_{AB}  f_{10}+2 f_5-f_6+51 f_9+\frac{81 f_{10}}{2}\right]\right)\right.\nonumber\\
&\left.+\nu^2\left(\frac{5 \bar{\gamma}_{AB} ^2}{9}-\frac{7 \bar{\gamma}_{AB} }{2}+4 F_{13}+2 F_{14}+2 F_{16}+2 F_{21}+2 F_{23}+4 F_{24}+10 F_{25}+5 F_{26}+5 F_{29}+10 F_{30}-4 F_{37}-4 F_{38}+4 F_{39}\right.\right.\nonumber\\
&\left.\left.+4 F_{40}+4 F_{47}+4 F_{48}-4 F_{49}-4 F_{50}-f_3-f_4+6 f_7+\frac{3 f_8}{2}+\frac{3 f_9}{2}+6 f_{10}-\frac{581}{144}\right.\right.\nonumber\\
&\left.\left.+\frac{m^0_B}{M} \left[2 F_{13}+2 F_{14}+2 F_{15}+2 F_{16}+2 F_{17}-2 F_{19}+5 F_{25}+5 F_{26}+5 F_{27}-2 f_1-3 f_2-3 f_3+f_6+\frac{3 f_9}{2}+\frac{3 f_{10}}{2}\right]\right.\right.\nonumber\\
&\left.\left.+\frac{m^0_A}{M} \left[-2 F_{18}+2 F_{20}+2 F_{21}+2 F_{22}+2 F_{23}+2 F_{24}+5 F_{28}+5 F_{29}+5 F_{30}+f_1-3 f_4-3 f_5-2 f_6+\frac{3 f_7}{2}+\frac{3 f_8}{2}\right]\right)\right)~,
\allowdisplaybreaks
\nonumber \\
h_{13}^{\text{3PN}}=&\frac{G_{AB}^3\nu}{\alpha_{AB}}\left(\frac{11}{24}[\bar{\gamma}_{AB} +2] \bar{\gamma}_{AB} -\frac{(\bar{\gamma}_{AB} +10) (\delta_A m^0_A+\delta_B m^0_B)}{(6 \bar{\gamma}_{AB} +12) M}\right)+G_{AB}^3\left(-\bar{\gamma}_{AB} ^3-\frac{29 \bar{\gamma}_{AB} ^2}{6}-\frac{47 \bar{\gamma}_{AB} }{6}-\frac{17}{4}-\frac{2 (f_6 m^0_A\bar{\beta}_{B} +f_1 m^0_B \bar{\beta}_{A})}{M}\right.\nonumber\\
&\left.+\frac{m^0_A}{M} [F_{36}-F_{54}-2 (3 \bar{\gamma}_{AB} +5) f_6+(5 \bar{\gamma}_{AB} +7) f_{13}]+\frac{m^0_B}{M} [F_{41}-F_{55}-2 (3 \bar{\gamma}_{AB} +5) f_1+(5 \bar{\gamma}_{AB} +7) f_{12}]+\frac{m^0_B {\epsilon}_A +m^0_A {\epsilon}_B}{12 M}\right.\nonumber\\
&\left.+[2 \bar{\gamma}_{AB} +3] \frac{m^0_A\bar{\beta}_{B} +m^0_B\bar{\beta}_{A}}{2 M}-[\frac{4 \bar{\gamma}_{AB} }{3}+2] \frac{m^0_A \delta_{A} +m^0_B \delta_{B}}{M}+\nu\left([\frac{7 \pi ^2}{256}-\frac{11}{24}]\bar{\gamma}_{AB} ^3-[\frac{5 \pi ^2}{64}+\frac{923}{72}] \bar{\gamma}_{AB} ^2-[\frac{1421}{36}+\frac{75 \pi ^2 }{128}]\bar{\gamma}_{AB} \right.\right.\nonumber\\
&\left.\left.-\frac{1531}{48}-\frac{41 \pi ^2}{64}-F_{36}-F_{41}+F_{54}+F_{55}+2 (3 \bar{\gamma}_{AB} +5) (f_1+f_6)-[5 \bar{\gamma}_{AB} +\frac{7}{2}](f_{12}+f_{13})+(5 \bar{\gamma}_{AB} +7) (f_{11}+f_{14})\right.\right.\nonumber\\
&\left.\left. +\frac{m^0_A}{M}[F_{33}-F_{35}-2 F_{36}+F_{46}-F_{52}-F_{53}+F_{54}+6 \bar{\gamma}_{AB}  (f_5+f_6-f_3)-11 f_3+\frac{21 f_5}{2}+8 f_6+\frac{3 f_{11}}{2}-\frac{7 f_{12}}{2}]\right.\right.\nonumber\\
&\left.\left.+\frac{m^0_B}{M} [F_{31}-2 F_{41}-F_{42}+F_{44}-F_{51}+F_{55}-F_{56}+6 \bar{\gamma}_{AB}  (f_1+ f_2- f_4)+8 f_1+\frac{21 f_2}{2}-11 f_4-\frac{7 f_{13}}{2}+\frac{3 f_{14}}{2}]\right.\right.\nonumber\\
&\left.\left.-[7 \bar{\gamma}_{AB} +\frac{37}{4}] \frac{m^0_A\bar{\beta}_{B}+m^0_B\bar{\beta}_{A}}{ M}+[\frac{5}{2}+\bar{\gamma}_{AB}]\frac{m^0_A\bar{\beta}_A+m^0_B\bar{\beta}_B}{M}-2f_1\frac{m^0_B\bar{\beta}_B}{M}-2f_6\frac{m^0_A\bar{\beta}_A}{M}+2[2f_6+f_5- f_3] \frac{m^0_A}{M}\bar{\beta}_{B} \right.\right.\nonumber\\
&\left.\left.+2[2 f_1+ f_2- f_4]\frac{m^0_B}{M}\bar{\beta}_{A}-[\frac{34}{9}+\frac{2}{3}\bar{\gamma}_{AB} -\frac{7}{64} \pi ^2 (\frac{\bar{\gamma}_{AB} }{2}+1)](\delta_{A}+\delta_{B})+[\frac{\bar{\gamma}_{AB}}{6}-\frac{7}{18}] \frac{\delta_{A} m^0_A+\delta_{B} m^0_B}{ M}+\frac{4 (\bar{\beta}_{A} \delta_{A}+\bar{\beta}_{B} \delta_{B})}{\bar{\gamma}_{AB} }-\frac{8 \bar{\beta}_{A} \bar{\beta}_{B}}{\bar{\gamma}_{AB}}\right.\right.\nonumber\\
&\left.\left.-\frac{m^0_A \epsilon_A+m^0_B \epsilon_B}{12 M}\right)+\nu^2\left(\frac{13 \bar{\gamma}_{AB} ^2}{2}+\frac{58 \bar{\gamma}_{AB} }{3}+\frac{157}{12}-F_{31}-F_{32}-F_{33}+F_{34}+F_{35}+F_{36}+F_{41}+F_{42}+F_{43}-F_{44}-F_{45}-F_{46}\right.\right.\nonumber\\
&\left.\left.+f_2+f_5+2 [f_1+ f_6+ f_{11}- f_{12}- f_{13}+ f_{14}]+2( \bar{\beta}_{B} - \bar{\beta}_{A}) [f_1+f_2+f_3-f_4-  f_5- f_6]-\frac{1}{2}[\bar{\beta}_A+\bar{\beta}_B]+\frac{4 \bar{\beta}_{A} \bar{\beta}_{B}}{\bar{\gamma}_{AB} }+\delta_{A}+\delta_{B}\right.\right.\nonumber\\
&\left.\left.+\frac{1}{12}[\epsilon_A+\epsilon_B]\right)\right)~,
\nonumber\\
h_{14}^{\text{3PN}}=&\frac{G_{AB}^3\nu}{\alpha_{AB}}\left(-\frac{11}{4}[1+\frac{\bar{\gamma}_{AB}}{2}]\bar{\gamma}_{AB}+\frac{(\bar{\gamma}_{AB} +10) (\delta_A m^0_A+\delta_B m^0_B)}{(2 \bar{\gamma}_{AB} +4) M}\right)+G_{AB}^3\left(\frac{\bar{\gamma}_{AB} ^3}{4}+\frac{11 \bar{\gamma}_{AB} ^2}{8}+\frac{5 \bar{\gamma}_{AB} }{2}+\frac{3}{2}\right.\nonumber\\
&\left.+\frac{m^0_A}{M} [2 F_{36}+3 (F_{40}+ F_{54})-(5+3\bar{\gamma}_{AB})(4f_6+6f_{10})-2(8+5\bar{\gamma}_{AB})f_{13}-2(2f_6+3f_{10})\bar{\beta}_B+(\frac{3}{2}+\bar{\gamma}_{AB})  \delta_A]\right.\nonumber\\
&\left.+\frac{m^0_B}{M} [2 F_{41}+3 (F_{47}+ F_{55})-(5+3\bar{\gamma}_{AB})(4f_1+6f_7)-2(8+5\bar{\gamma}_{AB})f_{12}-2(2f_1+3f_7)\bar{\beta}_A+(\frac{3}{2}+\bar{\gamma}_{AB})\delta_B]\right.\nonumber\\
&\left.+\nu\left([\frac{23}{8}-\frac{21 \pi ^2}{256}] \bar{\gamma}_{AB} ^3+[\frac{15 \pi ^2}{64}+\frac{713}{24}]\bar{\gamma}_{AB} ^2+[\frac{225 \pi ^2 }{128}+\frac{971}{12}]\bar{\gamma}_{AB} +\frac{123 \pi ^2}{64}+\frac{265}{4}-2[F_{36}+F_{41}]-3[F_{40}+F_{47}+F_{54}+F_{55}]\right.\right.\nonumber\\
&\left.\left.+[5+3\bar{\gamma}_{AB}](4[f_1+f_6]+6[f_7+f_{10}])-2[8+5\bar{\gamma}_{AB}](f_{11}+f_{14})+\frac{5}{2}[5+4\bar{\gamma}_{AB}](f_{12}+f_{13})\right.\right.\nonumber\\
&\left.\left.+\frac{m^0_A}{M} \left[2(F_{33}-F_{35}-2F_{36}-3F_{40}+F_{46})-3(F_{39}-F_{50}-F_{52}-F_{53}+F_{54})-(11+6\bar{\gamma}_{AB})(2f_3-3f_9)+\frac{3}{2}(15+8\bar{\gamma}_{AB})f_5\right.\right.\right.\nonumber\\
&\left.\left.\left.+(13+12\bar{\gamma}_{AB})f_6+3(7+6\bar{\gamma}_{AB})f_{10}-2 f_{11}+\frac{7 f_{12}}{2}+2(2 f_1-2f_6+3 f_7-3f_{10}) \bar{\beta}_A \right.\right.\right.\nonumber\\
&\left.\left.\left.+2(\frac{47}{8}+5\bar{\gamma}_{AB}-2f_3+2f_5+6f_6+3f_9+9f_{10})\bar{\beta}_B+(8+\frac{1}{2} \bar{\gamma}_{AB})  \delta_A+(\frac{77}{6}+\bar{\gamma}_{AB})\delta_B -\frac{3}{4} \epsilon_B\right]\right.\right.\nonumber\\
&\left.\left.+\frac{m^0_B}{M} \left[2(F_{31}-2F_{41}-F_{42}+F_{44}-3F_{47})+3(F_{37}-F_{48}+F_{51}-F_{55}+F_{56})+(13+12\bar{\gamma}_{AB})f_1+\frac{3}{2}(15+8\bar{\gamma}_{AB})f_2+3(7+6\bar{\gamma}_{AB})f_7\right.\right.\right.\nonumber\\
&\left.\left.\left.-(11+6\bar{\gamma}_{AB})(2f_4-3f_8)+\frac{7 f_{13}}{2}-2 f_{14}+2(\frac{47}{8}+5\bar{\gamma}_{AB}+6f_1+2f_2-2f_4+9f_7+3f_8)\bar{\beta}_A+2(-2f_1+2f_6-3f_7+3f_{10})\bar{\beta}_B\right.\right.\right.\nonumber\\
&\left.\left.\left.+(\frac{77}{6}+\bar{\gamma}_{AB})  \delta_A+(8+\frac{1}{2} \bar{\gamma}_{AB})  \delta_B-\frac{3}{4} \epsilon_A\right]-\frac{21}{64} \pi ^2[1+\frac{1}{2}\bar{\gamma}_{AB}] \delta_A-\frac{21}{64} \pi ^2[1+\frac{1}{2}\bar{\gamma}_{AB}] \delta_B-\frac{12 \bar{\beta}_A \delta_A}{\bar{\gamma}_{AB} }-\frac{12 \bar{\beta}_B \delta_B}{\bar{\gamma}_{AB} }+\frac{8 \bar{\beta}_A \bar{\beta}_B}{\bar{\gamma}_{AB} }\right) \right.\nonumber\\
&\left.+\nu^2\left(\frac{27 \bar{\gamma}_{AB} ^2}{2}+53 \bar{\gamma}_{AB}+\frac{197}{4}+2[-F_{31}- F_{32}- F_{33}+ F_{34}+ F_{35}+ F_{36}+ F_{41}+ F_{42}+ F_{43}- F_{44}- F_{45}- F_{46}]\right.\right.\nonumber\\
&\left.\left.+3[- F_{37}- F_{38}+ F_{39}+ F_{40}+ F_{47}+F_{48}- F_{49}- F_{50}]+7 f_1+2 f_2-3 f_3-3 f_4+2 f_5+7 f_6+9 f_7+3 f_8+3 f_9+9 f_{10}\right.\right.\nonumber\\
&\left.\left.+\frac{3}{2}[f_{12}+f_{13}-f_{11}-f_{14}]+2(\bar{\beta}_B-\bar{\beta}_A)[ 2 f_1+2 f_2+2 f_3-2f_4-2f_5-2f_6+3f_7+3f_8-3f_9-3f_{10}]-3[\bar{\beta}_A+\bar{\beta}_B]\right.\right.\nonumber\\
&\left.\left.+ \frac{12 \bar{\beta}_A \bar{\beta}_B}{\bar{\gamma}_{AB} }+3 [\delta_A+ \delta_B]+\frac{1}{4}[\epsilon_A+\epsilon_B]\right)\right)~,
\allowdisplaybreaks
\nonumber\\
h_{15}^{\text{3PN}}=&\frac{G_{AB}^4}{\alpha_{AB}}~\nu\left(\frac{11}{12}\bar{\gamma}_{AB}[2+\bar{\gamma}_{AB}]+\left(\frac{5-\bar{\gamma}_{AB}}{6+3\bar{\gamma}_{AB}}\right)\frac{m_A^0\delta_A+m_B^0\delta_B}{M}\right)+G_{AB}^4\left(\frac{3}{8}+\frac{1}{3}\bar{\gamma}_{AB}+\frac{1}{12}\bar{\gamma}_{AB}^2+(\frac{7}{6}+\frac{2}{3}\bar{\gamma}_{AB}+\frac{1}{6}\bar{\gamma}_{AB}^2)\frac{m^0_B\bar{\beta}_A+m^0_A\bar{\beta}_B}{M}\right.\nonumber\\
&\left.+\frac{1}{3}\frac{m^0_A{\delta}_A+m^0_B{\delta}_B}{M}+\frac{1}{6}\frac{m_A^0\epsilon_B+m_B^0\epsilon_A}{M}+\frac{1}{3}\frac{m_A^0\bar{\kappa}_B+m_B^0\bar{\kappa}_A}{M}+\frac{m^0_A}{M}\left[F_{54}-2(2+\bar{\gamma}_{AB})f_{13}+\frac{1}{2}\bar{\beta}_B^2+\frac{2}{3}\bar{\beta}_B\delta_A-2f_{13}\bar{\beta}_B\right]\right.\nonumber\\
&\left.+\frac{m^0_B}{M}\left[F_{55}-2(2+\bar{\gamma}_{AB})f_{12}+\frac{1}{2}\bar{\beta}_A^2+\frac{2}{3}\bar{\beta}_A\delta_B-2f_{12}\bar{\beta}_A\right]+\nu\left(\frac{641}{21}+\frac{799}{36}\bar{\gamma}_{AB}+\frac{49}{12}\bar{\gamma}_{AB}^2-\frac{11}{48}\bar{\gamma}_{AB}^3+2\bar{\beta}_A\bar{\beta}_B\left(1-\frac{8}{\bar{\gamma}_{AB}}\right)\right.\right.\nonumber\\
&\left.\left.+2\bar{\gamma}_{AB}[-f_{11}+f_{12}+f_{13}-f_{14}]-\frac{3}{2}\frac{m_A^0\bar{\beta}_A+m^0_B\bar{\beta}_B}{M}+(2f_{12}-2f_{13})\frac{m^0_A\bar{\beta}_A-m^0_B\bar{\beta}_B}{M}-\frac{4}{3M\bar{\gamma}_{AB}}\left(\bar{\beta}_A\delta_A+\bar{\beta}_B\delta_B\right)\right.\right.\nonumber\\
&\left.\left.+\left(\frac{209}{12}+\frac{20}{3}\bar{\gamma}_{AB}-\frac{1}{3}\bar{\gamma}_{AB}^2-2f_{11}-2f_{14}+\frac{16\bar{\beta}_A\bar{\beta}_B}{\bar{\gamma}_{AB}^2}\right)\frac{m_A^0\bar{\beta}_B+m^0_B\bar{\beta}_A}{M}+\left(\frac{5}{18}+\frac{1}{12}\bar{\gamma}_{AB}\right)\frac{m^0_A\delta_A+m^0_B\delta_B}{M}+\frac{19}{18}\frac{m^0_A\delta_B+m^0_B\delta_A}{M}\right.\right.\nonumber\\
&\left.\left.-\frac{1}{6}\frac{m^0_A\epsilon_A+m^0_B\epsilon_B}{M}+\frac{1}{6}\frac{m^0_A\epsilon_B+m^0_B\epsilon_A}{M}-\frac{1}{3}\frac{m^0_A\bar{\kappa}_A+m^0_B\bar{\kappa}_B}{M}-\frac{2}{3}\frac{m^0_A\bar{\kappa}_B+m^0_B\bar{\kappa}_A}{M}-\frac{2}{M~\bar{\gamma}_{AB}}\left(m^0_A\bar{\beta}_A\epsilon_B+m^0_B\bar{\beta}_B\epsilon_A\right)\right.\right.\nonumber\\
&\left.\left.+\frac{m^0_A}{M}\left[F_{52}+F_{53}-2F_{54}-F_{55}-5f_{11}+4f_{12}+3f_{13}-4f_{14}+4f_{13}\bar{\beta}_B-\frac{1}{2}\bar{\beta}_A^2+\bar{\beta}_B^2+\frac{2}{3}\left(\left[1-\frac{4}{\bar{\gamma}_{AB}}\right]\bar{\beta}_A-\bar{\beta}_B\right)\delta_A-\frac{2}{3}\bar{\beta}_A\delta_B\right]\right.\right.\nonumber\\
&\left.\left.+\frac{m^0_B}{M}\left[F_{51}-F_{54}-2F_{55}+F_{56}-4f_{11}+3f_{12}+4f_{13}-5f_{14}+4f_{12}\bar{\beta}_A+\bar{\beta}_A^2-\frac{1}{2}\bar{\beta}_B^2-\frac{2}{3}\bar{\beta}_B\delta_A+\frac{2}{3}\left(\left[1-\frac{4}{\bar{\gamma}_{AB}}\right]\bar{\beta}_B-\bar{\beta}_A\right)\delta_B\right]\right)\right)~.
\end{align*}
\end{widetext}

As a consistency check with the results of GR,  we find that our ST 3PN Hamiltonian reduces to  Hamiltonian of Ref.~\cite{Damour:1999cr} in
the GR limit by setting the $\mathcal{F}_{\rm 3PN}$ parameters as,
\begin{align*}
F_2&=F_{11}=2 F_{27}=2 F_{28}=6F_{16}=6F_{21}=-\frac{15}{8}~,\\
F_{3}&=F_{10}=2 F_6=2 F_7=\frac{2}{5}F_{34}=\frac{2}{5}F_{43}=-\frac{7}{4}~,\\F_{33}&=F_{44}=-F_{54}=-F_{55}=-\frac{2}{3}F_{15}=-\frac{2}{3}F_{22}=-\frac{3}{4}~,\\
F_5&=F_8=\frac{47}{8},\hspace{0.5cm}F_4=F_{9}=-\frac{23}{4},\hspace{0.5cm}F_{14}=F_{23}=\frac{5}{12}~,
\end{align*}
\begin{align*}
F_{17}&=F_{20}=\frac{41}{8},\hspace{0.5cm}F_{18}=F_{19}=-\frac{27}{16}~,\hspace{0.5cm}F_{31}=F_{46}=-\frac{113}{24}~,\\
F_{32}&=F_{45}=\frac{195}{16},\hspace{0.5cm}F_{36}=F_{41}=\frac{1}{8}~,\hspace{0.5cm}F_{37}=F_{50}=-\frac{91}{144}~,\\
F_{38}&=F_{49}=\frac{21}{16}~,\hspace{0.5cm}F_{51}=F_{52}=\frac{245}{18}-\frac{21}{32}\pi^2~,\\
F_{53}&=F_{56}=-\frac{1079}{72},
\end{align*}
with the remaining parameters of $\mathcal{F}_{\text{3PN}}$ set to zero and $\mathcal{F}_{\text{2PN}}$ 
coinciding with the expression in Ref.~\cite{Julie:2017pkb}.

\bibliography{local.bib,refs.bib}

\begin{thebibliography}{75}%
\makeatletter
\providecommand \@ifxundefined [1]{%
 \@ifx{#1\undefined}
}%
\providecommand \@ifnum [1]{%
 \ifnum #1\expandafter \@firstoftwo
 \else \expandafter \@secondoftwo
 \fi
}%
\providecommand \@ifx [1]{%
 \ifx #1\expandafter \@firstoftwo
 \else \expandafter \@secondoftwo
 \fi
}%
\providecommand \natexlab [1]{#1}%
\providecommand \enquote  [1]{``#1''}%
\providecommand \bibnamefont  [1]{#1}%
\providecommand \bibfnamefont [1]{#1}%
\providecommand \citenamefont [1]{#1}%
\providecommand \href@noop [0]{\@secondoftwo}%
\providecommand \href [0]{\begingroup \@sanitize@url \@href}%
\providecommand \@href[1]{\@@startlink{#1}\@@href}%
\providecommand \@@href[1]{\endgroup#1\@@endlink}%
\providecommand \@sanitize@url [0]{\catcode `\\12\catcode `\$12\catcode
  `\&12\catcode `\#12\catcode `\^12\catcode `\_12\catcode `\%12\relax}%
\providecommand \@@startlink[1]{}%
\providecommand \@@endlink[0]{}%
\providecommand \url  [0]{\begingroup\@sanitize@url \@url }%
\providecommand \@url [1]{\endgroup\@href {#1}{\urlprefix }}%
\providecommand \urlprefix  [0]{URL }%
\providecommand \Eprint [0]{\href }%
\providecommand \doibase [0]{https://doi.org/}%
\providecommand \selectlanguage [0]{\@gobble}%
\providecommand \bibinfo  [0]{\@secondoftwo}%
\providecommand \bibfield  [0]{\@secondoftwo}%
\providecommand \translation [1]{[#1]}%
\providecommand \BibitemOpen [0]{}%
\providecommand \bibitemStop [0]{}%
\providecommand \bibitemNoStop [0]{.\EOS\space}%
\providecommand \EOS [0]{\spacefactor3000\relax}%
\providecommand \BibitemShut  [1]{\csname bibitem#1\endcsname}%
\let\auto@bib@innerbib\@empty
\bibitem [{\citenamefont {Damour}\ and\ \citenamefont
  {Esposito-Farese}(1992)}]{Damour:1992we}%
  \BibitemOpen
  \bibfield  {author} {\bibinfo {author} {\bibfnamefont {T.}~\bibnamefont
  {Damour}}\ and\ \bibinfo {author} {\bibfnamefont {G.}~\bibnamefont
  {Esposito-Farese}},\ }\bibfield  {title} {\bibinfo {title} {{Tensor
  multiscalar theories of gravitation}},\ }\href
  {https://doi.org/10.1088/0264-9381/9/9/015} {\bibfield  {journal} {\bibinfo
  {journal} {Class. Quant. Grav.}\ }\textbf {\bibinfo {volume} {9}},\ \bibinfo
  {pages} {2093} (\bibinfo {year} {1992})}\BibitemShut {NoStop}%
\bibitem [{\citenamefont {Damour}\ and\ \citenamefont
  {Esposito-Farese}(1993)}]{Damour:1993hw}%
  \BibitemOpen
  \bibfield  {author} {\bibinfo {author} {\bibfnamefont {T.}~\bibnamefont
  {Damour}}\ and\ \bibinfo {author} {\bibfnamefont {G.}~\bibnamefont
  {Esposito-Farese}},\ }\bibfield  {title} {\bibinfo {title} {{Nonperturbative
  strong field effects in tensor - scalar theories of gravitation}},\ }\href
  {https://doi.org/10.1103/PhysRevLett.70.2220} {\bibfield  {journal} {\bibinfo
   {journal} {Phys. Rev. Lett.}\ }\textbf {\bibinfo {volume} {70}},\ \bibinfo
  {pages} {2220} (\bibinfo {year} {1993})}\BibitemShut {NoStop}%
\bibitem [{\citenamefont {Damour}\ and\ \citenamefont
  {Esposito-Farese}(1996)}]{Damour:1995kt}%
  \BibitemOpen
  \bibfield  {author} {\bibinfo {author} {\bibfnamefont {T.}~\bibnamefont
  {Damour}}\ and\ \bibinfo {author} {\bibfnamefont {G.}~\bibnamefont
  {Esposito-Farese}},\ }\bibfield  {title} {\bibinfo {title} {{Testing gravity
  to second postNewtonian order: A Field theory approach}},\ }\href
  {https://doi.org/10.1103/PhysRevD.53.5541} {\bibfield  {journal} {\bibinfo
  {journal} {Phys. Rev. D}\ }\textbf {\bibinfo {volume} {53}},\ \bibinfo
  {pages} {5541} (\bibinfo {year} {1996})},\ \Eprint
  {https://arxiv.org/abs/gr-qc/9506063} {arXiv:gr-qc/9506063} \BibitemShut
  {NoStop}%
\bibitem [{\citenamefont {Freire}\ \emph {et~al.}(2012)\citenamefont {Freire},
  \citenamefont {Wex}, \citenamefont {Esposito-Farese}, \citenamefont
  {Verbiest}, \citenamefont {Bailes}, \citenamefont {Jacoby}, \citenamefont
  {Kramer}, \citenamefont {Stairs}, \citenamefont {Antoniadis},\ and\
  \citenamefont {Janssen}}]{Freire:2012mg}%
  \BibitemOpen
  \bibfield  {author} {\bibinfo {author} {\bibfnamefont {P.~C.~C.}\
  \bibnamefont {Freire}}, \bibinfo {author} {\bibfnamefont {N.}~\bibnamefont
  {Wex}}, \bibinfo {author} {\bibfnamefont {G.}~\bibnamefont
  {Esposito-Farese}}, \bibinfo {author} {\bibfnamefont {J.~P.~W.}\ \bibnamefont
  {Verbiest}}, \bibinfo {author} {\bibfnamefont {M.}~\bibnamefont {Bailes}},
  \bibinfo {author} {\bibfnamefont {B.~A.}\ \bibnamefont {Jacoby}}, \bibinfo
  {author} {\bibfnamefont {M.}~\bibnamefont {Kramer}}, \bibinfo {author}
  {\bibfnamefont {I.~H.}\ \bibnamefont {Stairs}}, \bibinfo {author}
  {\bibfnamefont {J.}~\bibnamefont {Antoniadis}},\ and\ \bibinfo {author}
  {\bibfnamefont {G.~H.}\ \bibnamefont {Janssen}},\ }\bibfield  {title}
  {\bibinfo {title} {{The relativistic pulsar-white dwarf binary PSR J1738+0333
  II. The most stringent test of scalar-tensor gravity}},\ }\href
  {https://doi.org/10.1111/j.1365-2966.2012.21253.x} {\bibfield  {journal}
  {\bibinfo  {journal} {Mon. Not. Roy. Astron. Soc.}\ }\textbf {\bibinfo
  {volume} {423}},\ \bibinfo {pages} {3328} (\bibinfo {year} {2012})},\ \Eprint
  {https://arxiv.org/abs/1205.1450} {arXiv:1205.1450 [astro-ph.GA]}
  \BibitemShut {NoStop}%
\bibitem [{\citenamefont {Khalil}\ \emph
  {et~al.}(2022{\natexlab{a}})\citenamefont {Khalil}, \citenamefont {Mendes},
  \citenamefont {Ortiz},\ and\ \citenamefont {Steinhoff}}]{Khalil:2022sii}%
  \BibitemOpen
  \bibfield  {author} {\bibinfo {author} {\bibfnamefont {M.}~\bibnamefont
  {Khalil}}, \bibinfo {author} {\bibfnamefont {R.~F.~P.}\ \bibnamefont
  {Mendes}}, \bibinfo {author} {\bibfnamefont {N.}~\bibnamefont {Ortiz}},\ and\
  \bibinfo {author} {\bibfnamefont {J.}~\bibnamefont {Steinhoff}},\ }\bibfield
  {title} {\bibinfo {title} {{Effective-action model for dynamical
  scalarization beyond the adiabatic approximation}},\ }\href@noop {} {\
  (\bibinfo {year} {2022}{\natexlab{a}})},\ \Eprint
  {https://arxiv.org/abs/2206.13233} {arXiv:2206.13233 [gr-qc]} \BibitemShut
  {NoStop}%
\bibitem [{\citenamefont {Gautam}\ \emph {et~al.}(2022)\citenamefont {Gautam},
  \citenamefont {Freire}, \citenamefont {Batrakov}, \citenamefont {Kramer},
  \citenamefont {Miao}, \citenamefont {Parent},\ and\ \citenamefont
  {Zhu}}]{Gautam:2022cpb}%
  \BibitemOpen
  \bibfield  {author} {\bibinfo {author} {\bibfnamefont {T.}~\bibnamefont
  {Gautam}}, \bibinfo {author} {\bibfnamefont {P.~C.~C.}\ \bibnamefont
  {Freire}}, \bibinfo {author} {\bibfnamefont {A.}~\bibnamefont {Batrakov}},
  \bibinfo {author} {\bibfnamefont {M.}~\bibnamefont {Kramer}}, \bibinfo
  {author} {\bibfnamefont {C.~C.}\ \bibnamefont {Miao}}, \bibinfo {author}
  {\bibfnamefont {E.}~\bibnamefont {Parent}},\ and\ \bibinfo {author}
  {\bibfnamefont {W.~W.}\ \bibnamefont {Zhu}},\ }\bibfield  {title} {\bibinfo
  {title} {{Relativistic effects in a mildly recycled pulsar binary: PSR
  J1952+2630}},\ }\href@noop {} {\  (\bibinfo {year} {2022})},\ \Eprint
  {https://arxiv.org/abs/2210.03464} {arXiv:2210.03464 [astro-ph.HE]}
  \BibitemShut {NoStop}%
\bibitem [{\citenamefont {Doneva}\ \emph {et~al.}(2022)\citenamefont {Doneva},
  \citenamefont {Ramazano\u{g}lu}, \citenamefont {Silva}, \citenamefont
  {Sotiriou},\ and\ \citenamefont {Yazadjiev}}]{Doneva:2022ewd}%
  \BibitemOpen
  \bibfield  {author} {\bibinfo {author} {\bibfnamefont {D.~D.}\ \bibnamefont
  {Doneva}}, \bibinfo {author} {\bibfnamefont {F.~M.}\ \bibnamefont
  {Ramazano\u{g}lu}}, \bibinfo {author} {\bibfnamefont {H.~O.}\ \bibnamefont
  {Silva}}, \bibinfo {author} {\bibfnamefont {T.~P.}\ \bibnamefont
  {Sotiriou}},\ and\ \bibinfo {author} {\bibfnamefont {S.~S.}\ \bibnamefont
  {Yazadjiev}},\ }\bibfield  {title} {\bibinfo {title} {{Scalarization}},\
  }\href@noop {} {\  (\bibinfo {year} {2022})},\ \Eprint
  {https://arxiv.org/abs/2211.01766} {arXiv:2211.01766 [gr-qc]} \BibitemShut
  {NoStop}%
\bibitem [{\citenamefont {Abbott}\ \emph
  {et~al.}(2016{\natexlab{a}})\citenamefont {Abbott} \emph
  {et~al.}}]{Abbott:2016blz}%
  \BibitemOpen
  \bibfield  {author} {\bibinfo {author} {\bibfnamefont {B.~P.}\ \bibnamefont
  {Abbott}} \emph {et~al.} (\bibinfo {collaboration} {Virgo, LIGO
  Scientific}),\ }\bibfield  {title} {\bibinfo {title} {{Observation of
  Gravitational Waves from a Binary Black Hole Merger}},\ }\href
  {https://doi.org/10.1103/PhysRevLett.116.061102} {\bibfield  {journal}
  {\bibinfo  {journal} {Phys. Rev. Lett.}\ }\textbf {\bibinfo {volume} {116}},\
  \bibinfo {pages} {061102} (\bibinfo {year} {2016}{\natexlab{a}})},\ \Eprint
  {https://arxiv.org/abs/1602.03837} {arXiv:1602.03837 [gr-qc]} \BibitemShut
  {NoStop}%
\bibitem [{\citenamefont {Arun}\ \emph {et~al.}(2006)\citenamefont {Arun},
  \citenamefont {Iyer}, \citenamefont {Qusailah},\ and\ \citenamefont
  {Sathyaprakash}}]{Arun:2006yw}%
  \BibitemOpen
  \bibfield  {author} {\bibinfo {author} {\bibfnamefont {K.~G.}\ \bibnamefont
  {Arun}}, \bibinfo {author} {\bibfnamefont {B.~R.}\ \bibnamefont {Iyer}},
  \bibinfo {author} {\bibfnamefont {M.~S.~S.}\ \bibnamefont {Qusailah}},\ and\
  \bibinfo {author} {\bibfnamefont {B.~S.}\ \bibnamefont {Sathyaprakash}},\
  }\bibfield  {title} {\bibinfo {title} {{Testing post-Newtonian theory with
  gravitational wave observations}},\ }\href
  {https://doi.org/10.1088/0264-9381/23/9/L01} {\bibfield  {journal} {\bibinfo
  {journal} {Class. Quant. Grav.}\ }\textbf {\bibinfo {volume} {23}},\ \bibinfo
  {pages} {L37} (\bibinfo {year} {2006})},\ \Eprint
  {https://arxiv.org/abs/gr-qc/0604018} {arXiv:gr-qc/0604018} \BibitemShut
  {NoStop}%
\bibitem [{\citenamefont {Mishra}\ \emph {et~al.}(2010)\citenamefont {Mishra},
  \citenamefont {Arun}, \citenamefont {Iyer},\ and\ \citenamefont
  {Sathyaprakash}}]{Mishra:2010tp}%
  \BibitemOpen
  \bibfield  {author} {\bibinfo {author} {\bibfnamefont {C.~K.}\ \bibnamefont
  {Mishra}}, \bibinfo {author} {\bibfnamefont {K.~G.}\ \bibnamefont {Arun}},
  \bibinfo {author} {\bibfnamefont {B.~R.}\ \bibnamefont {Iyer}},\ and\
  \bibinfo {author} {\bibfnamefont {B.~S.}\ \bibnamefont {Sathyaprakash}},\
  }\bibfield  {title} {\bibinfo {title} {{Parametrized tests of post-Newtonian
  theory using Advanced LIGO and Einstein Telescope}},\ }\href
  {https://doi.org/10.1103/PhysRevD.82.064010} {\bibfield  {journal} {\bibinfo
  {journal} {Phys. Rev. D}\ }\textbf {\bibinfo {volume} {82}},\ \bibinfo
  {pages} {064010} (\bibinfo {year} {2010})},\ \Eprint
  {https://arxiv.org/abs/1005.0304} {arXiv:1005.0304 [gr-qc]} \BibitemShut
  {NoStop}%
\bibitem [{\citenamefont {Li}\ \emph {et~al.}(2012)\citenamefont {Li},
  \citenamefont {Del~Pozzo}, \citenamefont {Vitale}, \citenamefont {Van
  Den~Broeck}, \citenamefont {Agathos} \emph {et~al.}}]{Li:2011cg}%
  \BibitemOpen
  \bibfield  {author} {\bibinfo {author} {\bibfnamefont {T.}~\bibnamefont
  {Li}}, \bibinfo {author} {\bibfnamefont {W.}~\bibnamefont {Del~Pozzo}},
  \bibinfo {author} {\bibfnamefont {S.}~\bibnamefont {Vitale}}, \bibinfo
  {author} {\bibfnamefont {C.}~\bibnamefont {Van Den~Broeck}}, \bibinfo
  {author} {\bibfnamefont {M.}~\bibnamefont {Agathos}}, \emph {et~al.},\
  }\bibfield  {title} {\bibinfo {title} {{Towards a generic test of the strong
  field dynamics of general relativity using compact binary coalescence}},\
  }\href {https://doi.org/10.1103/PhysRevD.85.082003} {\bibfield  {journal}
  {\bibinfo  {journal} {Phys.Rev.}\ }\textbf {\bibinfo {volume} {D85}},\
  \bibinfo {pages} {082003} (\bibinfo {year} {2012})},\ \Eprint
  {https://arxiv.org/abs/1110.0530} {arXiv:1110.0530 [gr-qc]} \BibitemShut
  {NoStop}%
\bibitem [{\citenamefont {Agathos}\ \emph {et~al.}(2014)\citenamefont
  {Agathos}, \citenamefont {Del~Pozzo}, \citenamefont {Li}, \citenamefont
  {Broeck}, \citenamefont {Veitch} \emph {et~al.}}]{Agathos:2013upa}%
  \BibitemOpen
  \bibfield  {author} {\bibinfo {author} {\bibfnamefont {M.}~\bibnamefont
  {Agathos}}, \bibinfo {author} {\bibfnamefont {W.}~\bibnamefont {Del~Pozzo}},
  \bibinfo {author} {\bibfnamefont {T.~G.~F.}\ \bibnamefont {Li}}, \bibinfo
  {author} {\bibfnamefont {C.~V.~D.}\ \bibnamefont {Broeck}}, \bibinfo {author}
  {\bibfnamefont {J.}~\bibnamefont {Veitch}}, \emph {et~al.},\ }\bibfield
  {title} {\bibinfo {title} {{TIGER: A data analysis pipeline for testing the
  strong-field dynamics of general relativity with gravitational wave signals
  from coalescing compact binaries}},\ }\href
  {https://doi.org/10.1103/PhysRevD.89.082001} {\bibfield  {journal} {\bibinfo
  {journal} {Phys.Rev.}\ }\textbf {\bibinfo {volume} {D89}},\ \bibinfo {pages}
  {082001} (\bibinfo {year} {2014})},\ \Eprint
  {https://arxiv.org/abs/1311.0420} {arXiv:1311.0420 [gr-qc]} \BibitemShut
  {NoStop}%
\bibitem [{\citenamefont {Cornish}\ \emph {et~al.}(2011)\citenamefont
  {Cornish}, \citenamefont {Sampson}, \citenamefont {Yunes},\ and\
  \citenamefont {Pretorius}}]{Cornish:2011ys}%
  \BibitemOpen
  \bibfield  {author} {\bibinfo {author} {\bibfnamefont {N.}~\bibnamefont
  {Cornish}}, \bibinfo {author} {\bibfnamefont {L.}~\bibnamefont {Sampson}},
  \bibinfo {author} {\bibfnamefont {N.}~\bibnamefont {Yunes}},\ and\ \bibinfo
  {author} {\bibfnamefont {F.}~\bibnamefont {Pretorius}},\ }\bibfield  {title}
  {\bibinfo {title} {{Gravitational Wave Tests of General Relativity with the
  Parameterized Post-Einsteinian Framework}},\ }\href
  {https://doi.org/10.1103/PhysRevD.84.062003} {\bibfield  {journal} {\bibinfo
  {journal} {Phys. Rev. D}\ }\textbf {\bibinfo {volume} {84}},\ \bibinfo
  {pages} {062003} (\bibinfo {year} {2011})},\ \Eprint
  {https://arxiv.org/abs/1105.2088} {arXiv:1105.2088 [gr-qc]} \BibitemShut
  {NoStop}%
\bibitem [{\citenamefont {Berti}\ \emph {et~al.}(2015)\citenamefont {Berti}
  \emph {et~al.}}]{Berti:2015itd}%
  \BibitemOpen
  \bibfield  {author} {\bibinfo {author} {\bibfnamefont {E.}~\bibnamefont
  {Berti}} \emph {et~al.},\ }\bibfield  {title} {\bibinfo {title} {{Testing
  General Relativity with Present and Future Astrophysical Observations}},\
  }\href {https://doi.org/10.1088/0264-9381/32/24/243001} {\bibfield  {journal}
  {\bibinfo  {journal} {Class. Quant. Grav.}\ }\textbf {\bibinfo {volume}
  {32}},\ \bibinfo {pages} {243001} (\bibinfo {year} {2015})},\ \Eprint
  {https://arxiv.org/abs/1501.07274} {arXiv:1501.07274 [gr-qc]} \BibitemShut
  {NoStop}%
\bibitem [{\citenamefont {Yunes}\ \emph {et~al.}(2016)\citenamefont {Yunes},
  \citenamefont {Yagi},\ and\ \citenamefont {Pretorius}}]{Yunes:2016jcc}%
  \BibitemOpen
  \bibfield  {author} {\bibinfo {author} {\bibfnamefont {N.}~\bibnamefont
  {Yunes}}, \bibinfo {author} {\bibfnamefont {K.}~\bibnamefont {Yagi}},\ and\
  \bibinfo {author} {\bibfnamefont {F.}~\bibnamefont {Pretorius}},\ }\bibfield
  {title} {\bibinfo {title} {{Theoretical Physics Implications of the Binary
  Black-Hole Mergers GW150914 and GW151226}},\ }\href
  {https://doi.org/10.1103/PhysRevD.94.084002} {\bibfield  {journal} {\bibinfo
  {journal} {Phys. Rev. D}\ }\textbf {\bibinfo {volume} {94}},\ \bibinfo
  {pages} {084002} (\bibinfo {year} {2016})},\ \Eprint
  {https://arxiv.org/abs/1603.08955} {arXiv:1603.08955 [gr-qc]} \BibitemShut
  {NoStop}%
\bibitem [{\citenamefont {Abbott}\ \emph
  {et~al.}(2016{\natexlab{b}})\citenamefont {Abbott} \emph
  {et~al.}}]{TheLIGOScientific:2016src}%
  \BibitemOpen
  \bibfield  {author} {\bibinfo {author} {\bibfnamefont {B.~P.}\ \bibnamefont
  {Abbott}} \emph {et~al.} (\bibinfo {collaboration} {LIGO Scientific,
  Virgo}),\ }\bibfield  {title} {\bibinfo {title} {{Tests of general relativity
  with GW150914}},\ }\href {https://doi.org/10.1103/PhysRevLett.116.221101,
  10.1103/PhysRevLett.121.129902} {\bibfield  {journal} {\bibinfo  {journal}
  {Phys. Rev. Lett.}\ }\textbf {\bibinfo {volume} {116}},\ \bibinfo {pages}
  {221101} (\bibinfo {year} {2016}{\natexlab{b}})},\ \bibinfo {note} {[Erratum:
  Phys. Rev. Lett.121,no.12,129902(2018)]},\ \Eprint
  {https://arxiv.org/abs/1602.03841} {arXiv:1602.03841 [gr-qc]} \BibitemShut
  {NoStop}%
\bibitem [{\citenamefont {Aasi}\ \emph {et~al.}(2015)\citenamefont {Aasi} \emph
  {et~al.}}]{TheLIGOScientific:2014jea}%
  \BibitemOpen
  \bibfield  {author} {\bibinfo {author} {\bibfnamefont {J.}~\bibnamefont
  {Aasi}} \emph {et~al.} (\bibinfo {collaboration} {LIGO Scientific}),\
  }\bibfield  {title} {\bibinfo {title} {{Advanced LIGO}},\ }\href
  {https://doi.org/10.1088/0264-9381/32/7/074001} {\bibfield  {journal}
  {\bibinfo  {journal} {Class. Quant. Grav.}\ }\textbf {\bibinfo {volume}
  {32}},\ \bibinfo {pages} {074001} (\bibinfo {year} {2015})},\ \Eprint
  {https://arxiv.org/abs/1411.4547} {arXiv:1411.4547 [gr-qc]} \BibitemShut
  {NoStop}%
\bibitem [{\citenamefont {Acernese}\ \emph {et~al.}(2015)\citenamefont
  {Acernese} \emph {et~al.}}]{TheVirgo:2014hva}%
  \BibitemOpen
  \bibfield  {author} {\bibinfo {author} {\bibfnamefont {F.}~\bibnamefont
  {Acernese}} \emph {et~al.} (\bibinfo {collaboration} {VIRGO}),\ }\bibfield
  {title} {\bibinfo {title} {{Advanced Virgo: a second-generation
  interferometric gravitational wave detector}},\ }\href
  {https://doi.org/10.1088/0264-9381/32/2/024001} {\bibfield  {journal}
  {\bibinfo  {journal} {Class. Quant. Grav.}\ }\textbf {\bibinfo {volume}
  {32}},\ \bibinfo {pages} {024001} (\bibinfo {year} {2015})},\ \Eprint
  {https://arxiv.org/abs/1408.3978} {arXiv:1408.3978 [gr-qc]} \BibitemShut
  {NoStop}%
\bibitem [{\citenamefont {Akutsu}\ \emph {et~al.}(2021)\citenamefont {Akutsu}
  \emph {et~al.}}]{KAGRA:2020agh}%
  \BibitemOpen
  \bibfield  {author} {\bibinfo {author} {\bibfnamefont {T.}~\bibnamefont
  {Akutsu}} \emph {et~al.} (\bibinfo {collaboration} {KAGRA}),\ }\bibfield
  {title} {\bibinfo {title} {{Overview of KAGRA: Calibration, detector
  characterization, physical environmental monitors, and the geophysics
  interferometer}},\ }\href {https://doi.org/10.1093/ptep/ptab018} {\bibfield
  {journal} {\bibinfo  {journal} {PTEP}\ }\textbf {\bibinfo {volume} {2021}},\
  \bibinfo {pages} {05A102} (\bibinfo {year} {2021})},\ \Eprint
  {https://arxiv.org/abs/2009.09305} {arXiv:2009.09305 [gr-qc]} \BibitemShut
  {NoStop}%
\bibitem [{\citenamefont {Abbott}\ \emph
  {et~al.}(2018{\natexlab{a}})\citenamefont {Abbott} \emph
  {et~al.}}]{Aasi:2013wya}%
  \BibitemOpen
  \bibfield  {author} {\bibinfo {author} {\bibfnamefont {B.~P.}\ \bibnamefont
  {Abbott}} \emph {et~al.} (\bibinfo {collaboration} {VIRGO, KAGRA, LIGO
  Scientific}),\ }\bibfield  {title} {\bibinfo {title} {{Prospects for
  Observing and Localizing Gravitational-Wave Transients with Advanced LIGO,
  Advanced Virgo and KAGRA}},\ }\href
  {https://doi.org/10.1007/s41114-018-0012-9, 10.1007/lrr-2016-1} {\bibfield
  {journal} {\bibinfo  {journal} {Living Rev. Rel.}\ }\textbf {\bibinfo
  {volume} {21}},\ \bibinfo {pages} {3} (\bibinfo {year}
  {2018}{\natexlab{a}})},\ \bibinfo {note} {[Living Rev. Rel.19,1(2016)]},\
  \Eprint {https://arxiv.org/abs/1304.0670} {arXiv:1304.0670 [gr-qc]}
  \BibitemShut {NoStop}%
\bibitem [{\citenamefont {Abbott}\ \emph
  {et~al.}(2019{\natexlab{a}})\citenamefont {Abbott} \emph
  {et~al.}}]{LIGOScientific:2019fpa}%
  \BibitemOpen
  \bibfield  {author} {\bibinfo {author} {\bibfnamefont {B.~P.}\ \bibnamefont
  {Abbott}} \emph {et~al.} (\bibinfo {collaboration} {LIGO Scientific,
  Virgo}),\ }\bibfield  {title} {\bibinfo {title} {{Tests of General Relativity
  with the Binary Black Hole Signals from the LIGO-Virgo Catalog GWTC-1}},\
  }\href {https://doi.org/10.1103/PhysRevD.100.104036} {\bibfield  {journal}
  {\bibinfo  {journal} {Phys. Rev. D}\ }\textbf {\bibinfo {volume} {100}},\
  \bibinfo {pages} {104036} (\bibinfo {year} {2019}{\natexlab{a}})},\ \Eprint
  {https://arxiv.org/abs/1903.04467} {arXiv:1903.04467 [gr-qc]} \BibitemShut
  {NoStop}%
\bibitem [{\citenamefont {Abbott}\ \emph
  {et~al.}(2019{\natexlab{b}})\citenamefont {Abbott} \emph
  {et~al.}}]{LIGOScientific:2018dkp}%
  \BibitemOpen
  \bibfield  {author} {\bibinfo {author} {\bibfnamefont {B.~P.}\ \bibnamefont
  {Abbott}} \emph {et~al.} (\bibinfo {collaboration} {LIGO Scientific,
  Virgo}),\ }\bibfield  {title} {\bibinfo {title} {{Tests of General Relativity
  with GW170817}},\ }\href {https://doi.org/10.1103/PhysRevLett.123.011102}
  {\bibfield  {journal} {\bibinfo  {journal} {Phys. Rev. Lett.}\ }\textbf
  {\bibinfo {volume} {123}},\ \bibinfo {pages} {011102} (\bibinfo {year}
  {2019}{\natexlab{b}})},\ \Eprint {https://arxiv.org/abs/1811.00364}
  {arXiv:1811.00364 [gr-qc]} \BibitemShut {NoStop}%
\bibitem [{\citenamefont {Abbott}\ \emph
  {et~al.}(2021{\natexlab{a}})\citenamefont {Abbott} \emph
  {et~al.}}]{LIGOScientific:2020tif}%
  \BibitemOpen
  \bibfield  {author} {\bibinfo {author} {\bibfnamefont {R.}~\bibnamefont
  {Abbott}} \emph {et~al.} (\bibinfo {collaboration} {LIGO Scientific,
  Virgo}),\ }\bibfield  {title} {\bibinfo {title} {{Tests of general relativity
  with binary black holes from the second LIGO-Virgo gravitational-wave
  transient catalog}},\ }\href {https://doi.org/10.1103/PhysRevD.103.122002}
  {\bibfield  {journal} {\bibinfo  {journal} {Phys. Rev. D}\ }\textbf {\bibinfo
  {volume} {103}},\ \bibinfo {pages} {122002} (\bibinfo {year}
  {2021}{\natexlab{a}})},\ \Eprint {https://arxiv.org/abs/2010.14529}
  {arXiv:2010.14529 [gr-qc]} \BibitemShut {NoStop}%
\bibitem [{\citenamefont {Abbott}\ \emph
  {et~al.}(2021{\natexlab{b}})\citenamefont {Abbott} \emph
  {et~al.}}]{LIGOScientific:2021sio}%
  \BibitemOpen
  \bibfield  {author} {\bibinfo {author} {\bibfnamefont {R.}~\bibnamefont
  {Abbott}} \emph {et~al.} (\bibinfo {collaboration} {LIGO Scientific, VIRGO,
  KAGRA}),\ }\bibfield  {title} {\bibinfo {title} {{Tests of General Relativity
  with GWTC-3}},\ }\href@noop {} {\  (\bibinfo {year} {2021}{\natexlab{b}})},\
  \Eprint {https://arxiv.org/abs/2112.06861} {arXiv:2112.06861 [gr-qc]}
  \BibitemShut {NoStop}%
\bibitem [{\citenamefont {Maggiore}\ \emph {et~al.}(2020)\citenamefont
  {Maggiore} \emph {et~al.}}]{Maggiore:2019uih}%
  \BibitemOpen
  \bibfield  {author} {\bibinfo {author} {\bibfnamefont {M.}~\bibnamefont
  {Maggiore}} \emph {et~al.},\ }\bibfield  {title} {\bibinfo {title} {{Science
  Case for the Einstein Telescope}},\ }\href
  {https://doi.org/10.1088/1475-7516/2020/03/050} {\bibfield  {journal}
  {\bibinfo  {journal} {JCAP}\ }\textbf {\bibinfo {volume} {03}},\ \bibinfo
  {pages} {050}},\ \Eprint {https://arxiv.org/abs/1912.02622} {arXiv:1912.02622
  [astro-ph.CO]} \BibitemShut {NoStop}%
\bibitem [{\citenamefont {Evans}\ \emph {et~al.}(2021)\citenamefont {Evans}
  \emph {et~al.}}]{Evans:2021gyd}%
  \BibitemOpen
  \bibfield  {author} {\bibinfo {author} {\bibfnamefont {M.}~\bibnamefont
  {Evans}} \emph {et~al.},\ }\bibfield  {title} {\bibinfo {title} {{A Horizon
  Study for Cosmic Explorer: Science, Observatories, and Community}},\
  }\href@noop {} {\  (\bibinfo {year} {2021})},\ \Eprint
  {https://arxiv.org/abs/2109.09882} {arXiv:2109.09882 [astro-ph.IM]}
  \BibitemShut {NoStop}%
\bibitem [{\citenamefont {Palenzuela}\ \emph {et~al.}(2014)\citenamefont
  {Palenzuela}, \citenamefont {Barausse}, \citenamefont {Ponce},\ and\
  \citenamefont {Lehner}}]{Palenzuela:2013hsa}%
  \BibitemOpen
  \bibfield  {author} {\bibinfo {author} {\bibfnamefont {C.}~\bibnamefont
  {Palenzuela}}, \bibinfo {author} {\bibfnamefont {E.}~\bibnamefont
  {Barausse}}, \bibinfo {author} {\bibfnamefont {M.}~\bibnamefont {Ponce}},\
  and\ \bibinfo {author} {\bibfnamefont {L.}~\bibnamefont {Lehner}},\
  }\bibfield  {title} {\bibinfo {title} {{Dynamical scalarization of neutron
  stars in scalar-tensor gravity theories}},\ }\href
  {https://doi.org/10.1103/PhysRevD.89.044024} {\bibfield  {journal} {\bibinfo
  {journal} {Phys. Rev. D}\ }\textbf {\bibinfo {volume} {89}},\ \bibinfo
  {pages} {044024} (\bibinfo {year} {2014})},\ \Eprint
  {https://arxiv.org/abs/1310.4481} {arXiv:1310.4481 [gr-qc]} \BibitemShut
  {NoStop}%
\bibitem [{\citenamefont {Sennett}\ and\ \citenamefont
  {Buonanno}(2016)}]{Sennett:2016rwa}%
  \BibitemOpen
  \bibfield  {author} {\bibinfo {author} {\bibfnamefont {N.}~\bibnamefont
  {Sennett}}\ and\ \bibinfo {author} {\bibfnamefont {A.}~\bibnamefont
  {Buonanno}},\ }\bibfield  {title} {\bibinfo {title} {{Modeling dynamical
  scalarization with a resummed post-Newtonian expansion}},\ }\href
  {https://doi.org/10.1103/PhysRevD.93.124004} {\bibfield  {journal} {\bibinfo
  {journal} {Phys. Rev. D}\ }\textbf {\bibinfo {volume} {93}},\ \bibinfo
  {pages} {124004} (\bibinfo {year} {2016})},\ \Eprint
  {https://arxiv.org/abs/1603.03300} {arXiv:1603.03300 [gr-qc]} \BibitemShut
  {NoStop}%
\bibitem [{\citenamefont {Lang}(2014)}]{Lang:2013fna}%
  \BibitemOpen
  \bibfield  {author} {\bibinfo {author} {\bibfnamefont {R.~N.}\ \bibnamefont
  {Lang}},\ }\bibfield  {title} {\bibinfo {title} {{Compact binary systems in
  scalar-tensor gravity. II. Tensor gravitational waves to second
  post-Newtonian order}},\ }\href {https://doi.org/10.1103/PhysRevD.89.084014}
  {\bibfield  {journal} {\bibinfo  {journal} {Phys. Rev. D}\ }\textbf {\bibinfo
  {volume} {89}},\ \bibinfo {pages} {084014} (\bibinfo {year} {2014})},\
  \Eprint {https://arxiv.org/abs/1310.3320} {arXiv:1310.3320 [gr-qc]}
  \BibitemShut {NoStop}%
\bibitem [{\citenamefont {Lang}(2015)}]{Lang:2014osa}%
  \BibitemOpen
  \bibfield  {author} {\bibinfo {author} {\bibfnamefont {R.~N.}\ \bibnamefont
  {Lang}},\ }\bibfield  {title} {\bibinfo {title} {{Compact binary systems in
  scalar-tensor gravity. III. Scalar waves and energy flux}},\ }\href
  {https://doi.org/10.1103/PhysRevD.91.084027} {\bibfield  {journal} {\bibinfo
  {journal} {Phys. Rev. D}\ }\textbf {\bibinfo {volume} {91}},\ \bibinfo
  {pages} {084027} (\bibinfo {year} {2015})},\ \Eprint
  {https://arxiv.org/abs/1411.3073} {arXiv:1411.3073 [gr-qc]} \BibitemShut
  {NoStop}%
\bibitem [{\citenamefont {Bernard}(2018)}]{Bernard:2018hta}%
  \BibitemOpen
  \bibfield  {author} {\bibinfo {author} {\bibfnamefont {L.}~\bibnamefont
  {Bernard}},\ }\bibfield  {title} {\bibinfo {title} {{Dynamics of compact
  binary systems in scalar-tensor theories: Equations of motion to the third
  post-Newtonian order}},\ }\href {https://doi.org/10.1103/PhysRevD.98.044004}
  {\bibfield  {journal} {\bibinfo  {journal} {Phys. Rev. D}\ }\textbf {\bibinfo
  {volume} {98}},\ \bibinfo {pages} {044004} (\bibinfo {year} {2018})},\
  \Eprint {https://arxiv.org/abs/1802.10201} {arXiv:1802.10201 [gr-qc]}
  \BibitemShut {NoStop}%
\bibitem [{\citenamefont {Bernard}(2019)}]{Bernard:2018ivi}%
  \BibitemOpen
  \bibfield  {author} {\bibinfo {author} {\bibfnamefont {L.}~\bibnamefont
  {Bernard}},\ }\bibfield  {title} {\bibinfo {title} {{Dynamics of compact
  binary systems in scalar-tensor theories: II. Center-of-mass and conserved
  quantities to 3PN order}},\ }\href
  {https://doi.org/10.1103/PhysRevD.99.044047} {\bibfield  {journal} {\bibinfo
  {journal} {Phys. Rev. D}\ }\textbf {\bibinfo {volume} {99}},\ \bibinfo
  {pages} {044047} (\bibinfo {year} {2019})},\ \Eprint
  {https://arxiv.org/abs/1812.04169} {arXiv:1812.04169 [gr-qc]} \BibitemShut
  {NoStop}%
\bibitem [{\citenamefont {Bernard}(2020)}]{Bernard:2019yfz}%
  \BibitemOpen
  \bibfield  {author} {\bibinfo {author} {\bibfnamefont {L.}~\bibnamefont
  {Bernard}},\ }\bibfield  {title} {\bibinfo {title} {{Dipolar tidal effects in
  scalar-tensor theories}},\ }\href
  {https://doi.org/10.1103/PhysRevD.101.021501} {\bibfield  {journal} {\bibinfo
   {journal} {Phys. Rev. D}\ }\textbf {\bibinfo {volume} {101}},\ \bibinfo
  {pages} {021501} (\bibinfo {year} {2020})},\ \Eprint
  {https://arxiv.org/abs/1906.10735} {arXiv:1906.10735 [gr-qc]} \BibitemShut
  {NoStop}%
\bibitem [{\citenamefont {Sch\"on}\ and\ \citenamefont
  {Doneva}(2022)}]{Schon:2021pcv}%
  \BibitemOpen
  \bibfield  {author} {\bibinfo {author} {\bibfnamefont {O.}~\bibnamefont
  {Sch\"on}}\ and\ \bibinfo {author} {\bibfnamefont {D.~D.}\ \bibnamefont
  {Doneva}},\ }\bibfield  {title} {\bibinfo {title} {{Tensor-multiscalar
  gravity: Equations of motion to 2.5 post-Newtonian order}},\ }\href
  {https://doi.org/10.1103/PhysRevD.105.064034} {\bibfield  {journal} {\bibinfo
   {journal} {Phys. Rev. D}\ }\textbf {\bibinfo {volume} {105}},\ \bibinfo
  {pages} {064034} (\bibinfo {year} {2022})},\ \Eprint
  {https://arxiv.org/abs/2112.07388} {arXiv:2112.07388 [gr-qc]} \BibitemShut
  {NoStop}%
\bibitem [{\citenamefont {Sennett}\ \emph {et~al.}(2016)\citenamefont
  {Sennett}, \citenamefont {Marsat},\ and\ \citenamefont
  {Buonanno}}]{Sennett:2016klh}%
  \BibitemOpen
  \bibfield  {author} {\bibinfo {author} {\bibfnamefont {N.}~\bibnamefont
  {Sennett}}, \bibinfo {author} {\bibfnamefont {S.}~\bibnamefont {Marsat}},\
  and\ \bibinfo {author} {\bibfnamefont {A.}~\bibnamefont {Buonanno}},\
  }\bibfield  {title} {\bibinfo {title} {{Gravitational waveforms in
  scalar-tensor gravity at 2PN relative order}},\ }\href
  {https://doi.org/10.1103/PhysRevD.94.084003} {\bibfield  {journal} {\bibinfo
  {journal} {Phys. Rev. D}\ }\textbf {\bibinfo {volume} {94}},\ \bibinfo
  {pages} {084003} (\bibinfo {year} {2016})},\ \Eprint
  {https://arxiv.org/abs/1607.01420} {arXiv:1607.01420 [gr-qc]} \BibitemShut
  {NoStop}%
\bibitem [{\citenamefont {Bernard}\ \emph {et~al.}(2022)\citenamefont
  {Bernard}, \citenamefont {Blanchet},\ and\ \citenamefont
  {Trestini}}]{Bernard:2022noq}%
  \BibitemOpen
  \bibfield  {author} {\bibinfo {author} {\bibfnamefont {L.}~\bibnamefont
  {Bernard}}, \bibinfo {author} {\bibfnamefont {L.}~\bibnamefont {Blanchet}},\
  and\ \bibinfo {author} {\bibfnamefont {D.}~\bibnamefont {Trestini}},\
  }\bibfield  {title} {\bibinfo {title} {{Gravitational waves in scalar-tensor
  theory to one-and-a-half post-Newtonian order}},\ }\href
  {https://doi.org/10.1088/1475-7516/2022/08/008} {\bibfield  {journal}
  {\bibinfo  {journal} {JCAP}\ }\textbf {\bibinfo {volume} {08}}\bibfield
  {number} {\bibinfo  {number} { (08)},\ \bibinfo {pages} {008}},\ }\Eprint
  {https://arxiv.org/abs/2201.10924} {arXiv:2201.10924 [gr-qc]} \BibitemShut
  {NoStop}%
\bibitem [{\citenamefont {Brown}(2022)}]{Brown:2022kbw}%
  \BibitemOpen
  \bibfield  {author} {\bibinfo {author} {\bibfnamefont {S.~M.}\ \bibnamefont
  {Brown}},\ }\bibfield  {title} {\bibinfo {title} {{Tidal Deformability of
  Neutron Stars in Scalar-Tensor Theories of Gravity for Gravitational Wave
  Analysis}},\ }\href@noop {} {\  (\bibinfo {year} {2022})},\ \Eprint
  {https://arxiv.org/abs/2210.14025} {arXiv:2210.14025 [gr-qc]} \BibitemShut
  {NoStop}%
\bibitem [{\citenamefont {Hinderer}\ \emph {et~al.}(2010)\citenamefont
  {Hinderer}, \citenamefont {Lackey}, \citenamefont {Lang},\ and\ \citenamefont
  {Read}}]{Hinderer:2009ca}%
  \BibitemOpen
  \bibfield  {author} {\bibinfo {author} {\bibfnamefont {T.}~\bibnamefont
  {Hinderer}}, \bibinfo {author} {\bibfnamefont {B.~D.}\ \bibnamefont
  {Lackey}}, \bibinfo {author} {\bibfnamefont {R.~N.}\ \bibnamefont {Lang}},\
  and\ \bibinfo {author} {\bibfnamefont {J.~S.}\ \bibnamefont {Read}},\
  }\bibfield  {title} {\bibinfo {title} {{Tidal deformability of neutron stars
  with realistic equations of state and their gravitational wave signatures in
  binary inspiral}},\ }\href {https://doi.org/10.1103/PhysRevD.81.123016}
  {\bibfield  {journal} {\bibinfo  {journal} {Phys. Rev.}\ }\textbf {\bibinfo
  {volume} {D81}},\ \bibinfo {pages} {123016} (\bibinfo {year} {2010})},\
  \Eprint {https://arxiv.org/abs/0911.3535} {arXiv:0911.3535 [astro-ph.HE]}
  \BibitemShut {NoStop}%
\bibitem [{\citenamefont {Damour}\ \emph {et~al.}(2012)\citenamefont {Damour},
  \citenamefont {Nagar},\ and\ \citenamefont {Villain}}]{Damour:2012yf}%
  \BibitemOpen
  \bibfield  {author} {\bibinfo {author} {\bibfnamefont {T.}~\bibnamefont
  {Damour}}, \bibinfo {author} {\bibfnamefont {A.}~\bibnamefont {Nagar}},\ and\
  \bibinfo {author} {\bibfnamefont {L.}~\bibnamefont {Villain}},\ }\bibfield
  {title} {\bibinfo {title} {{Measurability of the tidal polarizability of
  neutron stars in late-inspiral gravitational-wave signals}},\ }\href
  {https://doi.org/10.1103/PhysRevD.85.123007} {\bibfield  {journal} {\bibinfo
  {journal} {Phys.Rev.}\ }\textbf {\bibinfo {volume} {D85}},\ \bibinfo {pages}
  {123007} (\bibinfo {year} {2012})},\ \Eprint
  {https://arxiv.org/abs/1203.4352} {arXiv:1203.4352 [gr-qc]} \BibitemShut
  {NoStop}%
\bibitem [{\citenamefont {Agathos}\ \emph {et~al.}(2015)\citenamefont
  {Agathos}, \citenamefont {Meidam}, \citenamefont {Del~Pozzo}, \citenamefont
  {Li}, \citenamefont {Tompitak}, \citenamefont {Veitch}, \citenamefont
  {Vitale},\ and\ \citenamefont {Broeck}}]{Agathos:2015uaa}%
  \BibitemOpen
  \bibfield  {author} {\bibinfo {author} {\bibfnamefont {M.}~\bibnamefont
  {Agathos}}, \bibinfo {author} {\bibfnamefont {J.}~\bibnamefont {Meidam}},
  \bibinfo {author} {\bibfnamefont {W.}~\bibnamefont {Del~Pozzo}}, \bibinfo
  {author} {\bibfnamefont {T.~G.~F.}\ \bibnamefont {Li}}, \bibinfo {author}
  {\bibfnamefont {M.}~\bibnamefont {Tompitak}}, \bibinfo {author}
  {\bibfnamefont {J.}~\bibnamefont {Veitch}}, \bibinfo {author} {\bibfnamefont
  {S.}~\bibnamefont {Vitale}},\ and\ \bibinfo {author} {\bibfnamefont
  {C.~V.~D.}\ \bibnamefont {Broeck}},\ }\bibfield  {title} {\bibinfo {title}
  {{Constraining the neutron star equation of state with gravitational wave
  signals from coalescing binary neutron stars}},\ }\href
  {https://doi.org/10.1103/PhysRevD.92.023012} {\bibfield  {journal} {\bibinfo
  {journal} {Phys. Rev.}\ }\textbf {\bibinfo {volume} {D92}},\ \bibinfo {pages}
  {023012} (\bibinfo {year} {2015})},\ \Eprint
  {https://arxiv.org/abs/1503.05405} {arXiv:1503.05405 [gr-qc]} \BibitemShut
  {NoStop}%
\bibitem [{\citenamefont {Abbott}\ \emph
  {et~al.}(2018{\natexlab{b}})\citenamefont {Abbott} \emph
  {et~al.}}]{Abbott:2018exr}%
  \BibitemOpen
  \bibfield  {author} {\bibinfo {author} {\bibfnamefont {B.~P.}\ \bibnamefont
  {Abbott}} \emph {et~al.} (\bibinfo {collaboration} {LIGO Scientific,
  Virgo}),\ }\bibfield  {title} {\bibinfo {title} {{GW170817: Measurements of
  neutron star radii and equation of state}},\ }\href
  {https://doi.org/10.1103/PhysRevLett.121.161101} {\bibfield  {journal}
  {\bibinfo  {journal} {Phys. Rev. Lett.}\ }\textbf {\bibinfo {volume} {121}},\
  \bibinfo {pages} {161101} (\bibinfo {year} {2018}{\natexlab{b}})},\ \Eprint
  {https://arxiv.org/abs/1805.11581} {arXiv:1805.11581 [gr-qc]} \BibitemShut
  {NoStop}%
\bibitem [{\citenamefont {Gamba}\ \emph {et~al.}(2021)\citenamefont {Gamba},
  \citenamefont {Bernuzzi},\ and\ \citenamefont {Nagar}}]{Gamba:2020ljo}%
  \BibitemOpen
  \bibfield  {author} {\bibinfo {author} {\bibfnamefont {R.}~\bibnamefont
  {Gamba}}, \bibinfo {author} {\bibfnamefont {S.}~\bibnamefont {Bernuzzi}},\
  and\ \bibinfo {author} {\bibfnamefont {A.}~\bibnamefont {Nagar}},\ }\bibfield
   {title} {\bibinfo {title} {{Fast, faithful, frequency-domain
  effective-one-body waveforms for compact binary coalescences}},\ }\href
  {https://doi.org/10.1103/PhysRevD.104.084058} {\bibfield  {journal} {\bibinfo
   {journal} {Phys. Rev. D}\ }\textbf {\bibinfo {volume} {104}},\ \bibinfo
  {pages} {084058} (\bibinfo {year} {2021})},\ \Eprint
  {https://arxiv.org/abs/2012.00027} {arXiv:2012.00027 [gr-qc]} \BibitemShut
  {NoStop}%
\bibitem [{\citenamefont {Buonanno}\ and\ \citenamefont
  {Damour}(1999)}]{Buonanno:1998gg}%
  \BibitemOpen
  \bibfield  {author} {\bibinfo {author} {\bibfnamefont {A.}~\bibnamefont
  {Buonanno}}\ and\ \bibinfo {author} {\bibfnamefont {T.}~\bibnamefont
  {Damour}},\ }\bibfield  {title} {\bibinfo {title} {{Effective one-body
  approach to general relativistic two-body dynamics}},\ }\href
  {https://doi.org/10.1103/PhysRevD.59.084006} {\bibfield  {journal} {\bibinfo
  {journal} {Phys. Rev.}\ }\textbf {\bibinfo {volume} {D59}},\ \bibinfo {pages}
  {084006} (\bibinfo {year} {1999})},\ \Eprint
  {https://arxiv.org/abs/gr-qc/9811091} {arXiv:gr-qc/9811091} \BibitemShut
  {NoStop}%
\bibitem [{\citenamefont {Buonanno}\ and\ \citenamefont
  {Damour}(2000)}]{Buonanno:2000ef}%
  \BibitemOpen
  \bibfield  {author} {\bibinfo {author} {\bibfnamefont {A.}~\bibnamefont
  {Buonanno}}\ and\ \bibinfo {author} {\bibfnamefont {T.}~\bibnamefont
  {Damour}},\ }\bibfield  {title} {\bibinfo {title} {{Transition from inspiral
  to plunge in binary black hole coalescences}},\ }\href
  {https://doi.org/10.1103/PhysRevD.62.064015} {\bibfield  {journal} {\bibinfo
  {journal} {Phys. Rev.}\ }\textbf {\bibinfo {volume} {D62}},\ \bibinfo {pages}
  {064015} (\bibinfo {year} {2000})},\ \Eprint
  {https://arxiv.org/abs/gr-qc/0001013} {arXiv:gr-qc/0001013} \BibitemShut
  {NoStop}%
\bibitem [{\citenamefont {Damour}\ \emph
  {et~al.}(2000{\natexlab{a}})\citenamefont {Damour}, \citenamefont
  {Jaranowski},\ and\ \citenamefont {Schaefer}}]{Damour:2000we}%
  \BibitemOpen
  \bibfield  {author} {\bibinfo {author} {\bibfnamefont {T.}~\bibnamefont
  {Damour}}, \bibinfo {author} {\bibfnamefont {P.}~\bibnamefont {Jaranowski}},\
  and\ \bibinfo {author} {\bibfnamefont {G.}~\bibnamefont {Schaefer}},\
  }\bibfield  {title} {\bibinfo {title} {{On the determination of the last
  stable orbit for circular general relativistic binaries at the third
  postNewtonian approximation}},\ }\href
  {https://doi.org/10.1103/PhysRevD.62.084011} {\bibfield  {journal} {\bibinfo
  {journal} {Phys. Rev.}\ }\textbf {\bibinfo {volume} {D62}},\ \bibinfo {pages}
  {084011} (\bibinfo {year} {2000}{\natexlab{a}})},\ \Eprint
  {https://arxiv.org/abs/gr-qc/0005034} {arXiv:gr-qc/0005034 [gr-qc]}
  \BibitemShut {NoStop}%
\bibitem [{\citenamefont {Damour}\ \emph {et~al.}(2008)\citenamefont {Damour},
  \citenamefont {Jaranowski},\ and\ \citenamefont
  {Sch{\"a}fer}}]{Damour:2008qf}%
  \BibitemOpen
  \bibfield  {author} {\bibinfo {author} {\bibfnamefont {T.}~\bibnamefont
  {Damour}}, \bibinfo {author} {\bibfnamefont {P.}~\bibnamefont {Jaranowski}},\
  and\ \bibinfo {author} {\bibfnamefont {G.}~\bibnamefont {Sch{\"a}fer}},\
  }\bibfield  {title} {\bibinfo {title} {{Effective one body approach to the
  dynamics of two spinning black holes with next-to-leading order spin-orbit
  coupling}},\ }\href {https://doi.org/10.1103/PhysRevD.78.024009} {\bibfield
  {journal} {\bibinfo  {journal} {Phys.Rev.}\ }\textbf {\bibinfo {volume}
  {D78}},\ \bibinfo {pages} {024009} (\bibinfo {year} {2008})},\ \Eprint
  {https://arxiv.org/abs/0803.0915} {arXiv:0803.0915 [gr-qc]} \BibitemShut
  {NoStop}%
\bibitem [{\citenamefont {Damour}\ \emph {et~al.}(2014)\citenamefont {Damour},
  \citenamefont {Jaranowski},\ and\ \citenamefont
  {Sch\"afer}}]{Damour:2014jta}%
  \BibitemOpen
  \bibfield  {author} {\bibinfo {author} {\bibfnamefont {T.}~\bibnamefont
  {Damour}}, \bibinfo {author} {\bibfnamefont {P.}~\bibnamefont {Jaranowski}},\
  and\ \bibinfo {author} {\bibfnamefont {G.}~\bibnamefont {Sch\"afer}},\
  }\bibfield  {title} {\bibinfo {title} {{Nonlocal-in-time action for the
  fourth post-Newtonian conservative dynamics of two-body systems}},\ }\href
  {https://doi.org/10.1103/PhysRevD.89.064058} {\bibfield  {journal} {\bibinfo
  {journal} {Phys. Rev. D}\ }\textbf {\bibinfo {volume} {89}},\ \bibinfo
  {pages} {064058} (\bibinfo {year} {2014})},\ \Eprint
  {https://arxiv.org/abs/1401.4548} {arXiv:1401.4548 [gr-qc]} \BibitemShut
  {NoStop}%
\bibitem [{\citenamefont {Damour}\ \emph {et~al.}(2015)\citenamefont {Damour},
  \citenamefont {Jaranowski},\ and\ \citenamefont
  {Sch\"afer}}]{Damour:2015isa}%
  \BibitemOpen
  \bibfield  {author} {\bibinfo {author} {\bibfnamefont {T.}~\bibnamefont
  {Damour}}, \bibinfo {author} {\bibfnamefont {P.}~\bibnamefont {Jaranowski}},\
  and\ \bibinfo {author} {\bibfnamefont {G.}~\bibnamefont {Sch\"afer}},\
  }\bibfield  {title} {\bibinfo {title} {{Fourth post-Newtonian effective
  one-body dynamics}},\ }\href {https://doi.org/10.1103/PhysRevD.91.084024}
  {\bibfield  {journal} {\bibinfo  {journal} {Phys. Rev. D}\ }\textbf {\bibinfo
  {volume} {91}},\ \bibinfo {pages} {084024} (\bibinfo {year} {2015})},\
  \Eprint {https://arxiv.org/abs/1502.07245} {arXiv:1502.07245 [gr-qc]}
  \BibitemShut {NoStop}%
\bibitem [{\citenamefont {Damour}\ \emph {et~al.}(2016)\citenamefont {Damour},
  \citenamefont {Jaranowski},\ and\ \citenamefont
  {Sch{\"a}fer}}]{Damour:2016abl}%
  \BibitemOpen
  \bibfield  {author} {\bibinfo {author} {\bibfnamefont {T.}~\bibnamefont
  {Damour}}, \bibinfo {author} {\bibfnamefont {P.}~\bibnamefont {Jaranowski}},\
  and\ \bibinfo {author} {\bibfnamefont {G.}~\bibnamefont {Sch{\"a}fer}},\
  }\bibfield  {title} {\bibinfo {title} {{Conservative dynamics of two-body
  systems at the fourth post-Newtonian approximation of general relativity}},\
  }\href {https://doi.org/10.1103/PhysRevD.93.084014} {\bibfield  {journal}
  {\bibinfo  {journal} {Phys. Rev.}\ }\textbf {\bibinfo {volume} {D93}},\
  \bibinfo {pages} {084014} (\bibinfo {year} {2016})},\ \Eprint
  {https://arxiv.org/abs/1601.01283} {arXiv:1601.01283 [gr-qc]} \BibitemShut
  {NoStop}%
\bibitem [{\citenamefont {Juli\'e}(2018)}]{Julie:2017ucp}%
  \BibitemOpen
  \bibfield  {author} {\bibinfo {author} {\bibfnamefont {F.-L.}\ \bibnamefont
  {Juli\'e}},\ }\bibfield  {title} {\bibinfo {title} {{Reducing the two-body
  problem in scalar-tensor theories to the motion of a test particle : a
  scalar-tensor effective-one-body approach}},\ }\href
  {https://doi.org/10.1103/PhysRevD.97.024047} {\bibfield  {journal} {\bibinfo
  {journal} {Phys. Rev. D}\ }\textbf {\bibinfo {volume} {97}},\ \bibinfo
  {pages} {024047} (\bibinfo {year} {2018})},\ \Eprint
  {https://arxiv.org/abs/1709.09742} {arXiv:1709.09742 [gr-qc]} \BibitemShut
  {NoStop}%
\bibitem [{\citenamefont {Juli\'e}\ and\ \citenamefont
  {Deruelle}(2017)}]{Julie:2017pkb}%
  \BibitemOpen
  \bibfield  {author} {\bibinfo {author} {\bibfnamefont {F.-L.}\ \bibnamefont
  {Juli\'e}}\ and\ \bibinfo {author} {\bibfnamefont {N.}~\bibnamefont
  {Deruelle}},\ }\bibfield  {title} {\bibinfo {title} {{Two-body problem in
  Scalar-Tensor theories as a deformation of General Relativity : an
  Effective-One-Body approach}},\ }\href
  {https://doi.org/10.1103/PhysRevD.95.124054} {\bibfield  {journal} {\bibinfo
  {journal} {Phys. Rev. D}\ }\textbf {\bibinfo {volume} {95}},\ \bibinfo
  {pages} {124054} (\bibinfo {year} {2017})},\ \Eprint
  {https://arxiv.org/abs/1703.05360} {arXiv:1703.05360 [gr-qc]} \BibitemShut
  {NoStop}%
\bibitem [{\citenamefont {Julie}\ \emph {et~al.}(2022)\citenamefont {Julie}
  \emph {et~al.}}]{Julieprep}%
  \BibitemOpen
  \bibfield  {author} {\bibinfo {author} {\bibfnamefont {F.-L.}\ \bibnamefont
  {Julie}} \emph {et~al.},\ }\href@noop {} {}\bibinfo {howpublished} {in
  preparation} (\bibinfo {year} {2022})\BibitemShut {NoStop}%
\bibitem [{\citenamefont {{Eardley}}(1975)}]{1975ApJ...196L..59E}%
  \BibitemOpen
  \bibfield  {author} {\bibinfo {author} {\bibfnamefont {D.~M.}\ \bibnamefont
  {{Eardley}}},\ }\bibfield  {title} {\bibinfo {title} {{Observable effects of
  a scalar gravitational field in a binary pulsar.}},\ }\href
  {https://doi.org/10.1086/181744} {\bibfield  {journal} {\bibinfo  {journal}
  {apjl}\ }\textbf {\bibinfo {volume} {196}},\ \bibinfo {pages} {L59} (\bibinfo
  {year} {1975})}\BibitemShut {NoStop}%
\bibitem [{\citenamefont {Schafer}(1984)}]{Schafer:1984mr}%
  \BibitemOpen
  \bibfield  {author} {\bibinfo {author} {\bibfnamefont {G.}~\bibnamefont
  {Schafer}},\ }\bibfield  {title} {\bibinfo {title} {{Acceleration-dependent
  lagrangians in general relativity}},\ }\href
  {https://doi.org/10.1016/0375-9601(84)90947-2} {\bibfield  {journal}
  {\bibinfo  {journal} {Phys. Lett. A}\ }\textbf {\bibinfo {volume} {100}},\
  \bibinfo {pages} {128} (\bibinfo {year} {1984})}\BibitemShut {NoStop}%
\bibitem [{\citenamefont {Damour}\ and\ \citenamefont
  {Schaefer}(1991)}]{Damour:1990jh}%
  \BibitemOpen
  \bibfield  {author} {\bibinfo {author} {\bibfnamefont {T.}~\bibnamefont
  {Damour}}\ and\ \bibinfo {author} {\bibfnamefont {G.}~\bibnamefont
  {Schaefer}},\ }\bibfield  {title} {\bibinfo {title} {{Redefinition of
  position variables and the reduction of higher order Lagrangians}},\ }\href
  {https://doi.org/10.1063/1.529135} {\bibfield  {journal} {\bibinfo  {journal}
  {J. Math. Phys.}\ }\textbf {\bibinfo {volume} {32}},\ \bibinfo {pages} {127}
  (\bibinfo {year} {1991})}\BibitemShut {NoStop}%
\bibitem [{\citenamefont {de~Andrade}\ \emph {et~al.}(2001)\citenamefont
  {de~Andrade}, \citenamefont {Blanchet},\ and\ \citenamefont
  {Faye}}]{deAndrade:2000gf}%
  \BibitemOpen
  \bibfield  {author} {\bibinfo {author} {\bibfnamefont {V.~C.}\ \bibnamefont
  {de~Andrade}}, \bibinfo {author} {\bibfnamefont {L.}~\bibnamefont
  {Blanchet}},\ and\ \bibinfo {author} {\bibfnamefont {G.}~\bibnamefont
  {Faye}},\ }\bibfield  {title} {\bibinfo {title} {{Third postNewtonian
  dynamics of compact binaries: Noetherian conserved quantities and equivalence
  between the harmonic coordinate and ADM Hamiltonian formalisms}},\ }\href
  {https://doi.org/10.1088/0264-9381/18/5/301} {\bibfield  {journal} {\bibinfo
  {journal} {Class. Quant. Grav.}\ }\textbf {\bibinfo {volume} {18}},\ \bibinfo
  {pages} {753} (\bibinfo {year} {2001})},\ \Eprint
  {https://arxiv.org/abs/gr-qc/0011063} {arXiv:gr-qc/0011063} \BibitemShut
  {NoStop}%
\bibitem [{\citenamefont {Damour}\ and\ \citenamefont
  {Sch\"afer}(1985)}]{Damour:1985mt}%
  \BibitemOpen
  \bibfield  {author} {\bibinfo {author} {\bibfnamefont {T.}~\bibnamefont
  {Damour}}\ and\ \bibinfo {author} {\bibfnamefont {G.}~\bibnamefont
  {Sch\"afer}},\ }\bibfield  {title} {\bibinfo {title} {{Lagrangians for$n$
  point masses at the second post-Newtonian approximation of general
  relativity}},\ }\href {https://doi.org/10.1007/BF00773685} {\bibfield
  {journal} {\bibinfo  {journal} {Gen. Rel. Grav.}\ }\textbf {\bibinfo {volume}
  {17}},\ \bibinfo {pages} {879} (\bibinfo {year} {1985})}\BibitemShut
  {NoStop}%
\bibitem [{\citenamefont {Damour}\ \emph
  {et~al.}(2000{\natexlab{b}})\citenamefont {Damour}, \citenamefont
  {Jaranowski},\ and\ \citenamefont {Schaefer}}]{Damour:1999cr}%
  \BibitemOpen
  \bibfield  {author} {\bibinfo {author} {\bibfnamefont {T.}~\bibnamefont
  {Damour}}, \bibinfo {author} {\bibfnamefont {P.}~\bibnamefont {Jaranowski}},\
  and\ \bibinfo {author} {\bibfnamefont {G.}~\bibnamefont {Schaefer}},\
  }\bibfield  {title} {\bibinfo {title} {{Dynamical invariants for general
  relativistic two-body systems at the third postNewtonian approximation}},\
  }\href {https://doi.org/10.1103/PhysRevD.62.044024} {\bibfield  {journal}
  {\bibinfo  {journal} {Phys. Rev. D}\ }\textbf {\bibinfo {volume} {62}},\
  \bibinfo {pages} {044024} (\bibinfo {year} {2000}{\natexlab{b}})},\ \Eprint
  {https://arxiv.org/abs/gr-qc/9912092} {arXiv:gr-qc/9912092} \BibitemShut
  {NoStop}%
\bibitem [{\citenamefont {Blanchet}\ and\ \citenamefont
  {Faye}(2001)}]{Blanchet:2000ub}%
  \BibitemOpen
  \bibfield  {author} {\bibinfo {author} {\bibfnamefont {L.}~\bibnamefont
  {Blanchet}}\ and\ \bibinfo {author} {\bibfnamefont {G.}~\bibnamefont
  {Faye}},\ }\bibfield  {title} {\bibinfo {title} {{General relativistic
  dynamics of compact binaries at the third postNewtonian order}},\ }\href
  {https://doi.org/10.1103/PhysRevD.63.062005} {\bibfield  {journal} {\bibinfo
  {journal} {Phys. Rev. D}\ }\textbf {\bibinfo {volume} {63}},\ \bibinfo
  {pages} {062005} (\bibinfo {year} {2001})},\ \Eprint
  {https://arxiv.org/abs/gr-qc/0007051} {arXiv:gr-qc/0007051} \BibitemShut
  {NoStop}%
\bibitem [{\citenamefont {Hinderer}\ \emph {et~al.}(2013)\citenamefont
  {Hinderer} \emph {et~al.}}]{Hinderer:2013uwa}%
  \BibitemOpen
  \bibfield  {author} {\bibinfo {author} {\bibfnamefont {T.}~\bibnamefont
  {Hinderer}} \emph {et~al.},\ }\bibfield  {title} {\bibinfo {title}
  {{Periastron advance in spinning black hole binaries: comparing
  effective-one-body and Numerical Relativity}},\ }\href
  {https://doi.org/10.1103/PhysRevD.88.084005} {\bibfield  {journal} {\bibinfo
  {journal} {Phys. Rev.}\ }\textbf {\bibinfo {volume} {D88}},\ \bibinfo {pages}
  {084005} (\bibinfo {year} {2013})},\ \Eprint
  {https://arxiv.org/abs/1309.0544} {arXiv:1309.0544 [gr-qc]} \BibitemShut
  {NoStop}%
\bibitem [{\citenamefont {Bl\"umlein}\ \emph {et~al.}(2022)\citenamefont
  {Bl\"umlein}, \citenamefont {Maier}, \citenamefont {Marquard},\ and\
  \citenamefont {Sch\"afer}}]{Blumlein:2021txe}%
  \BibitemOpen
  \bibfield  {author} {\bibinfo {author} {\bibfnamefont {J.}~\bibnamefont
  {Bl\"umlein}}, \bibinfo {author} {\bibfnamefont {A.}~\bibnamefont {Maier}},
  \bibinfo {author} {\bibfnamefont {P.}~\bibnamefont {Marquard}},\ and\
  \bibinfo {author} {\bibfnamefont {G.}~\bibnamefont {Sch\"afer}},\ }\bibfield
  {title} {\bibinfo {title} {{The fifth-order post-Newtonian Hamiltonian
  dynamics of two-body systems from an effective field theory approach}},\
  }\href {https://doi.org/10.1016/j.nuclphysb.2022.115900} {\bibfield
  {journal} {\bibinfo  {journal} {Nucl. Phys. B}\ }\textbf {\bibinfo {volume}
  {983}},\ \bibinfo {pages} {115900} (\bibinfo {year} {2022})},\ \Eprint
  {https://arxiv.org/abs/2110.13822} {arXiv:2110.13822 [gr-qc]} \BibitemShut
  {NoStop}%
\bibitem [{\citenamefont {Damour}(2016)}]{Damour:2016gwp}%
  \BibitemOpen
  \bibfield  {author} {\bibinfo {author} {\bibfnamefont {T.}~\bibnamefont
  {Damour}},\ }\bibfield  {title} {\bibinfo {title} {{Gravitational scattering,
  post-Minkowskian approximation and Effective One-Body theory}},\ }\href
  {https://doi.org/10.1103/PhysRevD.94.104015} {\bibfield  {journal} {\bibinfo
  {journal} {Phys. Rev.}\ }\textbf {\bibinfo {volume} {D94}},\ \bibinfo {pages}
  {104015} (\bibinfo {year} {2016})},\ \Eprint
  {https://arxiv.org/abs/1609.00354} {arXiv:1609.00354 [gr-qc]} \BibitemShut
  {NoStop}%
\bibitem [{\citenamefont {Damour}(2018)}]{Damour:2017zjx}%
  \BibitemOpen
  \bibfield  {author} {\bibinfo {author} {\bibfnamefont {T.}~\bibnamefont
  {Damour}},\ }\bibfield  {title} {\bibinfo {title} {{High-energy gravitational
  scattering and the general relativistic two-body problem}},\ }\href
  {https://doi.org/10.1103/PhysRevD.97.044038} {\bibfield  {journal} {\bibinfo
  {journal} {Phys. Rev.}\ }\textbf {\bibinfo {volume} {D97}},\ \bibinfo {pages}
  {044038} (\bibinfo {year} {2018})},\ \Eprint
  {https://arxiv.org/abs/1710.10599} {arXiv:1710.10599 [gr-qc]} \BibitemShut
  {NoStop}%
\bibitem [{\citenamefont {Damour}(2020)}]{Damour:2019lcq}%
  \BibitemOpen
  \bibfield  {author} {\bibinfo {author} {\bibfnamefont {T.}~\bibnamefont
  {Damour}},\ }\bibfield  {title} {\bibinfo {title} {{Classical and quantum
  scattering in post-Minkowskian gravity}},\ }\href
  {https://doi.org/10.1103/PhysRevD.102.024060} {\bibfield  {journal} {\bibinfo
   {journal} {Phys. Rev. D}\ }\textbf {\bibinfo {volume} {102}},\ \bibinfo
  {pages} {024060} (\bibinfo {year} {2020})},\ \Eprint
  {https://arxiv.org/abs/1912.02139} {arXiv:1912.02139 [gr-qc]} \BibitemShut
  {NoStop}%
\bibitem [{\citenamefont {Antonelli}\ \emph {et~al.}(2019)\citenamefont
  {Antonelli}, \citenamefont {Buonanno}, \citenamefont {Steinhoff},
  \citenamefont {van~de Meent},\ and\ \citenamefont
  {Vines}}]{Antonelli:2019ytb}%
  \BibitemOpen
  \bibfield  {author} {\bibinfo {author} {\bibfnamefont {A.}~\bibnamefont
  {Antonelli}}, \bibinfo {author} {\bibfnamefont {A.}~\bibnamefont {Buonanno}},
  \bibinfo {author} {\bibfnamefont {J.}~\bibnamefont {Steinhoff}}, \bibinfo
  {author} {\bibfnamefont {M.}~\bibnamefont {van~de Meent}},\ and\ \bibinfo
  {author} {\bibfnamefont {J.}~\bibnamefont {Vines}},\ }\bibfield  {title}
  {\bibinfo {title} {{Energetics of two-body Hamiltonians in post-Minkowskian
  gravity}},\ }\href {https://doi.org/10.1103/PhysRevD.99.104004} {\bibfield
  {journal} {\bibinfo  {journal} {Phys. Rev.}\ }\textbf {\bibinfo {volume}
  {D99}},\ \bibinfo {pages} {104004} (\bibinfo {year} {2019})},\ \Eprint
  {https://arxiv.org/abs/1901.07102} {arXiv:1901.07102 [gr-qc]} \BibitemShut
  {NoStop}%
\bibitem [{\citenamefont {Antonelli}\ \emph {et~al.}(2020)\citenamefont
  {Antonelli}, \citenamefont {van~de Meent}, \citenamefont {Buonanno},
  \citenamefont {Steinhoff},\ and\ \citenamefont {Vines}}]{Antonelli:2019fmq}%
  \BibitemOpen
  \bibfield  {author} {\bibinfo {author} {\bibfnamefont {A.}~\bibnamefont
  {Antonelli}}, \bibinfo {author} {\bibfnamefont {M.}~\bibnamefont {van~de
  Meent}}, \bibinfo {author} {\bibfnamefont {A.}~\bibnamefont {Buonanno}},
  \bibinfo {author} {\bibfnamefont {J.}~\bibnamefont {Steinhoff}},\ and\
  \bibinfo {author} {\bibfnamefont {J.}~\bibnamefont {Vines}},\ }\bibfield
  {title} {\bibinfo {title} {{Quasicircular inspirals and plunges from
  nonspinning effective-one-body Hamiltonians with gravitational self-force
  information}},\ }\href {https://doi.org/10.1103/PhysRevD.101.024024}
  {\bibfield  {journal} {\bibinfo  {journal} {Phys. Rev.}\ }\textbf {\bibinfo
  {volume} {D101}},\ \bibinfo {pages} {024024} (\bibinfo {year} {2020})},\
  \Eprint {https://arxiv.org/abs/1907.11597} {arXiv:1907.11597 [gr-qc]}
  \BibitemShut {NoStop}%
\bibitem [{\citenamefont {Khalil}\ \emph
  {et~al.}(2022{\natexlab{b}})\citenamefont {Khalil}, \citenamefont {Buonanno},
  \citenamefont {Steinhoff},\ and\ \citenamefont {Vines}}]{Khalil:2022ylj}%
  \BibitemOpen
  \bibfield  {author} {\bibinfo {author} {\bibfnamefont {M.}~\bibnamefont
  {Khalil}}, \bibinfo {author} {\bibfnamefont {A.}~\bibnamefont {Buonanno}},
  \bibinfo {author} {\bibfnamefont {J.}~\bibnamefont {Steinhoff}},\ and\
  \bibinfo {author} {\bibfnamefont {J.}~\bibnamefont {Vines}},\ }\bibfield
  {title} {\bibinfo {title} {{Energetics and scattering of gravitational
  two-body systems at fourth post-Minkowskian order}},\ }\href@noop {} {\
  (\bibinfo {year} {2022}{\natexlab{b}})},\ \Eprint
  {https://arxiv.org/abs/2204.05047} {arXiv:2204.05047 [gr-qc]} \BibitemShut
  {NoStop}%
\bibitem [{\citenamefont {Damour}\ and\ \citenamefont
  {Rettegno}(2022)}]{Damour:2022ybd}%
  \BibitemOpen
  \bibfield  {author} {\bibinfo {author} {\bibfnamefont {T.}~\bibnamefont
  {Damour}}\ and\ \bibinfo {author} {\bibfnamefont {P.}~\bibnamefont
  {Rettegno}},\ }\bibfield  {title} {\bibinfo {title} {{Strong-field scattering
  of two black holes: Numerical Relativity meets Post-Minkowskian gravity}},\
  }\href@noop {} {\  (\bibinfo {year} {2022})},\ \Eprint
  {https://arxiv.org/abs/2211.01399} {arXiv:2211.01399 [gr-qc]} \BibitemShut
  {NoStop}%
\bibitem [{\citenamefont {Bini}\ \emph
  {et~al.}(2020{\natexlab{a}})\citenamefont {Bini}, \citenamefont {Damour},\
  and\ \citenamefont {Geralico}}]{Bini:2020nsb}%
  \BibitemOpen
  \bibfield  {author} {\bibinfo {author} {\bibfnamefont {D.}~\bibnamefont
  {Bini}}, \bibinfo {author} {\bibfnamefont {T.}~\bibnamefont {Damour}},\ and\
  \bibinfo {author} {\bibfnamefont {A.}~\bibnamefont {Geralico}},\ }\bibfield
  {title} {\bibinfo {title} {{Sixth post-Newtonian local-in-time dynamics of
  binary systems}},\ }\href {https://doi.org/10.1103/PhysRevD.102.024061}
  {\bibfield  {journal} {\bibinfo  {journal} {Phys. Rev. D}\ }\textbf {\bibinfo
  {volume} {102}},\ \bibinfo {pages} {024061} (\bibinfo {year}
  {2020}{\natexlab{a}})},\ \Eprint {https://arxiv.org/abs/2004.05407}
  {arXiv:2004.05407 [gr-qc]} \BibitemShut {NoStop}%
\bibitem [{\citenamefont {Bini}\ \emph
  {et~al.}(2020{\natexlab{b}})\citenamefont {Bini}, \citenamefont {Damour},\
  and\ \citenamefont {Geralico}}]{Bini:2020hmy}%
  \BibitemOpen
  \bibfield  {author} {\bibinfo {author} {\bibfnamefont {D.}~\bibnamefont
  {Bini}}, \bibinfo {author} {\bibfnamefont {T.}~\bibnamefont {Damour}},\ and\
  \bibinfo {author} {\bibfnamefont {A.}~\bibnamefont {Geralico}},\ }\bibfield
  {title} {\bibinfo {title} {{Sixth post-Newtonian nonlocal-in-time dynamics of
  binary systems}},\ }\href@noop {} {\  (\bibinfo {year}
  {2020}{\natexlab{b}})},\ \Eprint {https://arxiv.org/abs/2007.11239}
  {arXiv:2007.11239 [gr-qc]} \BibitemShut {NoStop}%
\bibitem [{\citenamefont {Nagar}\ \emph {et~al.}(2020)\citenamefont {Nagar},
  \citenamefont {Riemenschneider}, \citenamefont {Pratten}, \citenamefont
  {Rettegno},\ and\ \citenamefont {Messina}}]{Nagar:2020pcj}%
  \BibitemOpen
  \bibfield  {author} {\bibinfo {author} {\bibfnamefont {A.}~\bibnamefont
  {Nagar}}, \bibinfo {author} {\bibfnamefont {G.}~\bibnamefont
  {Riemenschneider}}, \bibinfo {author} {\bibfnamefont {G.}~\bibnamefont
  {Pratten}}, \bibinfo {author} {\bibfnamefont {P.}~\bibnamefont {Rettegno}},\
  and\ \bibinfo {author} {\bibfnamefont {F.}~\bibnamefont {Messina}},\
  }\bibfield  {title} {\bibinfo {title} {{Multipolar effective one body
  waveform model for spin-aligned black hole binaries}},\ }\href
  {https://doi.org/10.1103/PhysRevD.102.024077} {\bibfield  {journal} {\bibinfo
   {journal} {Phys. Rev. D}\ }\textbf {\bibinfo {volume} {102}},\ \bibinfo
  {pages} {024077} (\bibinfo {year} {2020})},\ \Eprint
  {https://arxiv.org/abs/2001.09082} {arXiv:2001.09082 [gr-qc]} \BibitemShut
  {NoStop}%
\bibitem [{\citenamefont {Schmidt}\ \emph {et~al.}(2021)\citenamefont
  {Schmidt}, \citenamefont {Breschi}, \citenamefont {Gamba}, \citenamefont
  {Pagano}, \citenamefont {Rettegno}, \citenamefont {Riemenschneider},
  \citenamefont {Bernuzzi}, \citenamefont {Nagar},\ and\ \citenamefont
  {Del~Pozzo}}]{Schmidt:2020yuu}%
  \BibitemOpen
  \bibfield  {author} {\bibinfo {author} {\bibfnamefont {S.}~\bibnamefont
  {Schmidt}}, \bibinfo {author} {\bibfnamefont {M.}~\bibnamefont {Breschi}},
  \bibinfo {author} {\bibfnamefont {R.}~\bibnamefont {Gamba}}, \bibinfo
  {author} {\bibfnamefont {G.}~\bibnamefont {Pagano}}, \bibinfo {author}
  {\bibfnamefont {P.}~\bibnamefont {Rettegno}}, \bibinfo {author}
  {\bibfnamefont {G.}~\bibnamefont {Riemenschneider}}, \bibinfo {author}
  {\bibfnamefont {S.}~\bibnamefont {Bernuzzi}}, \bibinfo {author}
  {\bibfnamefont {A.}~\bibnamefont {Nagar}},\ and\ \bibinfo {author}
  {\bibfnamefont {W.}~\bibnamefont {Del~Pozzo}},\ }\bibfield  {title} {\bibinfo
  {title} {{Machine Learning Gravitational Waves from Binary Black Hole
  Mergers}},\ }\href {https://doi.org/10.1103/PhysRevD.103.043020} {\bibfield
  {journal} {\bibinfo  {journal} {Phys. Rev. D}\ }\textbf {\bibinfo {volume}
  {103}},\ \bibinfo {pages} {043020} (\bibinfo {year} {2021})},\ \Eprint
  {https://arxiv.org/abs/2011.01958} {arXiv:2011.01958 [gr-qc]} \BibitemShut
  {NoStop}%
\bibitem [{\citenamefont {Riemenschneider}\ \emph {et~al.}(2021)\citenamefont
  {Riemenschneider}, \citenamefont {Rettegno}, \citenamefont {Breschi},
  \citenamefont {Albertini}, \citenamefont {Gamba}, \citenamefont {Bernuzzi},\
  and\ \citenamefont {Nagar}}]{Riemenschneider:2021ppj}%
  \BibitemOpen
  \bibfield  {author} {\bibinfo {author} {\bibfnamefont {G.}~\bibnamefont
  {Riemenschneider}}, \bibinfo {author} {\bibfnamefont {P.}~\bibnamefont
  {Rettegno}}, \bibinfo {author} {\bibfnamefont {M.}~\bibnamefont {Breschi}},
  \bibinfo {author} {\bibfnamefont {A.}~\bibnamefont {Albertini}}, \bibinfo
  {author} {\bibfnamefont {R.}~\bibnamefont {Gamba}}, \bibinfo {author}
  {\bibfnamefont {S.}~\bibnamefont {Bernuzzi}},\ and\ \bibinfo {author}
  {\bibfnamefont {A.}~\bibnamefont {Nagar}},\ }\bibfield  {title} {\bibinfo
  {title} {{Assessment of consistent next-to-quasicircular corrections and
  postadiabatic approximation in effective-one-body multipolar waveforms for
  binary black hole coalescences}},\ }\href
  {https://doi.org/10.1103/PhysRevD.104.104045} {\bibfield  {journal} {\bibinfo
   {journal} {Phys. Rev. D}\ }\textbf {\bibinfo {volume} {104}},\ \bibinfo
  {pages} {104045} (\bibinfo {year} {2021})},\ \Eprint
  {https://arxiv.org/abs/2104.07533} {arXiv:2104.07533 [gr-qc]} \BibitemShut
  {NoStop}%
\bibitem [{\citenamefont {Gamba}\ \emph {et~al.}(2022)\citenamefont {Gamba},
  \citenamefont {Ak\c{c}ay}, \citenamefont {Bernuzzi},\ and\ \citenamefont
  {Williams}}]{Gamba:2021ydi}%
  \BibitemOpen
  \bibfield  {author} {\bibinfo {author} {\bibfnamefont {R.}~\bibnamefont
  {Gamba}}, \bibinfo {author} {\bibfnamefont {S.}~\bibnamefont {Ak\c{c}ay}},
  \bibinfo {author} {\bibfnamefont {S.}~\bibnamefont {Bernuzzi}},\ and\
  \bibinfo {author} {\bibfnamefont {J.}~\bibnamefont {Williams}},\ }\bibfield
  {title} {\bibinfo {title} {{Effective-one-body waveforms for precessing
  coalescing compact binaries with post-Newtonian twist}},\ }\href
  {https://doi.org/10.1103/PhysRevD.106.024020} {\bibfield  {journal} {\bibinfo
   {journal} {Phys. Rev. D}\ }\textbf {\bibinfo {volume} {106}},\ \bibinfo
  {pages} {024020} (\bibinfo {year} {2022})},\ \Eprint
  {https://arxiv.org/abs/2111.03675} {arXiv:2111.03675 [gr-qc]} \BibitemShut
  {NoStop}%
\bibitem [{\citenamefont {Gamba}\ and\ \citenamefont
  {Bernuzzi}(2022)}]{Gamba:2022mgx}%
  \BibitemOpen
  \bibfield  {author} {\bibinfo {author} {\bibfnamefont {R.}~\bibnamefont
  {Gamba}}\ and\ \bibinfo {author} {\bibfnamefont {S.}~\bibnamefont
  {Bernuzzi}},\ }\bibfield  {title} {\bibinfo {title} {{Resonant tides in
  binary neutron star mergers: analytical-numerical relativity study}},\
  }\href@noop {} {\  (\bibinfo {year} {2022})},\ \Eprint
  {https://arxiv.org/abs/2207.13106} {arXiv:2207.13106 [gr-qc]} \BibitemShut
  {NoStop}%
\end{thebibliography}%

\end{document}